\documentclass[acmsmall,nonacm]{acmart}

\AtBeginDocument{%
  \providecommand\BibTeX{{%
    \normalfont B\kern-0.5em{\scshape i\kern-0.25em b}\kern-0.8em\TeX}}}

\usepackage{todonotes}

\usepackage[mathscr]{euscript}
\usepackage[capitalise]{cleveref}

\AtBeginEnvironment{lemma}{\crefalias{theorem}{lemma}}
\AtBeginEnvironment{corollary}{\crefalias{theorem}{corollary}}
\AtBeginEnvironment{proposition}{\crefalias{theorem}{proposition}}
\AtBeginEnvironment{definition}{\crefalias{theorem}{definition}}
\AtBeginEnvironment{example}{\crefalias{theorem}{example}}

\usepackage{paralist}

\usepackage{tikz}
\usetikzlibrary{decorations.pathreplacing,calligraphy}

\usepackage{verbatim}
\usepackage[utf8x]{inputenc}
\usepackage{fancyvrb}
\usepackage{textcomp}
\usepackage{pmboxdraw} %

\graphicspath{ {./figures/} }

\newcommand{\hubie}[1]{\todo[inline,color=red!50]{Hubie: #1}}
\newcommand{\fabio}[1]{\todo[inline,color=green!50]{Fabio: #1}}
\newcommand{\sam}[1]{\todo[inline]{Sam: #1}}

\newcommand{\para}[1]{\medskip \noindent \textbf{#1.}}

\renewcommand{\hubie}[1]{} \renewcommand{\fabio}[1]{} \renewcommand{\sam}[1]{}

\newenvironment{sketch}{%
  \proof}{\endproof}

\newenvironment{claimproof}{%
  \proof}{\hfill $\lhd$}

\newtheorem{observation}{Observation}
\newtheorem{claim}{Claim}

\theoremstyle{definition}
\newtheorem{notation}{Notation}

\mathchardef\mhyphen="2D

\newcommand{\proofsubparagraph}[1]{\textbf{\emph{#1}}}

\newcommand{\subs}{\sigma}

\newcommand{\domain}{\mathsf{dom}}
\newcommand{\range}{\mathsf{ran}}
\newcommand{\cq}{\ifmmode{\mathsf{CQ}}\else{\textsf{CQ}}\xspace\fi}
\newcommand{\ucq}{\ifmmode{\mathsf{UCQ}}\else{\textsf{UCQ}}\xspace\fi}

\newcommand{\union}{\mathrel{\cup}}
\newcommand{\intersect}{\mathrel{\cap}}
\newcommand{\bigland}{\bigwedge}

\newcommand{\df}{:=}

\newcommand{\regions}{\mathsf{regions}}
\newcommand{\tw}{\mathsf{tw}}
\newcommand{\logwidth}{\mathsf{logwidth}}
\newcommand{\width}{\mathsf{width}}
\newcommand{\free}{\mathsf{free}}

\title{Tight Width Minimisation of Self-Join-Free Quantified Conjunctive Queries}

\author{Hubie Chen}
\orcid{0009-0005-4025-8086}
\email{hubie.chen@kcl.ac.uk}
\affiliation{%
  \institution{King's College London}
  \city{London}
  \country{UK}
}

\author{Fabio Mogavero}
\orcid{0000-0002-5140-5783}
\email{fabio.mogavero@unina.it}
\affiliation{%
  \institution{Università degli Studi di Napoli “Federico II”}
  \city{Napoli}
  \country{Italy}
}

\author{Sam M. Thompson}
\orcid{0000-0002-3476-6739}
\email{S.M.Thompson2@liverpool.ac.uk}
\affiliation{%
  \institution{University of Liverpool}
  \city{Liverpool}
  \country{UK}
}

\begin{document}

\begin{abstract}
\emph{Query minimisation} is a classical problem in database theory and the
\emph{width} of a formula, defined as the maximum number of free variables over
all subformulae, is a central measure for query evaluation.
We study \emph{width minimisation} for a subclass of \emph{quantified
conjunctive queries} (\emph{QCQs}), namely \emph{self-join-free QCQs}.
For this subclass, we show that rewriting optimality coincides with semantic
optimality: given a query $\phi$, the minimum-width query obtainable from $\phi$
by applying natural and established rewriting rules has minimum width among all
first-order formulae logically equivalent to $\phi$.
As a consequence, the width of a self-join-free QCQ can be optimised by a
logical-equivalence-preserving rewriting algorithm.
\end{abstract}



\maketitle

\section{Introduction}

\para{Background}
The problem of \emph{minimising} a query while preserving \emph{logical
equivalence} is a classical and central topic in database theory.
Indeed, the notions of equivalence, containment, and minimisation were already
identified in the 1970s~\cite{CM77} as giving rise to fundamental algorithmic
problems underlying \emph{conjunctive queries} (\emph{CQs}).
Since then, a rich line of work has studied these questions in close connection
with \emph{constraint satisfaction}, \emph{homomorphisms}, and
\emph{finite-variable
logics}~\cite{KV00,DKV02,ASU79,ASU79a,IR95,CR00,Gro07,BLR14,GSS01,ChenMueller15-fineclass,ChenMueller17-hierarchy,KLNOZ25}.
In this line of research, a central structural measure is the~\emph{width} of a
formula $\varphi$, denoted by $\width(\varphi)$, which is the maximum number of
free variables over all
subformulae~\cite{BC14,BC17,Wil14,GSS01,Gro07,AW12,CM26}.
Using standard variable renaming, in ordinary one-sorted first-order logic, the
minimum width of a sentence $\phi$ over all formulae logically equivalent to $\phi$, which we
call the \emph{logical width} and denote by $\logwidth(\phi)$, coincides with
the minimum number of variables needed to express~$\phi$~\cite{Imm82a}.
This coincidence does not hold in general in many-sorted variants, where sorts
can restrict variable reuse.
Width is therefore a more general structural parameter in such settings,
particularly for lower-bound arguments, and we take it as our reference measure
throughout this work.
The natural bottom-up evaluation algorithm for first-order logic (on finite
structures) has an exponential dependence on the width~\cite{Wil14} and can be
performed in polynomial time for sentences of \emph{bounded width}.
Moreover, when the arity of relations is fixed, there are dichotomy theorems
showing that bounded width is the exclusive explanation for the tractability of
classes of formulae~\cite{GSS01,Gro07,ChenMueller15-fineclass,ChenMueller17-hierarchy,Che17}.
Thus, width is a fundamental measure in query~evaluation.

A strongly related measure associated with CQs, and generalisations thereof, is
the treewidth~\cite{RS86} of the \emph{query hypergraph}: each atom induces a
hyperedge on its variables.
This measure has been used %
to identify structurally tractable classes of formulae and related evaluation
problems~\cite{KV00,DKV02,GLS02}.
The interaction between width and treewidth in this context, as well as the
study of other measures of the query hypergraph, has proved central to modern
approaches to \emph{query evaluation}, \emph{semantic optimisation}, and
\emph{query rewriting}~\cite{BPR17,CM26,CR00,GLS02,GSS01,Gro07,Mar10}.
However, \emph{width minimisation} of a first-order query up to logical
equivalence is noncomputable in general~\cite{AW12}, and already noncomputable
in positive logic~\cite{BC14}.
This motivates the search for exact minimisation results on robust subclasses of
queries~\cite{BC19}, as well as %
characterisations of the minimum width obtainable under rewritings~\cite{CM26}.

\para{Quantified Conjunctive Queries}
The aim of this article is to go beyond the purely existential case by
considering width minimisation for \emph{quantified conjunctive queries}
(\emph{QCQs}).
These naturally extend CQs by allowing both existential and universal
quantifiers, while keeping conjunction as the only Boolean connective.\footnote{We do not assume QCQs are necessarily in prenex normal form.}
In this sense, QCQs are closely related to the \emph{quantified constraint
satisfaction problem} (\emph{QCSP}) and, more generally, to \emph{positive Horn
logic}~\cite{Che08,BBCJK09}.
This connection has been fruitfully exploited in the study of \emph{containment
problems}: for QCSPs, \emph{inclusion between templates} admits homomorphic and
algebraic characterisations~\cite{CMM08,CMM15}; for QCQs, \emph{entailment}
admits a Chandra--Merlin style characterisation via Skolem homomorphisms, is
NP-complete, and coincides with its finite variant~\cite{BM17}.
The \emph{model-checking problem} for QCQs has also been studied through
parameters that generalise treewidth to account for the interaction between the
quantifier prefix and the \emph{query hypergraph}~\cite{CD16}.
Thus, QCQs combine the structural discipline of CQs with the additional
expressive power brought by the presence of both existential and universal
quantification and the resulting quantifier alternation.
This makes minimisation questions for QCQs particularly compelling.
In the existential-positive setting,
when the goal is to optimise a given existential-positive query, optimising
within the class of existential-positive queries is equivalent to optimising
within the class of all first-order queries~\cite{BC19}.\footnote{In contrast, existing work shows that analysing the complexity of
equivalence, entailment, and minimisation becomes more subtle when one moves
from \emph{conjunctive queries} to existential-positive logic~\cite{BC15}.}
This raises the question of whether a similar phenomenon holds for QCQs
and whether the logical width of QCQs can be analysed.

We take a first step in this direction by considering the width minimisation of
\emph{self-join-free QCQs}, namely QCQs where no relation symbol is repeated.
This setting captures the essential interaction between width and quantifier
alternation, while providing a clean stepping stone towards understanding
general QCQs.
Minimisation within the class of CQs is a classical issue~\cite{CM77}; for width
minimisation, after passing to the \emph{core}, the minimal width is controlled
by treewidth~\cite{BC19}.
In contrast, even for self-join-free QCQs, prior to this work, there was no
transparent method for minimising width within the class of QCQs itself, let
alone over the class of first-order formulae.

\para{Rewriting}
To address the problem of width minimisation for QCQs, we consider a
\emph{rewriting framework} developed recently by Chen and Mengel~\cite{CM26}.
Informally, this framework is based on a set of rewriting rules $\mathbb{A}$
that are known to preserve logical equivalence, and in the case of QCQs, contain
rules that reorder quantifiers of the same kind, push existential quantifiers
towards the atoms,
split universal quantification over conjunctions, and remove vacuous
quantifiers.
These rules organise a QCQ into alternating \emph{regions}: \emph{$\{ \exists,
\land \}$-regions}, built from existential quantifiers and conjunctions, and
\emph{$\{ \forall \}$-regions}, built from universal quantifiers only.
Each region $\psi$ has an associated hypergraph~$H_{\psi}$.
Its hyperedges record how $\psi$ interacts both with its immediate subregions
and the context above it: for each (potentially atomic) child region, there is a
hyperedge containing the variables of said child, and there is a further hyperedge
containing all the free variables of $\psi$.

This rewriting framework~\cite{CM26} provides an algorithm that, upon input of a
positive first-order formula~$\varphi$, outputs a formula $\varphi^{+}$ of
minimum width within the class of all formulae obtainable from~$\varphi$ via the
rewriting rules; this class is denoted by $[\varphi]_{\mathbb{A}}$.
The width of this output can be characterised (in a certain precise sense) as
the maximum treewidth over all regions.

The following examples suggest why, in the presence of universal quantifiers, a
measure more sophisticated than the treewidth of the query is needed to
characterise its logical width, and also anticipate some of the complexity
involved in showing width lower bounds.
First, consider %
\[
  \varphi_{1}
=
  \exists x_{1} \cdots \exists x_{l} \exists x {\bigwedge}_{i = 1}^{l} R_{i}(x,
  x_{i})
\quad \equiv \quad
  \exists x \exists x_{1} \cdots \exists x_{l} {\bigwedge}_{i = 1}^{l} R_{i}(x,
  x_{i})
\quad \equiv \quad
  \exists x {\bigwedge}_{i = 1}^{l} \exists x_{i} R_{i}(x, x_{i}).
\]
Its query hypergraph is clearly a star, and hence has treewidth $1$; as shown
above, the query can be rewritten into one of width $2$, by reordering the
quantifiers and pushing each $\exists x_{i}$ in front of the unique atom in
which $x_{i}$ occurs.
This is the kind of behaviour captured by the existential-positive theory of
variable minimisation~\cite{BC19}.
Next, consider instead the QCQ
\[
  \varphi_{2}
=
  \forall y_{1} \cdots \forall y_{m} \, \psi_{\exists} \,,
\;\text{ where }\,
  \psi_{\exists}
=
  \exists x {\bigwedge}_{j = 1}^{m}
  S_{j}(x, y_{j}) \,.
\]
Although the query hypergraph is again a star, the universal variables $y_{j}$,
intuitively speaking, must remain simultaneously available when the existential
witness $x$ is chosen.
There is a hypergraph associated with the formula $\psi_{\exists}$; this
hypergraph accounts for the universal variables by including a top-boundary
hyperedge that contains all of its free variables, namely, $\{ y_{1}, \ldots,
y_{m} \}$.
The relevant treewidth is no longer that of the query hypergraph, but of a
larger hypergraph that includes this boundary hyperedge.
Indeed, this article shows this formula to have logical width $m + 1$.
Finally, the combination of the above two situations gives formulae, for $m >
0$, of the form
\[
  \varphi_{3}
=
  \exists x_{1} \cdots \exists x_{l} \, \psi_{\forall} \,,
\;\text{ where }\,
  \psi_{\forall}
=
  \forall y_{1} \cdots \forall y_{m} \, \psi_{\exists}
\,\text{ and }\,
  \psi_{\exists}
=
  \exists x \left( {\bigwedge}_{i = 1}^{l} R_{i}(x, x_{i}) \wedge {\bigwedge}_{j
  = 1}^{m} S_{j}(x, y_{j}) \right).
\]
Intuitively speaking, here the universal variables $y_j$ depend on the
existential variables $x_i$, and so all of these variables must be kept in scope
when considering how to assign $x$.
This is accounted for in the hypergraph associated to $\psi_{\exists}$, where
the top-boundary hyperedge contains all variables but $x$; as this article
shows, for these formulae the logical width is $l + m + 1$.
These examples point to why the theory of CQs does not transfer directly.

\para{Contributions}
Our main result shows that, for self-join-free QCQs,
the minimum width over all formulae in $[\varphi]_{\mathbb{A}}$ is in fact the
logical width of $\varphi$.

\begin{theorem}[Main Result]
\label{thm:width-treewidth-intro}
  Let $\varphi$ be a self-join-free QCQ and $\varphi^{+}$ a self-join-free QCQ
  of minimum width among all formulae in $[\varphi]_{\mathbb{A}}$.
  Then, $\logwidth(\varphi) = \width(\varphi^{+})$.
  Moreover, there is an algorithm that, given a self-join-free QCQ $\varphi$, computes such a QCQ $\varphi^{+}$
  in polynomial time, assuming access to an oracle for computing minimum-width
  tree decompositions.
\end{theorem}

This result also gives a complexity description of width computation.
Given a self-join-free QCQ $\varphi$ and a number $k \in \mathbb{N}$, deciding
whether $\logwidth(\varphi) \leq k$ is NP-complete.
Moreover, computing the logical width of such a query is polynomial-time Turing equivalent to
computing graph treewidth.  See~\cref{cor:complexity} for details on these results.

To reformulate this main theorem (\cref{thm:width-treewidth-intro}), for self-join-free QCQs, logical width admits exact algorithmic
optimisation via syntactic rewriting, and thus the width found by rewriting is
not merely an artefact, but rather the true logical width of the query.
An alternative reading of this theorem is that no width
blow-up is incurred when passing from full first-order logic back to
self-join-free QCQs: when a first-order formula of width $k$ is equivalent to a
self-join-free QCQ, it is equivalent to a self-join-free QCQ of width $k$.
This extends a tightness phenomenon previously known for existential-positive
queries~\cite{BC19,Ros25}.

Regarding the theorem's statement, the inequality $\logwidth(\varphi) \leq \width(\varphi^{+})$ is immediate, since~$\varphi^{+}$ is logically equivalent to $\varphi$.
The difficult part is the reverse inequality, since the rewriting algorithm only
gives an \emph{a priori} syntactic upper bound: it constructs an equivalent QCQ
whose width is governed by the maximum value of the treewidth (plus one) over
its regions; the algorithm does not, by itself, rule out the possibility of a
completely different first-order formula of smaller width.
To prove semantic optimality, we focus on a region having maximum treewidth, and
use it to show a lower bound on finite-variable definability.
Here \emph{pebble games}~\cite{Imm82a}, a variant of
\emph{Ehrenfeucht-Fra\"iss\'e games}~\cite{EF95,GKLMSVVW05}, enter the picture:
by the classical characterisation of finite-variable first-order logic, to show
that $\logwidth(\varphi) > k$, it suffices to construct two finite structures
$\mathcal{A}$ and $\mathcal{B}$ that are separated by $\varphi$, but
indistinguishable by first-order formulae using only $k$ variables; this is the
route we take.
We first carry out these constructions in variable-sorted logic, where each
query variable has its own domain.
This lets us choose these domains independently and forces pebble-game responses
to remain in the domain associated with the relevant variable.
The resulting lower bounds are then transferred back to ordinary one-sorted
structures (see~\cref{lemma:var-sorted}).

\para{On the role of self-join-freeness}
One may wonder whether a result like~\cref{thm:width-treewidth-intro} holds without the self-join-free assumption.
However, even for conjunctive queries, a result analogous to~\cref{thm:width-treewidth-intro} does not hold.
For example, the conjunctive query
\[\phi_1 = \exists x_1 \exists x_2 \exists x_3 \bigwedge_{i, j \in \{ 1, 2, 3 \}} E(x_i,x_j)\]
is logically equivalent to the conjunctive query
\[\phi_2 = \exists x E(x,x);\]
this latter query $\phi_2$ has width $1$, but the minimum width among all formulae in $[\phi_1]_{\mathbb{A}}$ can be verified to be $3$.\footnote{In brief, this can be proved in the following way.  Up to the renaming of variables, when a sequence of rewriting steps is applied to $\phi_1$, each quantification $\exists x_i$ will remain above each atom where $x_i$ occurs (and hence no quantification can ever be removed).  Viewing the resulting query as a tree, there must be a quantification $\exists x_k$ with no other quantification below it, and the subformula immediately below this quantification must contain each atom where $x_k$ occurs, and thus has $3$ free variables.}
Bova and Chen~\cite{BC19} establish that the logical width of a CQ $\phi$ is equal to the treewidth of the \emph{core} of $\phi$ plus one;
equivalently, one can apply the rewriting methods described in the present article to minimise the width of the core.
Paralleling this situation for CQs, we conjecture that the logical width of an arbitrary QCQ $\phi$ can be obtained similarly:
pass to an appropriate notion of core of $\phi$, and then apply the rewriting methods.
This would involve introducing a notion of core for QCQs, see~\cref{sec:conclusions} for more details.
Moreover, we use the assumption that the query is self-join-free for many proofs along the way to establishing our main result; more details are provided throughout the paper.

\para{Organisation}
\cref{sec:prelims} provides the requisite preliminaries.
In~\cref{sec:rewrite}, we detail the rewriting framework established
in~\cite{CM26} needed to formalise our main results, which are stated
in~\cref{sec:main}.
The rest of the article is dedicated to establishing our main result.
We handle the case where a $\{ \forall \}$-region is the region with maximum
treewidth directly in~\cref{sec:main}, leaving the more complicated case of when
a $\{ \exists, \land \}$-region has maximum treewidth.
\cref{sec:qsg} establishes the logical width lower bound on \emph{simple QCQs},
a syntactic restriction on QCQs where the lower-bound argument can be carried
out explicitly via pebble games.
Finally, a sequence of reductions, sketched in~\cref{sec:reductions}, reduces
the general $\{ \exists, \land\}$ case to the simple QCQ case, in effect lifting
the lower bound.

\section{Preliminaries}
\label{sec:prelims}

Let $\mathbb{N} \df \{ 0, 1, \ldots \}$ and $\mathbb{N}_{+} \df \mathbb{N}
\setminus \{ 0 \}$.
We use $\sqcup$ for disjoint union.
For $i, j \in \mathbb{N}$, let $[i, j] \df \{ i, i + 1, \ldots, j \}$ and $[j]
\df [1,j]$, with $[i, j] \df \emptyset$, if $i > j$.
When $f \colon A \rightarrow B$ is a (partial) function, we denote by
$\domain(f)$ the domain and by $\range(f)$ the range of $f$; in addition, we use
$f [ x \mapsto a]$ to denote the (partial) function equal to $f$ everywhere
except $x$, where it is mapped to $a$.
The restriction of $f$ to $A'$, for some $A' \subseteq \domain(f)$, is denoted
by $f |_{A'}$.
For a tuple $\vec t = (t_1, \ldots, t_r)$, let $\pi_i(\vec t) = t_i$, for every
$i \in [r]$.
When $A$ is a set, $\binom{A}{2}$ denotes the set of 2-element subsets of $A$.

\para{Graphs and Hypergraphs}
A \emph{graph} $G \df (V, E)$ is a pair consisting of a finite set $V$ of
\emph{vertices} and a set $E \subseteq \binom{V}{2}$ of \emph{edges}.
A \emph{walk} in $G$ is a sequence $(v_1, \ldots, v_m) \in V^m$, with $m \geq 2$,
such that $\{ v_i, v_{i + 1} \} \in E$, for all $i \in [m - 1]$.
For a vertex $v \in V$, we use $E_v$ for the set of edges incident on~$v$.

A \emph{tree decomposition} of a graph $G \df (V, E)$ is a tree $T$ where each
node $u$ of $T$ is associated with a non-empty subset $B_u$ of $V$ called a
\emph{bag} such that:
\begin{enumerate}[a)]
\item
  for all edges $e \in E$, there is some node $u$ of $T$ such that $e \subseteq
  B_u$;
\item
  for all vertices $v \in V$, the nodes $u$ of $T$ with $v \in B_u$ form a
  non-empty connected subtree of $T$.
\end{enumerate}
The \emph{width} of a tree decomposition $C \df (T, \{ B_u \}_{u \in V(T)})$ is
defined by $\max_{u \in V(T)} |B_u| - 1$ and the \emph{treewidth}
$\mathsf{tw}(G)$ of a graph $G = (V,E)$ is the minimum width over all tree
decompositions of $G$.

For a graph $G = (V,E)$ and two sets $U, U' \subseteq V$, we say that $U$ and
$U'$ \emph{touch} if $U \intersect U' \neq \emptyset$ or there exist $u \in U$
and $u' \in U'$ such that $\{ u, u' \} \in E$.
A set $\mathcal{M}$ of pairwise touching connected subsets of $V$ is called a
\emph{bramble} of $G$.
A subset $H$ of $V$ is a \emph{hitting set} of the bramble $\mathcal{M}$ if $H
\intersect M \neq \emptyset$, for all $M \in \mathcal{M}$.
The \emph{order} of a bramble is the smallest size over all hitting sets.
We shall use the following known result (see~\cite[Theorem 10.41]{BM08b} for
more information).

\begin{theorem}
\label{thm:bramble}
  A graph has treewidth $\geq \kappa \in \mathbb{N}_{+}$ if and only if it has a
  bramble of order $\geq \kappa + 1$.
\end{theorem}

A \emph{hypergraph} $H \df (V, E)$ is a pair consisting of a set of vertices $V$
and a set of \emph{hyperedges} $E \subseteq 2^V$, \emph{i.e.}, a set
of sets of vertices; here, $2^X$ denotes the powerset of $X$.
The \emph{primal graph} of a hypergraph $H \df (V, E)$ is the graph $G \df (V,
E')$, where $E'$ contains a pair $e' \in \binom{V}{2}$ if and only if there
exists a hyperedge $e \in E$ such that $e' \subseteq e$.
We say that a hypergraph is \emph{connected} if its primal graph is connected.
The treewidth of a hypergraph is its primal graph's treewidth.

For a graph $G = (V,E)$ and a set $X \subseteq V$, we define the \emph{induced
subgraph of $G$ on $X$}, denoted $G |_{X}$, to be the graph with vertices $X$
and edge set $E \intersect \binom{X}{2}$.
Likewise, for a hypergraph $H = (V,E)$ and a set $X \subseteq V$, let $H |_{X}
\df (X, \{ e \cap X \mid e \in E \} )$ be the \emph{induced hypergraph of $H$ on
$X$}.

\para{Structures and Logic}
A \emph{signature} $\tau$ is a set where each element $R \in \tau$ is called a
\emph{relation symbol} and has an associated arity $\mathsf{ar}(R) \in
\mathbb{N}$.
For a signature $\tau$, a \emph{$\tau$-structure} consists of a non-empty set
$A$ called the \emph{universe} and a relation $R^{\mathcal A} \subseteq
A^{\mathsf{ar}(R)}$, for each $R \in \tau$.
We shall tend to use $\mathcal{A}, \mathcal{B}, \ldots$ for structures and $A,
B, \ldots$ for their respective universes.
In this work, we assume all structures are finite, \emph{i.e.}, have finite
signature and universe.
A $\tau$-structure $\mathcal{B}$ is a \emph{substructure} of the
$\tau$-structure $\mathcal{A}$ if $B \subseteq A$ and, for each $R \in \tau$,
we have that $R^\mathcal{B} \subseteq R^\mathcal{A}$.
If $\mathcal{A}$ is a $\tau$-structure and $B \subseteq A$, then we use
$\mathcal{A} |_B$ to denote the structure with universe $B$ and $R^{\mathcal{A}
|_B} \df R^\mathcal{A} \intersect B^{\mathsf{ar}(R)}$, for each $R \in \tau$.

An \emph{atom} (over a signature $\tau$) is either an equality of variables $(x
= y)$ or has the form $R(x_1, \ldots, x_r)$, where $R$ is a relational symbol,
$x_1, \ldots, x_r$ are variables, and $\mathsf{ar}(R) = r$.
\emph{Formulae} in \emph{first-order logic} are defined by combining atoms with
conjunction ($\land$), disjunction ($\lor$), negation ($\neg)$, universal
quantification ($\forall$), and existential quantification ($\exists$).
For a structure $\mathcal{A}$, an assignment of variables~$\subs$
in~$\mathcal{A}$, and a formula $\phi$, we write $(\mathcal{A}, \subs) \models
\phi$ to denote that $\phi$  is true in $\mathcal{A}$ under $\subs$.
We assume the reader is familiar with the semantics of first-order logic (we
refer to \cite{EF95,Lib04,GKLMSVVW05} for more details).
We write $\mathsf{free}(\phi)$ to denote the set of free variables of a
first-order formula $\phi$.
Two formulae $\phi$ and $\psi$ are \emph{logically equivalent}
if $\mathsf{free}(\phi) = \mathsf{free}(\psi)$
and for every finite structure $\mathcal{A}$ and every assignment~$\subs$ we have
$(\mathcal{A}, \subs) \models \phi \iff (\mathcal{A}, \subs) \models \psi$.
The \emph{width} of a formula~$\phi$, denoted by $\mathsf{width}(\phi)$, is the
maximum number of free variables over all subformulae of $\phi$.
The \emph{logical width} of a formula $\phi$, denoted by
$\mathsf{logwidth}(\phi)$, is the minimum width over all first-order formulae
that are logically equivalent to~$\phi$ on all finite structures over the same
signature.

A \emph{quantified conjunctive query} (\emph{QCQ}, for short) is a sentence
(\emph{i.e.}, a formula without free variables) that combines non-equality (relation) atoms
with conjunction and both universal and existential quantifiers.
A QCQ $\varphi$ is in \emph{prenex normal form} if $\varphi = Q_1 x_1 \, Q_2 x_2
\cdots Q_n x_n (\psi)$, where $Q_i \in \{ \forall, \exists \}$ and $\psi$ is a
conjunction of non-equality atoms, and $\varphi$ is \emph{self-join-free} if no
relation symbol appears more than once in $\varphi$.
A QCQ $\varphi$ is \emph{standardised} if, for each occurrence $Q x$ of a
quantifier, $x$ is not quantified elsewhere.
By simply renaming variables where necessary, we often assume that QCQs are
standardised.
Throughout this article, we use \emph{evaluation games}; see, for
example,~\cite{Gra02a} for details.
We adapt the notion for our particular purposes, namely QCQs.

\begin{definition}
\label{defn:eval-game}
Let $\varphi = Q_1 x_1 \cdots Q_n x_n ( \psi )$ be a standardised QCQ in prenex
normal form, where~$\psi$ is a conjunction of atoms.
Let $\mathcal{C}$ be a structure over the signature of $\varphi$, with
universe~$C$.
The \emph{evaluation game} over $\mathcal{C}$ and $\varphi$ is a two-player game
played by \emph{Universal} and \emph{Existential}.
A \emph{position} of the game is $(i, \subs)$, where $i \in [n]$
and $\subs : \{ x_{1}, \ldots, x_{i - 1} \} \to C$ is a partial assignment.
The initial position is $(1, \emptyset)$.
For a position $(i, \subs)$ with $1 \leq i \leq n$, a round of the game proceeds
as~follows:
\begin{itemize}[$\bullet$]
\item
  if $Q_i = \forall$, then Universal picks a value $c \in C$ and the new
  position is $(i + 1, \subs[x_{i} \mapsto c])$;
\item
  if $Q_i = \exists$, then Existential picks a value $c \in C$ and the new
  position is $(i + 1, \subs[x_{i} \mapsto c])$.
\end{itemize}
When a final position $(n + 1, \subs)$ is reached, Existential wins if and
only if $(\mathcal{C}, \subs) \models \psi$.
Otherwise Universal wins.
Clearly, Existential has a winning strategy if and only if $\mathcal{C} \models
\varphi$.
\end{definition}

\para{Pebble Games}
An \emph{isomorphism} between %
structures $\mathcal{A}$ and $\mathcal{B}$ (over the same signature) is a
bijective function $h \colon A \rightarrow B$ such that, for all relation
symbols $R$ in the signature and tuples $\bar a \in A^{\mathsf{ar}(R)}$, we have
that $\bar a \in R^\mathcal{A}$ if and only if $h(\bar a) \in R^\mathcal{B}$,
where $h(\bar a)$ denotes the tuple obtained by applying $h$ coordinate-wise.
A \emph{partial isomorphism} between
structures $\mathcal{A}$ and $\mathcal{B}$ (over the same signature) is a
partial function $f \colon A \rightarrow B$ where $f$ is an isomorphism between
$\mathcal{A} |_{\mathsf{dom}(f)}$ and $\mathcal{B} |_{\mathsf{ran}(f)}$.

\emph{The $k$-pebble game} is played by two players: \emph{Spoiler} and
\emph{Duplicator}.
They play over two structures $\mathcal{A}$ and $\mathcal{B}$ over the same
signature.
There is a size-$k$ set $P$ of pebbles, and a position of the game is a pair of
partial functions $(\alpha, \beta)$ where $\alpha \colon P \rightarrow A$ and
$\beta \colon  P \rightarrow B$.
Unless stated otherwise, we assume the initial position of any game to have
$\domain(\alpha) = \domain(\beta) = \emptyset$.
Each round of the game, starting in some position $(\alpha, \beta)$, is as
follows: Spoiler selects a pebble $p \in P$ and some element $a \in
A$ or $b \in B$; Duplicator then responds picking some $b \in
B$ or $a \in A$, respectively.
The new position is defined as $(\alpha[p \mapsto a], \beta[p \mapsto b])$.
Duplicator wins the game if $\{ (\alpha(p), \beta(p)) \mid p \in \domain(\alpha)
\}$ is a partial isomorphism, between $\mathcal{A}$ and $\mathcal{B}$, for each
position $(\alpha, \beta)$ that occurs.
We assume the game goes on for an indefinite number of rounds.
We denote that Duplicator has a winning strategy for the $k$-pebble game over
$\mathcal{A}$ and $\mathcal{B}$ by writing $\mathcal{A} \equiv^k \mathcal{B}$.
See, for example, \cite{EF95,Lib04,GKLMSVVW05} for more details.

Two structures $\mathcal{A}$ and $\mathcal{B}$ are \emph{$k$-indistinguishable}
if for every first-order sentence $\phi$ that uses at most $k$ distinct
variables,
we have that $\mathcal{A} \models \phi$ if and only if $\mathcal{B} \models
\phi$.
We now state the $k$-pebble game characterisation of the expressive power of the
$k$-variable fragment of first-order logic.

\begin{theorem}[\cite{Imm82a}]
\label{thm:pebble-game-char}
Let $\mathcal{A}$ and $\mathcal{B}$ be relational structures over the same
signature and $k \in \mathbb{N}$.
Then, $\mathcal{A}$ and $\mathcal{B}$ are $k$-indistinguishable if and only if
$\mathcal{A} \equiv^k \mathcal{B}$.
\end{theorem}

In ordinary one-sorted first-order logic, if a formula has a width of $k$, then
it can be rewritten into an equivalent formula using at most $k$ variables by
simply renaming variables.\footnote{Note that this property does not hold in
general for many-sorted variants, where variable renaming is constrained by the
need to respect sorts.}
Moreover, a formula that uses at most $k$ variables obviously has a width of at
most $k$.
Thus, in order to establish that $\mathsf{logwidth}(\varphi) > k$, it suffices
to define two structures $\mathcal{A}$ and $\mathcal{B}$ such that $\mathcal{A}
\models \varphi$ and $\mathcal{B} \not\models \varphi$, while at the same time
$\mathcal{A} \equiv^k \mathcal{B}$.

\para{QCQs Decomposition and Regions}
This subsection introduces some requisite notions of formulae adapted
from~\cite{CM24} to the case of QCQs.

\paragraph{Holey QCQs.}

A \emph{holey QCQ} is a formula defined recursively as follows: each natural
number $i \geq 1$ is a holey QCQ, and if $\varphi, \phi$ are holey QCQs then so
are $\varphi \land \phi$, $\exists x \varphi$, and $\forall x \varphi$, where
$x$ is a variable.
As a convention, we shall require that no natural number occurs more than once
in any holey QCQ.
The natural numbers that occur in a holey QCQ are called \emph{holes}.
A $\{ \exists, \land\}$-holey formula consists entirely of natural numbers,
existential quantification, and conjunction.
A $\{ \forall\}$-holey formula is of the form $\forall x_1 \forall x_2 \ldots
\forall x_n i$ for a natural number $i$.
When $\phi$ has holes among $1, \ldots, k$, and $\varphi_1, \ldots, \varphi_k$
are QCQs, then $\phi \llbracket \varphi_1, \ldots, \varphi_k \rrbracket$ denotes
the QCQ that is obtained by replacing $i$ with $\varphi_i$ for each $i \in [k]$.
We call such $\varphi_i$ a \emph{child region} of $\phi$.
An \emph{association} for a holey QCQ $\phi$ is a partial mapping $a$ that maps
at least each number occurring in $\phi$ to a set of variables; thus, whenever
$a(i)$ is defined, it is a set of variables.
Given such an association $a$, we set $\mathsf{free}_a(i) = a(i)$ and define
$\mathsf{free}_a(\phi)$ for any holey formula $\phi$ using the usual inductive definitions.

\paragraph*{Organised QCQs.}

We define $\{ \exists, \land \}$-organised and $\{ \forall \}$-organised
formulae by mutual induction:
\begin{itemize}[$\bullet$]
\item
  if $\phi$ is a non-atomic holey $\{ \exists, \land \}$-formula and
  $\varphi_{1}, \ldots, \varphi_{k}$ are either atomic formulae or a $\{ \forall
  \}$-organised formulae, then $\phi \llbracket \varphi_{1}, \ldots, \varphi_{k}
  \rrbracket$ is an $\{ \exists, \land \}$-organised formula;
\item
  if $\phi$ is a non-atomic holey $\{ \forall \}$-formula and $\varphi$ is an atomic formula or an
   $\{ \exists, \land
  \}$-organised formula, then $\phi \llbracket \varphi
  \rrbracket$ is a $\{ \forall \}$-organised formula.
\end{itemize}

A QCQ is \emph{organised} if it is either an $\{ \exists, \land\}$-organised formula or a $\{ \forall \}$-organised formula.

\begin{example}
\label{example:organised}
Consider the QCQ $\varphi \df \forall x \forall x' \exists y \exists y' \bigl( R(x,y) \land \exists z ( T(x',y',z) ) \bigr)$.
By defining
$\alpha \df \forall x \forall x' 1$ and
$\beta \df \exists y \exists y' (1 \land \exists z 2)$,
we can view $\varphi$ as the organised QCQ $\alpha \llbracket \beta \llbracket R(x,y), T(x',y',z) \rrbracket \rrbracket$.
\end{example}

Via the following result, we may view each QCQ as organised.

\begin{proposition}[{\cite[follows from Proposition 5.5]{CM26}}]
Each non-atomic QCQ is an organised QCQ.
In particular, each non-atomic QCQ starting with $\exists$ or $\land$ is an $\{
\exists, \land \}$-organised formula and each non-atomic QCQ starting with
$\forall$ is an $\{ \forall \}$-organised formula.
\end{proposition}

Recursively, we define the \emph{regions} of an organised formula: if $\varphi
= \phi \llbracket \psi_{1}, \ldots, \psi_{k} \rrbracket$ is an organised
formula, then $\mathsf{regions}(\varphi) = \{ \phi \} \union \bigcup_{i \text{
is a hole in } \phi} \mathsf{regions}(\psi_{i})$; we say that $\phi$ is the
\emph{top region} of $\varphi$, and each other region of $\varphi$
is a \emph{non-top region} of $\varphi$.

\begin{example}
For the organised QCQ $\varphi = \alpha \llbracket \beta \llbracket R(x,y), T(x',y',z) \rrbracket \rrbracket$ from~\cref{example:organised} we have that
$\mathsf{regions}(\varphi) = \{ \alpha, \beta, R(x,y), T(x',y',z)\}$ where $\alpha$ is the top region of $\varphi$.
\end{example}

When $\phi$ is a holey formula and $a$ is an association for $\phi$, then the
hypergraph for $\phi$, denoted by $H_{\phi, a}$ (or $H_{\phi}$, when $a$ is
clear from context) is the hypergraph with the vertex set $\bigcup_{i} a(i)$ and
with the edge set $\{ a(i) \mid i \text{ occurs in } \phi \} \union \{
\mathsf{free}_a(\phi) \}$.

\begin{example}
Consider again the formulas from \cref{example:organised}.
Looking at the holey formula $\beta$, the association $a$ suggested by
the original formula $\varphi$ would have $a(1) = \{ x, y \}$
and $a(2) = \{ x', y', z \}$.  We then would have
$\mathsf{free}_a(\beta) = \{ x, x' \}$, and so the hypergraph
$H_{\beta,a}$ would have vertex set $\{ x, x', y, y', z \}$
and edge set $\{ \{ x, y \}, \{ x', y', z \}, \{ x, x' \}  \}$.
\end{example}

\para{Variable-Sorted Logic}
Informally, \emph{variable-sorted logic} is a \emph{many-sorted} first-order
logic, where each variable has its own finite domain (or sort).
Each relation symbol has an arity that is an ordered sequence of variables,
and must be interpreted by a cross product of the domains corresponding
to this arity.
In this setting, the ability to rewrite so that width becomes number of
variables (discussed above) does not hold in general, since variables of
different sorts cannot be identified by renaming.
At the same time, the separation into variable sorts lets us define the domains
associated with different variables independently and constrains pebble-game
responses to the corresponding domains.
These features simplify the lower-bound constructions used later in the paper.

Let $\mathcal{V}$ be a finite set of variables.
A \emph{variable-sorted signature} is the pair $(\tau, \mathcal{V})$, where
$\tau$ is a set of relation symbols such that each $R \in \tau$ has an
associated arity $\mathsf{ar}(R) \in \mathcal{V}^*$, where $\mathcal{V}^*$ is
the set of finite strings generated by $\mathcal{V}$.
A \emph{variable-sorted structure} $\mathcal{A}$ consists of a nonempty finite set
$A_{x}$ for every $x \in \mathcal{V}$, called the \emph{domain of $x$}, along
with, for each $R \in \tau$ with $\mathsf{ar}(R) = x_1 \cdots x_k$, an
interpretation $R^\mathcal{A} \subseteq A_{x_{1}} \times \cdots \times A_{x_{k}}$.
We always assume the domains of two distinct variables are disjoint.
When convenient, we write $A$ for $\bigsqcup_{x \in \mathcal{V}} A_{x}$.
Formulae over $(\tau, \mathcal{V})$ are built from atomic formulae $R(x_{1},
\ldots, x_{k})$, where $R \in \tau$ and $\mathsf{ar}(R) = x_{1} \cdots x_{k}$,
along with usual first-order connectives and quantifiers (which range over the
domain of the particular variable being quantified).
For a variable-sorted structure $\mathcal{A}$, a substitution $\subs$ that maps
variables to their domains, and a formula~$\varphi$, we use $(\mathcal{A},
\subs) \models_{\mathsf{var}} \varphi$ to denote that $\varphi$ is true in
$\mathcal{A}$ under $\subs$.
We use the term \emph{one-sorted structures} when we need to distinguish usual
relational structures from variable-sorted~structures.

\begin{example}
Consider the variable-sorted signature $(\{ R \}, \{ x, y \})$, where
$\mathsf{ar}(R) = x y$.
Let $\mathcal{A}$ be a structure over this signature with $A_{x} = \{ a, b \}$,
$A_{y} = \{ c, d \}$, and $R^\mathcal{A} = \{ (a, c), (b,c) \}$.
Then, $\mathcal{A} \models_\mathsf{var} \forall x \exists y \, R(x,y)$ but
$\mathcal{A} \not\models_\mathsf{var} \forall x \forall y \, R(x, y)$.
\end{example}

For variable-sorted structures $\mathcal{A}$ and $\mathcal{B}$ over the same
signature, an isomorphism is a bijection $h \colon A \rightarrow B$ such that
$h(A_x) = B_x$, for each $x \in \mathcal{V}$, and for every $R \in \tau$ with
$\mathsf{ar}(R) = x_{1} \cdots x_{k}$ and $\bar a \in A_{x_{1}} \times \cdots
A_{x_{k}}$, we have that $\bar a \in R^\mathcal{A}$ if and only if $h(\bar a)
\in R^\mathcal{B}$.
Partial isomorphisms are defined as expected, hence the definition of pebble
games transfers directly to the variable-sorted setting, with the caveat that if
Spoiler places a pebble in $A_x$, then Duplicator must respond by pebbling an
element of $B_x$, and conversely.
If Duplicator has a winning strategy for the $k$-pebble game over two
variable-sorted structures $\mathcal{A}$ and $\mathcal{B}$, we write
$\mathcal{A} \equiv^k_\mathsf{var} \mathcal{B}$.
\emph{Throughout, we flexibly view any self-join-free QCQ $\phi$ as a formula in
variable-sorted logic; this is done naturally by considering any atom
$R(x_1,\ldots,x_k)$ that occurs in $\phi$ to have $\mathsf{ar}(R) = x_1 \cdots
x_k$.}
We will make use of the following lemma that links variable-sorted structures to
one-sorted structures.

\begin{lemma}
\label{lemma:var-sorted}
Let $\varphi$ be a standardised self-join-free QCQ and $k \in \mathbb{N}_{+}$.
If there are two variable-sorted structures $\mathcal{A}$ and $\mathcal{B}$ such
that $\mathcal{A} \models_{\mathsf{var}} \varphi$, $\mathcal{B}
\not\models_{\mathsf{var}} \varphi$, and $\mathcal{A} \equiv^k_\mathsf{var}
\mathcal{B}$, then there are one-sorted structures $\bar{\mathcal{A}}$ and
$\bar{\mathcal{B}}$ such that $\bar{\mathcal{A}} \models \varphi$,
$\bar{\mathcal{B}} \not\models \varphi$, and $\bar{\mathcal{A}} \equiv^k
\bar{\mathcal{B}}$.
\end{lemma}

This lemma ensures that lower bounds proved in the variable-sorted setting
transfer back to the usual one-sorted setting.
This allows us to conveniently work in the variable-sorted setting which shall help simplify subsequent proofs (see~\cref{sec:qsg}).
We note that the assumption that $\varphi$ is standardised and self-join-free is essential to be able to transfer lower bounds in this way.
Informally speaking, the fact that $\varphi$ is self-join-free allows us to associate each relation symbol $R$  that appears in $\varphi$ with a particular tuple of variables $\vec x$
where $R(\vec x)$  is the single application of $R$ that appears in $\varphi$.
Moreover, since $\varphi$ is standardised, each variable in $\vec x$ can be associated with its particular quantifier.
This then allows us to construct $R^{\bar{\mathcal A}}$ and $R^{\bar{\mathcal B}}$ from $R^{\mathcal{A}}$ and $R^{\mathcal B}$ respectively;
this is done by adding those tuples where at least one component corresponding to a universally quantified variable does not fall into the respective domain.

\section{Rewriting}
\label{sec:rewrite}

In this section, we recall the notions of term rewriting for first-order logic
that will be used throughout the paper.
The definitions are adapted from~\cite{CM26} and the references therein.
While~\cite{CM26} develops the framework for positive first-order logic, we
specialise the relevant concepts and results to the setting of QCQs.

\subsection{Rewriting systems}

A \emph{rewriting system}, or just \emph{system}, is a pair $(D, \rightarrow)$
consisting of a set $D$ and a binary relation $\rightarrow$ over $D$, called a
\emph{reduction}.
The \emph{inverse} of $\rightarrow$ is denoted by $\leftarrow$, thus, $a
\rightarrow b$ if and only if $b \leftarrow a$.
Let $\leftrightarrow$ denote $\rightarrow \union \leftarrow$ and
$\overset{*}{\rightarrow}$, $\overset{*}{\leftarrow}$, and
$\overset{*}{\leftrightarrow}$ the reflexive-transitive closures of
$\rightarrow$, $\leftarrow$, and $\leftrightarrow$, respectively.
Relative to a system $(D, \rightarrow)$, we say that $d \in D$ is a \emph{normal
form} if $d \rightarrow d'$ does not hold for any $d' \in D$.
Moreover, for $d, e \in D$, we say $e$ is a \emph{normal form of $d$} if $d
\overset{*}{\rightarrow} e$ and $e$ is a normal form.

For a set $D$ and an equivalence relation $\equiv$ on it, we define
$D / \equiv$ as the set of $\equiv$-equivalence classes.
If $(D, \rightarrow)$ is a system, then $(D, \rightarrow) / \equiv$ denotes the
system $(D / \equiv, \rightarrow)$, where for $\equiv$-classes $C$ and $C'$, we
have that $C \rightarrow C'$ if and only if there is some $d \in C$ and $d' \in
C'$ with $d \rightarrow d'$.

\subsection{Rules}

Consider the following rewriting rules, all of which %
preserve
logical equivalence under the stated side conditions~\cite{CM26}.
Each rule may be applied to an arbitrary subformula of a formula.

\smallskip
\begin{tabular}{@{} l p{0.725\linewidth} @{}}
  Associativity
&
  $(\varphi_1 \land (\varphi_2 \land \varphi_3)) \rightarrow_A ((\varphi_1 \land
  \varphi_2) \land \varphi_3)$;
  \enspace
  $((\varphi_1 \land \varphi_2) \land \varphi_3) \rightarrow_A (\varphi_1 \land
  (\varphi_2 \land \varphi_3))$.
\\
  Commutativity
&
  $(\varphi_1 \land \varphi_2) \rightarrow_C (\varphi_2 \land \varphi_1)$.
\\
  Reordering
&
  $Q x Q y \varphi \rightarrow_O Q y Q x \varphi$, where $Q \in \{ \forall,
  \exists \}$.
\\
  Pushdown
&
  $\exists x (\varphi_1 \land \varphi_2) \rightarrow_{P \downarrow} (\exists x
  \varphi_1) \land \varphi_2$, whenever $x \not\in \mathsf{free}(\varphi_2)$.
\\
  Splitdown
&
  $\forall x (\varphi_1 \land \varphi_2) \rightarrow_{S \downarrow} (\forall x
  \varphi_1) \land (\forall x \varphi_2)$.
\\
  Removal
&
  $Qx \varphi \rightarrow_M \varphi$, whenever $x \not\in
  \mathsf{free}(\varphi)$.
\\
  Renaming
&
  $Q x \varphi \rightarrow_N Q y \varphi'$, for $Q \in \{ \forall, \exists \}$,
  where $y$ is assumed not to occur in $\varphi$ and $\varphi'$ is obtained by
  replacing all free occurrences of $x$ in $\varphi$ with $y$.
\end{tabular}
\smallskip

We further define the \emph{splitup} rule $\rightarrow_{S \uparrow}$ and
\emph{pushup} rule $\rightarrow_{P \uparrow}$ as the inverses of the splitdown
and pushdown rules, respectively.

For a set $\mathcal{R}$ of subscripts for rewriting rules, we let
$\rightarrow_\mathcal{R} = \bigcup_{R \in \mathcal{R}} \rightarrow_R$.
For example, $\rightarrow_{\{ A, C \}}$ denotes $\rightarrow_{A} \union
\rightarrow_{C}$, and to avoid notational clutter we write
$\rightarrow_{AC}$ for $\rightarrow_{\{ A, C \}}$.
The same conventions are also applied to $\leftrightarrow$,
$\overset{*}{\rightarrow}$, etc.
Given a formula $\varphi$, we define $[ \varphi ]_\mathcal{R}$ to be the
$\overset{*}{\leftrightarrow}_\mathcal{R}$-equivalence class of $\varphi$.
For example, $[ \varphi ]_{ACO}$ denotes
the~$\overset{*}{\leftrightarrow}_{ACO}$-equivalence class of $\varphi$.

\begin{notation}
Let us fix some notation to be used throughout this article.
We define the set of rewriting rules $\mathbb{T} \df \{A,C,O,P\downarrow, N,
P\uparrow \}$, as well as $\mathbb{A} \df \{  S\downarrow, S\uparrow, M \}
\union \mathbb{T}$.
Thus, $\mathbb{A}$ contains all the rewriting rules introduced above, whereas
$\mathbb{T}$ contains all of them except splitdown, splitup, and removal.
Furthermore, we shall make use of the rewriting system
$\mathscr{R} \df (\mathsf{QCQ} , \rightarrow_{\{ S \downarrow, M \}} ) /
\overset{*}{\leftrightarrow}_{\mathbb{T}}$.
\end{notation}

\subsection{Width Minimisation Algorithm}

Let us now state a version of Theorem 4.1 of~\cite{CM26} that is specialised to
QCQs; our formulation here draws both from that theorem and its proof.

\begin{theorem}[\cite{CM26}]
\label{lemma:algo}
There exists an algorithm that, given a QCQ $\phi$, outputs a QCQ $\phi^+$,
where:
\begin{itemize}[$\bullet$]
\item
  $\phi^+$ has minimum width among all formulae in $[ \phi]_{\mathbb{A}}$;
\item
   $[\phi^+]_\mathbb{T}$ is a normal form of $[\phi]_\mathbb{T}$ in the system
  $\mathscr{R}$;
\item
  $\width(\phi^+) = \max_{\psi \in \regions(\phi^{+})} (\tw(H_{\psi}) + 1)$,
  where $H_{\psi}$ denotes the hypergraph of the region $\psi$.
\end{itemize}
With oracle access to an algorithm for computing tree decompositions of minimum
width, this algorithm can be implemented in polynomial time.
\end{theorem}

The algorithm roughly works as follows: given a QCQ $\phi$, it first computes a
QCQ $\phi'$, where $[\phi']_{\mathbb{T}}$ is a normal form of $[\phi]_\mathbb{T}$ in the system
$\mathscr{R}$.
Then, the output $\phi^{+}$ is chosen inside $[\phi']_{\mathbb{T}}$ by
minimising each region independently, invoking an oracle for computing tree
decompositions of minimum width.
The argument in~\cite{CM26} shows that the output is a width-minimal QCQ inside
$[\phi]_{\mathbb{A}}$.

\section{Main Results}
\label{sec:main}

This section states our main results and outlines the proof strategy.
The first theorem is the semantic core of the paper: for self-join-free QCQs,
the logical width coincides with the optimal width obtainable by the rewriting
rules $\mathbb{A}$.

\begin{theorem}
\label{thm:width-treewidth}
  Let $\phi$ be a self-join-free QCQ and $\phi^{+}$ a self-join-free QCQ of
  minimum width among all formulae in $[\phi]_{\mathbb{A}}$. Then,
  $\logwidth(\phi) = \width(\phi^{+})$.
\end{theorem}

The difficult direction of~\cref{thm:width-treewidth} is the lower bound
$\logwidth(\phi) \geq \width(\phi^{+})$.
The converse inequality $\logwidth(\phi) \leq \width(\phi^{+})$ is immediate,
since the rules in $\mathbb{A}$ are known to preserve logical
equivalence~\cite{CM26}.
Combining~\cref{thm:width-treewidth} and~\cref{lemma:algo}, we immediately derive:

\begin{theorem}
\label{main:thm}
  There is an algorithm that, given a self-join-free QCQ $\phi$, outputs a
  self-join-free QCQ $\phi^{+}$, obtained from $\phi$ via the rewriting rules in
  $\mathbb{A}$, such that $\width(\phi^{+}) = \logwidth(\phi)$.
  Moreover, $\phi^{+}$ can be computed in polynomial time given access to an
  oracle for computing minimum tree decompositions.
\end{theorem}

From this theorem, we can readily derive the following result on computing the width of a given self-join-free QCQ.

\begin{corollary}
\label{cor:complexity}
  Given a self-join-free QCQ $\phi$ and a number $k \in \mathbb{N}$, deciding
  whether $\logwidth(\phi) \leq k$ is NP-complete.
  Moreover, the problems of computing the logical width of a self-join-free QCQ
  and the treewidth of a graph are polynomial-time Turing equivalent.
\end{corollary}

Since self-join-free QCQs form a subclass of QCQs, the same reduction shows that
deciding whether $\logwidth(\phi) \leq k$ for arbitrary QCQs is NP-hard.
Determining the precise complexity of this problem remains open.

We say that a region $\theta$ of a QCQ $\phi$ \emph{has maximum treewidth} if
$\tw(H_{\theta}) = \max_{\psi \in \mathsf{regions}(\phi)} \tw(H_{\psi})$.
The proof of~\cref{thm:width-treewidth} is split into two cases, according to
where the maximum treewidth is attained.
The first case is when there is a $\{\forall \}$-region with maximum treewidth,
and no $\{ \exists, \land \}$-region has such treewidth; this case is rather
straightforward and is handled directly.
The second case is when there is a $\{ \exists, \land \}$-region with maximum
treewidth.
To handle it, we first establish~\cref{thm:width-treewidth} for so-called
\emph{simple QCQs} in~\cref{sec:qsg} and then lift the result to arbitrary
self-join-free QCQs in~\cref{sec:reductions}.

\subsection{$\{ \forall \}$-Regions of Maximum Treewidth}

Observe that $\{ \forall \}$-regions do not contain any conjunctions. Thus, $\{
\forall\}$-organised formulae can be assumed to be in the form $(\forall y_{1}
\cdots \forall y_{k} \, 1) \llbracket \theta \rrbracket$, where $\theta$ is
either an atom or a non-atomic $\{ \exists, \land \}$-organised formula.

\begin{theorem}
\label{thm:forall-width-treewidth}
  Let $\phi$ be a self-join-free QCQ and $\phi^{+}$ a self-join-free QCQ of
  minimum width among all formulae in $[\phi]_{\mathbb{A}}$.
  Suppose there is a $\{\forall \}$-region of $\phi^{+}$ with maximum treewidth,
  and no $\{ \exists, \land \}$-region of $\phi^{+}$ has maximum treewidth.
  Then, $ \logwidth(\phi) = \width(\phi^{+})$.
\end{theorem}

The above result is established by showing that, if $\psi = (\forall y_{1}
\cdots \forall y_{k} \, 1) \llbracket \theta \rrbracket$ with $\theta$
non-atomic, then $H_{\theta}$ has treewidth at least equal to $\tw(H_{\psi})$.
Hence, if $\psi$ has maximum treewidth and no $\{ \exists, \land\}$-region has
the same treewidth, then the child region $\theta$ must be necessarily an atom.
In this case, the desired lower bound follows by a direct pebble-game argument.

\subsection{$\{\exists, \land \}$-Regions of Maximum Treewidth}

Thanks to~\cref{thm:forall-width-treewidth}, in order to
prove~\cref{thm:width-treewidth} it suffices to show the following result.

\begin{theorem}
\label{thm:main-exists-and}
  Let $\phi$ be a self-join-free QCQ and $\phi^{+}$ a self-join-free QCQ of
  minimum width
  in $[\phi]_{\mathbb{A}}$.
  Suppose there is a $\{\exists, \land \}$-region of $\phi^+$ with maximum
  treewidth.
  Then, $\logwidth(\phi) = \width(\phi^{+})$.
\end{theorem}

The rest of this article is dedicated to establishing precisely this theorem.
Our first step in doing so is to consider some properties that must hold true
for an $\{ \exists, \land \}$-region.
Let $\phi$ be an $\{ \exists, \land \}$-holey formula, where $X$ is the set of
variables existentially quantified in it.
With respect to the implicit association, we say that $\phi$ satisfies the
\emph{connectivity property} if the primal graph of $H_{\phi} |_X$ is connected.
Furthermore, we say that $\phi$ satisfies the \emph{tightness property} if every
hyperedge of $H_{\phi}$, except the hyperedge containing all the free variables,
contains a variable from $X$.
The connectivity and tightness properties naturally lift to $\{ \exists, \land
\}$-regions of QCQs.

\begin{lemma}
\label{lemma:connected-and-tight}
  Suppose $\phi$ is a QCQ such that $[\phi]_{\mathbb{T}}$ is a normal form in
  the distinguished system $\mathscr{R}$.
  Then, every non-top $\{ \exists, \land \}$-region of $\phi$ satisfies both the
  connectivity and tightness property.
\end{lemma}

The main idea behind~\cref{lemma:connected-and-tight} is simple: if an $\{
\exists, \land \}$-region of $\phi$ fails connectivity or tightness, then one
surely finds a formula in $[\phi]_{\mathbb T}$ to which the splitdown rule
applies, contradicting the fact that $[\phi]_{\mathbb T}$ is a normal form in
the system $\mathscr{R}$.

\section{Simple QCQs}
\label{sec:qsg}

This section considers a special class of QCQs, which we call \emph{simple
QCQs}, and establishes a width lower bound in terms of the treewidth of an $\{
\exists, \land \}$-region.
Although simple QCQs may appear rather restrictive, they capture the core
difficulty in proving~\cref{thm:main-exists-and}.
In~\cref{sec:reductions}, we shall sketch a sequence of reductions showing how
the general case of~\cref{thm:main-exists-and} follows from the lower bound
proved for simple QCQs.

\begin{definition}
\label{defn:rb-pointed}
A QCQ $\varphi$ is \emph{simple} if it is self-join-free, standardised, and of
the form
\begin{center}
$\varphi = Q_{1} v_{1} Q_{2} v_{2} \cdots Q_{m - 1} v_{m - 1} \; \forall v_{m}
\; \exists x_{1} \cdots \exists x_{k} (\phi)$,
\end{center}
where $m, k \geq 1$, $\phi$ is a conjunction of atoms, $Q_i \in \{ \forall,
\exists \}$ for $i \in [m-1]$, and the following conditions hold, writing
$\theta\llbracket \alpha_{1}, \ldots, \alpha_n \rrbracket$ for the $\{ \exists,
\wedge \}$-organised formula equal to $\exists x_{1} \cdots \exists x_{k}
(\phi)$: \begin{enumerate}[(1)]
\item\label{rbp1}\label{rbp2}
  each atom of $\phi$ consists of a binary relation symbol applied to two
  distinct variables;
\item\label{rbp3}\label{rbp4}
  the $\{ \exists, \land \}$-region $\theta$ satisfies both the connectivity and
  tightness properties;
\item\label{rbp5}
  each $v_{i}$ and $x_{j}$ occurs in some atom of $\phi$, over $i \in [m]$ and
  $j \in [k]$.
\end{enumerate}
We call the region $\theta$ of $\varphi$ the \emph{inner region of $\varphi$}.
\end{definition}

Our aim in this section is to show, for every simple QCQ $\varphi$, that
$\logwidth(\varphi) \!\geq\! \tw(H_{\theta}) \!+\! 1$, where~$\theta$ is the
inner region of~$\varphi$.
To this end, we show that there exist two variable-sorted structures
$\mathcal{A}$ and~$\mathcal{B}$ such that $\mathcal{A} \models_\mathsf{var}
\varphi$, $\mathcal{B} \not\models_\mathsf{var} \varphi$, but $\mathcal{A} \equiv^{\tw(H_\theta)} \mathcal{B}$.
Working in the variable-sorted setting lets us define the domain of each query
variable independently in the construction below.

\begin{example}
\label{example:running}
Consider the simple QCQ $\bar\varphi \df \forall v_{1} \forall v_{2} \exists
x_{1} \exists x_{2} (R_{1}(v_{1}, x_{1}) \land ( R_{2}(v_{2},x_{2}) \land
R_{3}(x_{1}, x_{2}) ) )$, and let $\bar\theta(v_{1}, v_{2}) = \exists x_{1}
\exists x_{2} \bigl( 1 \land (2 \land 3) \bigr) $ be its inner region.
The hypergraph $H_{\bar\theta}$ has hyperedges $\{ v_{1}, x_{1} \}, \{ v_{2},
x_{2}\}, \{ x_{1}, x_{2} \} , \{ v_{1}, v_{2}\}$ because the child regions of $\bar\theta$ are
$R_1(v_1, x_1)$, $R_2(v_2, x_2)$ and $R_3(x_1,x_2)$, and because $\free(\bar\theta) = \{ v_{1}, v_{2} \}$.
It is easy to verify that $\tw( H_{\bar\theta}) = 2$, since $H_{\bar\theta}$ is
a cycle of four vertices.
Thus, there is a bramble of order $3$: for example, consider~$\{ \{ x_{1} ,
x_{2} \}, \{ v_{1} \}, \{ v_{2} \} \}$.

Now, notice that $\bar\varphi \rightarrow_{P \downarrow} \bar\varphi'$, where
$\bar\varphi' = \forall v_{1} \forall v_{2} \exists x_{1} \bigl( R_{1}(v_{1},
x_{1}) \land \exists x_{2} (R_{2}(v_{2}, x_{2}) \land R_{3}(x_{1}, x_{2}) )
\bigr)$, and that $\mathsf{width}(\bar\varphi') = \tw(H_{\bar\theta}) +1 =  3$.
By establishing that the logical width of every simple QCQ is at least the
treewidth of the hypergraph for the inner region plus one, we obtain that
$\bar\varphi'$ is a first-order sentence that is logically
equivalent to $\bar \varphi$ and has width equal to $\mathsf{logwidth}(\bar \varphi)$.
\end{example}

\begin{notation}
\label{notation:conventions}
The following notation is fixed throughout this section.
\begin{enumerate}[(1)]
\item
  Let $\varphi \df Q_{1} v_{1} \, Q_{2} v_{2} \dots Q_{m - 1} v_{m - 1} \forall
  v_{m} \exists x_{1} \dots \exists x_{k} ( \phi )$ be the simple QCQ under
  discussion and $\theta$ its inner region; for notational convenience, set
  $Q_{m} \df \forall$.
\item
  Let $\tau$ be the set of relation symbols that occur in $\varphi$.
\item
  Let $\mathcal{V} \df \{ v_{1}, \dots, v_{m} \}$, $\mathcal{X} \df \{ x_{1},
  \dots, x_{k} \}$, and $\mathcal{Z} \df \mathcal{V} \union \mathcal{X}$.
\item
  Set $s \in \mathcal{X}$ to be an arbitrary but fixed variable.
\item
  Let $G_{\theta} \df (\mathcal{Z}, E_{\theta})$ be the primal graph of
  $H_{\theta}$.
\item
  Let $\rho \in \mathbb{N}$ denote the treewidth of $G_{\theta}$.
\item
  Set $\mathcal{M}$ to be an arbitrary but fixed bramble of $G_{\theta}$ with
  order $> \rho$.
\end{enumerate}
Let us recall that, by definition, the hypergraph $H_{\theta}$ contains $\{
v_{1}, \ldots, v_{m} \}$ as a hyperedge, as well as a hyperedge
$\free(\alpha)$ for each atom $\alpha$ of $\phi$.
\end{notation}

\subsection{Defining $\mathcal{A}$ and $\mathcal{B}$}
\label{sec:defining-A-B}

Our next focus is to define the variable-sorted structures $\mathcal{A}$ and
$\mathcal{B}$.
For each $v \in \mathcal{Z}$, recall that~$E_{v}$ denotes the set of edges
incident on $v$.
We say that a function $f_{v} \colon E_{v} \rightarrow \{ 0, 1 \}$ has
\emph{even parity}, if $\sum_{e \in E_{v}} f_{v}(e) \equiv 0 \pmod{2}$, and
\emph{odd parity}, otherwise.
When convenient, we view $\phi$ as a set of atomic formulae, and use notation
such as $R(x,y) \in \phi$ to indicate that $R(x,y)$ occurs in $\phi$.
We define $\mathcal{A}$ and $\mathcal{B}$ as variable-sorted structures over the
signature $(\tau, \mathcal{Z})$.
\begin{definition}[Domains for $\mathcal{A}$]
\label{defn:dom-A}
For  $z \in \mathcal Z$, let $\mathcal F_{z} \df \{\, (z, f_{z}) \mid f_{z}
\colon E_{z} \to \{ 0, 1 \} \text{ has even parity} \}$.
Then, the domains for each variable are as follows.
\begin{itemize}[$\bullet$]
\item
  For $1 \leq i < m$, set $A_{v_{i}} \df \mathcal{F}_{v_{i}}$.
\item
  For $v_{m}$, set $A_{v_{m}} \df \mathcal{F}_{v_{m}} \union A_{v_{m}}'$, where
  \begin{center}
    $A_{v_{m}}' \df \{\, (v_{r}, \{ v_j, v_r \}) \mid v_{r} \text{ is existentially
    quantified and }
j < r  < m \}$.
  \end{center}
\item
  For $1 \le i \le k$, set $A_{x_{i}} \df \mathcal F_{x_{i}} \cup A_{x_{i}}'$,
  where
  \begin{center}
    $A_{x_{i}}' \df \{\, (v_{r}, f_{v_{r}}, \{ v_j, v_r \}, x_{i}) \mid (v_{r}, f_{v_{r}})
    \in \mathcal F_{v_{r}} %
\text{ and }
j < r\leq m \}$.
  \end{center}
\end{itemize}
In the latter two set definitions here, it holds that $\{ v_j, v_r \} \in E_{\theta}$.
\end{definition}

Recall from~\cref{notation:conventions} that $s \in \mathcal{X}$ is arbitrary
but fixed.
The difference in the definitions of the domains, between the two structures,
lies precisely in the domain of variable $s$.
\begin{definition}[Domains for $\mathcal{B}$]
\label{defn:dom-B}
For each $z \in \mathcal{Z} \setminus \{ s \}$, define $B_{z} \df A_{z}$;
for each $z \in \mathcal{X} \union \{ v_{m} \}$, define $B_{z}' \df A_{z}'$.
Define $B_s \df \mathcal{F}_{s}' \cup B_{s}'$, where $ \mathcal{F}'_{s} \df
\bigl\{  (s, f_{s}) \mid f_{s} \colon E_{s} \rightarrow \{ 0, 1 \} \text{ has
odd parity} \}$.
\end{definition}

Before defining the interpretation of relation symbols, let us explain the roles of the various domain elements.
When all of the domains are restricted to the elements in the sets
$\mathcal{F}_{z}$, we essentially obtain a previous construction~\cite{BC19}
used to obtain bounds for the logical width of conjunctive queries.
In this previous construction, atoms were used to ensure that different
functions $f_{u} \colon E_{u }\to \{ 0, 1 \}$, $f_{v} \colon E_{v} \to \{ 0, 1
\}$ match on the edge $\{ u, v \}$---if this edge is in the primal graph.
The new elements, namely, the elements of $A_{v_{m}}' = B_{v_{m}}'$ and
$A_{x_{i}}' = B_{x_{i}}'$, are used to account for the edges of $G_{\theta}$
between variables $v_{1}, \ldots, v_{m}$ which \emph{are not} witnessed by
atoms, but \emph{are} taken into account to determine treewidth.
The new elements provide novel machinery needed to ensure that each pair of
functions $f_{u}, f_{v}$ match on the relevant edges.

Consider the evaluation game.
Let us say that a player plays \emph{inconsistently} in the $r$-th round (with
$r \in [m]$) if there is a previous index $j$, with $j < r$, such that $v_{j}$
and $v_{r}$ are set to pairs $(v_{j}, f_{v_{j}}), (v_{r}, f_{v_{r}})$ that do
not match on the common edge, that is, when $f_{v_{j}}( \{ v_{j}, v_{r} \}) \neq
f_{v_{r}}( \{ v_{j}, v_{r} \})$.
The structures are defined so that as soon as one player plays inconsistently,
the other has a winning strategy.
\begin{itemize}[$\bullet$]
\item
  If Existential is the first to play inconsistently by setting $v_{r}$ as
  above, then Universal can win by assigning $v_{m}$ to the element $(v_{r}, \{
  v_{j}, v_{r} \} )$ of $A_{v_{m}}'$; this will leave Existential with no way
  to win, and in particular forces Existential to set each variable $x_{i} \in
  \mathcal{X}$ to a value in~$A'_{x_{i}}$.
  In fact, the structures shall be defined such that when Universal assigns
  $v_{m}$ to any pair $(v_{r}, \{ v_{j}, v_{r} \})$ of $A'_{v_{m}}$, Existential
  has a winning strategy from this point if and only if there is a match on the
  edge $\{ v_{j}, v_{r} \}$.
  Observe that, among the variables $v_{1}, \ldots, v_{m}$, the only one that
  could be set to a value outside of a set $\mathcal{F}_{z}$ is $v_{m}$.
  Thus, $v_{m}$, which is the last universally quantified variable in the
  quantifier prefix, has a special role: it polices the Existential player.
\item
  If Universal is the first to play inconsistently by setting $v_{r}$ as above,
  then Existential can win by playing consistently from that point on, and then
  (assuming $v_{m}$ is not set to a pair in $A'_{v_{m}}$) can set its variables
  $x_{i} \in \mathcal{X}$ to values in~$A'_{x_{i}}$ that point out a mismatch
  on the edge $e = \{ v_{j}, v_{r} \}$.
  If $v_{m}$ is set to a pair in $A'_{v_{m}}$, Existential will win due to
  having played consistently.
  Thus, each variable $x_{i} \in \mathcal{X}$ can be used to police the
  Universal player.
\end{itemize}

When $\subs$ is a variable assignment that maps each $v_{i}$, for $i \in [m]$,
to some $(v_{i}, f_{v_{i}}) \in \mathcal{F}_{v_{i}}$, and each pair of functions
match on their shared edge (when it exists), $\mathcal{A}$ and $\mathcal{B}$
shall be defined such that $(\mathcal{A}, \subs) \models_\mathsf{var} \theta$
and $(\mathcal{B}, \subs) \not\models_\mathsf{var} \theta$.

The interpretations of relation symbols are defined based on how they appear as atoms in $\varphi$:
\begin{enumerate}[(1)]
\item\label{item:rel-1}
  $R(x_{i}, x_{j}) \in \phi$, for some $R \in \tau$ and $x_{i}, x_{j} \in
  \mathcal{X}$;
\item \label{item:rel-2}
  $P(v_{t}, x_{i}) \in \phi$, for some $P \in \tau$, $v_{t} \in \mathcal{V}$ and
  $x_{i} \in \mathcal{X}$.
\end{enumerate}
Up to reversing the order of the variables, these cover all cases due to our
assumption that $\varphi$ is simple (see~\cref{defn:rb-pointed}).
Note that for every $v_{t} \in \mathcal{V}$, there is some atom $P(v_{t}, x_{i})
\in \phi$ (or $P(x_{i}, v_{t}) \in \phi$) for some relation symbol $P$ and some
$x_{i} \in \mathcal{X}$: this is because the tightness condition ensures each
atom of $\phi$ contains at least one variable from $\mathcal{X}$, and we have
assumed each $v_{t} \in \mathcal{V}$ appears in some atom of $\phi$.

\begin{definition}
\label{defn:R-inter}
Let $R$ be a binary relation symbol of arity $\mathsf{ar}(R)= x_{i} \, x_{j}$,
where $x_{i},x_{j}\in\mathcal{X}$.
Then~$R^\mathcal{A}$ consists exactly of the tuples $(a_i,a_j)$ satisfying one
of the following:
\begin{enumerate}[(1)]
\item\label{R-inter:one}
  $a_i = (x_{i}, f_{x_{i}}) \in \mathcal{F}_{x_{i}}$ and $a_j = (x_{j},
  f_{x_{j}}) \in \mathcal{F}_{x_{j}}$, with $f_{x_{i}}(\{ x_{i}, x_{j} \}) =
  f_{x_{j}}(\{ x_{i}, x_{j} \})$;
\item\label{R-inter:two}
  $a_i = (v_{r}, f_{v_{r}}, e, x_{i}) \in A_{x_{i}}'$ and $a_j = (v_{r},
  f_{v_{r}}, e, x_{j}) \in A_{x_{j}}'$, for some shared $v_{r}$, $f_{v_{r}}$,
  $e$.
\end{enumerate}
The relation $R^{\mathcal{B}}$ is defined identically, but such that
$\mathcal{F}_{s}$ is replaced with~$\mathcal{F}'_s$, whenever used.
\end{definition}

Notice that $R^\mathcal{A}$ is a disjoint union of two parts.
The first is a subset of $\mathcal{F}_{x_{i}} \times \mathcal{F}_{x_{j}}$ where
the two functions agree on the common edge, and the second relates two elements
of $A_{x_{i}}'$ and $A_{x_{j}}'$ that share all but the last component.
This holds symmetrically for $R^\mathcal{B}$.

We now move on to the more complicated case, that is, the interpretation of
relation symbols $P$ where $P(v_{t}, x_{i}) \in \phi$ for $v_{t} \in
\mathcal{V}$ and $x_{i} \in \mathcal{X}$.

\begin{definition}
\label{defn:P-inter}
Let $P$ be a binary relation symbol of arity $\mathsf{ar}(P) = v_{t} \, x_{i}$,
for $v_{t} \in\mathcal V$ and $x_{i} \in \mathcal X$.
Then~$P^\mathcal{A}$ consists exactly of the pairs $(a, a')$ satisfying one of
the following:
\begin{enumerate}[(1)]
\item\label{item:2.1}
  $a = (v_{t}, g_{v_{t}}) \in \mathcal F_{v_{t}}$ and $a' = (x_{i}, f_{x_{i}})
  \in \mathcal F_{x_{i}}$, with $g_{v_{t}}(\{ v_{t}, x_{i} \}) = f_{x_{i}}(\{
  v_{t}, x_{i} \})$.
\item\label{item:2.5}
  $t = m$, $a = (v_{r}, e) \in A'_{v_{m}}$, and $a' = (v_{r}, f_{v_{r}}, e,
  x_{i}) \in A'_{x_{i}}$, for some shared $v_{r}$.

\item\label{item:2.3}
  $a= (v_{t} , g_{v_{t}}) \in \mathcal F_{v_{t}}$, with $t \in [m-1]$, and
  $a' = (v_{r}, f_{v_{r}}, e, x_{i})\in A_{x_{i}}'$, where $v_{r}$ is
  existentially quantified, and
  \begin{enumerate}[(a)]
  \item
    $t=r$ implies $g_{v_{t}}(e) = f_{v_{r}}(e)$, and
  \item
    $e = \{ v_{t}, v_{r} \}$ implies $g_{v_{t}}(e) = f_{v_{r}}(e)$.
  \end{enumerate}

\item\label{item:2.2}
  $a = (v_{t}, g_{v_{t}}) \in \mathcal F_{v_{t}}$ and $a' = (v_{r}, f_{v_{r}},
  e, x_{i}) \in A_{x_{i}}'$, where $v_{r}$ is universally quantified, and
  \begin{enumerate}[(a)]
  \item\label{U-inconsistent-1}
    $t = r$ implies $g_{v_{t}} = f_{v_{r}}$; and
  \item \label{U-inconsistent-2}
    $e = \{ v_{t}, v_{r} \}$ implies $g_{v_{t}}(e) \neq f_{v_{r}}(e)$.
  \end{enumerate}

\end{enumerate}
The relation $P^{\mathcal{B}}$ is defined identically, but such that
$\mathcal{F}_s$ is replaced with~$\mathcal{F}'_s$, whenever used.
\end{definition}

Let us revisit the intuition of $\mathcal{A}$ and $\mathcal{B}$.
Suppose Existential is the first to make an inconsistent pick by assigning
$v_r$ to $(v_r, f_{v_r})$ such that $f_{v_r}( \{ v_j, v_r \}) \neq f_{v_j}( \{ v_j, v_r \} )$
where $j < r$ and $v_j$ has previously been assigned $(v_j, f_{v_j})$.
If Universal picks $( v_r, \{ v_r, v_j \} ) \in A_{v_m}'$ for $v_m$, then
Existential is forced to assign every $x \in \mathcal{X}$ to the value
$(v_{r}, f_{v_{r}}', \{ v_{r}, v_{j} \}, x) \in A'_{x}$, each of which
shares the first three components (or Existential will lose).
This is because $( (v_{r}, \{ v_{r}, v_{j} \}), a') \in P_{1}^\mathcal{A}$ can
only hold when $a' \in A_{x_{i}}'$ (see~\cref{defn:P-inter}(\ref{item:2.5}))
along with some further reasoning that is deferred to the appendix.
Note that there are atoms $P_{1}(v_{m}, x_{i}) , P_{2}(v_{r}, x_\ell),
P_{3}(v_{j}, x_\jmath) \in \phi$, for some $x_{i}, x_\ell, x_\jmath \in
\mathcal{X}$, since $\varphi$ is a simple QCQ (see~\cref{defn:rb-pointed}).
However, for both $( (v_{r}, f_{v_{r}}), (v_{r}, f_{v_{r}}', \{ v_{r}, v_{j}
\}, x_\ell) )$ and $( (v_{j}, f_{v_{j}}), (v_{r}, f_{v_{r}}', \{ v_{r}, v_{j}
\}, x_\jmath) )$ to belong to $P_{2}^{\mathcal{A}}$ and $P_{3}^{\mathcal{A}}$,
respectively, \cref{defn:P-inter}(\ref{item:2.3}) requires that
$f_{v_{r}}(\{ v_{j}, v_{r} \}) = f_{v_{r}}'(\{ v_{j}, v_{r} \}) = f_{v_{j}}(\{
v_{j}, v_{r} \})$, which is impossible by our assumption.
Consequently, Universal has a winning strategy from the position where
Existential makes an inconsistent pick.

On the other hand, if Universal assigns $v_{m}$ to some element of $A'_{v_{m}}$
that does not point to an inconsistency, then Existential has a winning
strategy from that point.
This follows by the above reasoning, with the exception that
$f_{v_{r}}(\{ v_{j}, v_{r} \}) = f'_{v_{r}}(\{ v_{j}, v_{r} \}) = f_{v_{j}}(\{
v_{j}, v_{r} \})$ does indeed hold, along with some further reasoning.

Now suppose that Universal is the first to make an inconsistent pick by
assigning $v_{r}$ to $(v_{r}, f_{v_{r}})$ such that $f_{v_{r}}( \{ v_{j}, v_{r}
\}) \neq f_{v_{j}} (\{ v_{j}, v_{r} \})$, for some $j < r$.
In this case, Existential can play consistently until round $m$, and then
(assuming Universal does not assign $v_{m}$ to some element of $A_{v_{m}}'$, in
which case Existential can win, as described above) can assign each $x_{i} \in
\mathcal{X}$ to $(v_{r}, f_{v_{r}}, \{ v_{j}, v_{r} \}, x_{i}) \in A'_{x_{i}}$.
\begin{itemize}[$\bullet$]
\item
  For $P_{2}( v_{r}, x_{\ell} )$, we have $( (v_{r}, f_{v_{r}}), ( v_{r},
  f_{v_{r}}, \{ v_{j}, v_{r} \}, x_{\ell} ) ) \in P_{2}^{\mathcal{A}}$ by
  \cref{defn:P-inter}(\ref{U-inconsistent-1}).
\item
  For $P_{3} ( v_{j}, x_{\jmath} )$, we have $( (v_{j}, f_{v_{j}}), (v_{r},
  f_{v_{r}}, \{ v_{j}, v_{r} \}, x_{\jmath} ) ) \in P_{3}^{\mathcal{A}}$,
  by~\cref{defn:P-inter}(\ref{U-inconsistent-2}).
\item
  For every atom $P( v_q , x_{i'})$ with $v_q  \in \mathcal{V} \setminus \{
  v_{r}, v_{j} \}$, we have $( (v_q, f_{v_q}), (v_{r}, f_{v_{r}}, \{ v_{j},
  v_{r} \}, x_{i'}) ) \in P^\mathcal{A}$ by the fact that the antecedents for
  \cref{defn:P-inter}(\ref{U-inconsistent-1}) and
  \cref{defn:P-inter}(\ref{U-inconsistent-2}) are both false.
\end{itemize}
From this, we can show Existential has a winning strategy from such a position.
Therefore, when Universal makes an inconsistent pick first, Existential has a
winning strategy from that point.

\begin{example}
\label{example:incoherent}
Recall $\bar\varphi$ defined in~\cref{example:running}.
In $\mathcal{A}$, each variable $v$ of $\bar\varphi$ has a domain containing
$(v, f_{v})$ with $f_{v} \colon E_{v} \rightarrow \{ 0,1 \}$ having even parity.
Suppose $\subs(v_{1}) = (v_{1}, f_{v_{1}})$ with $f_{v_{1}}(\{ v_{1}, v_{2} \})
= 0, f_{v_{1}}(\{ v_{1}, x_{1} \}) = 0$ and $\subs(v_{2}) = (v_{2}, f_{v_{2}} )$
with $f_{v_{2}}(\{ v_{1}, v_{2} \}) = 1, f_{v_{2}}(\{ v_{2}, x_{2}\})=1$.
Then, let $\subs(u) = (v_{2}, f_{v_{2}}, \{ v_{1}, v_{2} \}, u)$ for $u \in \{
x_{1}, x_{2} \}$ and notice by following the definitions we get:
\begin{itemize}[$\bullet$]
\item
  $( (v_{1}, f_{v_{1}}), (v_{2}, f_{v_{2}}, \{ v_{1}, v_{2} \}, x_{1}) ) \in
  R_{1}^\mathcal{A}$, since $f_{v_{1}}(\{ v_{1}, v_{2} \}) \neq f_{v_{2}} (\{
  v_{1}, v_{2}\})$;
\item
  $( (v_{2}, f_{v_{2}}), (v_{2}, f_{v_{2}}, \{ v_{1}, v_{2} \}, x_{2}) ) \in
  R_{2}^\mathcal{A}$; and
\item
  $(  (v_{2}, f_{v_{2}}, \{ v_{1}, v_{2} \}, x_{1}),  (v_{2}, f_{v_{2}}, \{
  v_{1}, v_{2} \}, x_{2}) ) \in R_{3}^\mathcal{A}$.
\end{itemize}
Hence $(\mathcal{A}, \subs) \models R_{1}(v_{1}, x_{1}) \land R_{2}(v_{2},
x_{2}) \land R_{3}(x_{1}, x_{2})$.
\end{example}

\subsection{$\mathcal{A}$ and $\mathcal{B}$ disagree on $\varphi$}

The following notion allows us to rule out some strategies for Universal and
Existential in the evaluation games for $\varphi$ over $\mathcal A$ and
$\mathcal B$.

\begin{definition}
Consider the evaluation game for $\varphi$ over $\mathcal A$, or $\mathcal B$.
For a position $(r+1, \subs)$ of the evaluation game with $r \leq m$, we say that
Universal (Existential) has \emph{played inconsistently} if there exist indices 
$j < i \leq r$ where $Q_i = \forall$ (or $Q_i = \exists$, respectively) such that
$\subs(v_i) = (v_i, f_{v_i}) \in \mathcal{F}_{v_i}$ and 
$\subs(v_j) = (v_j, f_{v_j}) \in \mathcal{F}_{v_j}$
but where $f_{v_{i}}(\{ v_i, v_j \}) \neq f_{v_{j}}( \{ v_i, v_j \})$.
If a player has not played inconsistently, then they have \emph{played consistently}.
\end{definition}

We show that as soon as one player plays inconsistently, the other has a winning
strategy.
\begin{lemma}
\label{lem:first-inconsistent-loses}
Let $(r+1, \subs)$ be a position of the evaluation game for $\varphi$ over $\mathcal{A}$ or $\mathcal{B}$.
If one player played consistently while the other player played inconsistently, then the player that
played consistently has a winning strategy from the position $(r+1, \subs)$.
\end{lemma}

Thus, this proves that the only viable strategy for either Universal or
Existential is to play consistently.
Next, we rule out another potential strategy.

\begin{definition}
Let $(m + 1, \subs)$ be a position of the game for $\varphi$ over $\mathcal{A}$, where both players have played consistently.
Then, Universal \emph{played dishonestly} if $\subs(v_{m}) \in A'_{v_{m}}$,
and \emph{played honestly} otherwise.

Let $(m + r + 1, \subs)$ be a position of the game for $\varphi$ over $\mathcal{A}$, where $r \in [k]$, both players have
played consistently in the first $m$ rounds, and Universal has played honestly.
Existential \emph{played dishonestly} if $\subs(x_{r}) \in A'_{x_{r}}$,
and \emph{played honestly} otherwise.

We define having played honestly/dishonestly in the game for $\varphi$ over
$\mathcal{B}$ analogously.
\end{definition}

The following result shows us that for Universal (or Existential) to have any
chance of winning, they must play honestly.

\begin{lemma}
\label{lem:U-dishonest}
Consider the evaluation game for $\varphi$ over $\mathcal{A}$ (respectively,
over $\mathcal{B}$).
Then,
\begin{enumerate}[(1)]
\item\label{item-one-dishonest}
  If $(m + 1, \subs)$ is a position of the game where both players have played
  consistently, but Universal has played dishonestly, then Existential has a
  winning strategy from this position.
\item\label{item-two-dishonest}
  If $(m + r + 1, \subs)$ is a position of the game, where $r \in [k]$, such
  that both players have played consistently in the first $m$ rounds, Universal
  has played honestly, and Existential has played dishonestly, then Existential
  will lose from this position.
\end{enumerate}
\end{lemma}

We are now ready to show that the structure $\mathcal{A}$
satisfies $\varphi$, whereas the structure
$\mathcal{B}$ does not.

\begin{lemma}
\label{lemma:disagree}
$\mathcal{A} \models_\mathsf{var} \varphi$ and $\mathcal{B}
\not\models_\mathsf{var} \varphi$.
\end{lemma}
\begin{sketch}
Fix $(m + 1, \subs)$ to be a position of the evaluation game where both players
play consistently and Universal plays honestly.
We show that Existential has a winning strategy from this position for
$\mathcal{A}$, and does not have a winning strategy from this position for
$\mathcal{B}$.

\proofsubparagraph{Regarding $\mathcal{A} \models_\mathsf{var} \varphi$:}
Recall that $\subs(v_{m}) = (v_{m}, f_{v_{m}}) \in \mathcal{F}_{v_{m}}$ and
$f_{v_{i}}(\{ v_{i}, v_{j} \}) = f_{v_{j}}(\{ v_{i}, v_{j} \})$ for all $1 \leq
j < i \leq m$, since both players have played consistently and honestly.

Let $S \subseteq \mathcal{X}$, where $x \in S$ if and only if $\sum_{v_{i} \in
\mathcal{V}\,\cap\, N_{G_{\theta}}(x)} f_{v_{i}}(\{v_{i},x\}) \equiv 1 \pmod 2$.
Here, $N_{G_{\theta}}(x)$ denotes the neighbourhood of $x$ in the primal graph
$G_{\theta}$.
We can establish that $|S|$ is even.
Then by applying a graph-theoretic lemma, we get a subgraph of $G_{\theta}
|_\mathcal{X}$ whose set of odd-degree vertices is exactly $S$.
From this, we construct an extension of $\subs$ to all variables of $\varphi$
that satisfies $\phi$.

\proofsubparagraph{Regarding $\mathcal{B} \not\models_\mathsf{var} \varphi$:}
Let $\subs'$ be an assignment that extends $\subs$ and is defined on all
variables of $\varphi$.
We show that assuming $(\mathcal{B}, \subs') \models_\mathsf{var} \phi$ leads to
a contradiction.
For each $z$, we have $\subs'(z) = (z, f_{z})$ with $f_{z} \colon E_{z}
\rightarrow \{ 0,1 \}$, since play has been consistent and honest.
Moreover, all pairs of functions must agree on a shared edge.
Thus, we can consider every edge of $G_{\theta}$ to be labelled $0$ or $1$.
Summing the Boolean labels two ways, edgewise and vertexwise, we get a
mismatch in the parity of the two summations due to the fact that the labels
contributed by $s$ have odd parity.
Indeed, every edge contributes twice to the vertexwise sum, whereas the
vertexwise parities are even at all vertices except for the distinguished vertex
$s$.
\end{sketch}

\subsection{$\mathcal{A}$ and $\mathcal{B}$ are $\rho$-indistinguishable}

This section shows that Duplicator wins the variable-sorted $\rho$-pebble game
over $\mathcal{A}$ and $\mathcal{B}$, as defined in~\cref{sec:defining-A-B}.
Recall that $\rho = \tw(H_{\theta})$, where $\theta$ is the inner region of
$\varphi$.
The proof generalises the pebble game argument given in~\cite{BC19}.
A full proof is given in the appendix to deal with
the differences between the structures defined in this paper and those
defined in~\cite{BC19}.

\begin{definition}
Let $W = (a_{1}, \dots, a_{\ell})$ be a walk in the graph $G_{\theta}$.
For $e \in E_{\theta}$, let
\[
  \mathsf{occ}_W(e)
\df
  | \{ i \in [\ell-1] \mid e = \{ a_i, a_{i+1} \} \} |.
\]
That is, $\mathsf{occ}_W(e)$ is the number of times the edge $e$ appears in the
walk $W$.
Recall that $\mathcal{M}$ is a fixed bramble of $G_{\theta}$ with order $>
\rho$.
Now, for each $S \subseteq \mathcal{Z}$, let
\[
  \mathsf{avoid}_\mathcal{M}(S)
\df
  \bigcup_{\{ M \in \mathcal{M} \mid M \intersect S = \emptyset \} } M.
\]
Thus $\mathsf{avoid}_\mathcal{M}(S)$ denotes the union of the bramble sets in
$\mathcal{M}$ that are disjoint from $S$.
\end{definition}

Recall that for a tuple $\vec t = (t_{1}, \dots, t_{r})$, we set $\pi_i(\vec t)
= t_i$, for every $i \in [r]$.

We remind the reader of the definition of the domains of $\mathcal{A}$ and
$\mathcal{B}$, see~\cref{defn:dom-A} and~\cref{defn:dom-B}.
Observe that, with the exception of elements from $A_{v_{m}}'$ and $B_{v_{m}}'$,
every element of $\mathcal{A}$ and $\mathcal{B}$ is of the following form: the
first component is a vertex of $G_{\theta}$ and the second component is a
function that labels edges incident to the first component in $G_{\theta}$.
More formally, for $a \in A$, we have that $\pi_{1}(a) \in \mathcal{Z}$ and if
$a \notin A_{v_{m}}'$ also holds, then $\pi_{2}(a)$ is a function that maps
$E_{\pi_{1}(a)}$ to $\{ 0, 1 \}$.

Our aim is to show that Duplicator has a winning strategy for the $\rho$-pebble
game over $\mathcal{A}$ and $\mathcal{B}$.
To do so, Duplicator fixes a bramble $\mathcal{M}$ of $G_{\theta}$, and
maintains the following invariant:

\emph{
Suppose that $i \leq \rho$ pebbles are placed on $a_{1}, \dots, a_i \in A$
and on the corresponding elements $b_{1}, \dots, b_i \in B$.
Then, there exists a walk $W$ in $G_{\theta}$ from the distinguished element $s$
to some vertex in $\mathsf{avoid}_\mathcal{M}( \{ \pi_{1}(a_{1}), \dots,
\pi_{1}(a_i) \})$ such that:
\begin{itemize}[$\bullet$]
\item
  For every $j \in \{ 1, \dots, i \}$, if $\pi_{2}(b_j)$ is a function, then for
  every edge $e$ incident to $\pi_1(a_j)$, we have that $\pi_{2}(b_j)(e)$ is
  equal to $\pi_{2}(a_j)(e) + \mathsf{occ}_W(e) \pmod{2}$.
  If $\pi_{2}(b_j)$ is not a function, then $a_j = b_j$.
\end{itemize}
}
We note that this is essentially the same as Duplicator's strategy given in~\cite{BC19}.
To formalise the above intuition, let us introduce the following definition.

\begin{definition}
Let $W$ be a walk in $G_{\theta}$.
For every function $f \colon E_{v} \rightarrow \{ 0, 1 \}$, for some vertex $v$
of $G_{\theta}$, let $f^W \colon E_{v} \rightarrow \{ 0, 1 \}$ denote $f^W(e)
= f(e) + \mathsf{occ}_W(e) \pmod{2}$, for every $e \in E_{v}$.
Define $\Psi_W \colon A \rightarrow B$ as:
\begin{itemize}[$\bullet$]
\item
  if $a \in A_{v_{m}}'$, then $\Psi_W(a) = a$, and
\item
  otherwise, $\Psi_W(a)$ is obtained by replacing the second component
  $\pi_{2}(a)$ by $\pi_{2}(a)^W$ and keeping all other components unchanged.
\end{itemize}
\end{definition}

\begin{example}
Recall $\bar\varphi$ defined in~\cref{example:running}.
The hypergraph of the inner region $\theta$ contains the edges  $ \{
v_{1}, v_{2} \} , \{ v_{1}, x_{1} \}, \{ v_{2}, x_{2} \}, \{ x_{1}, x_{2} \} $.
Let $f_{x_{1}} \colon E_{x_{1}} \rightarrow \{0 ,1 \}$ be defined by $f_{x_{1}}
( \{ v_{1}, x_{1} \}) = 0$ and $f_{x_{1}} ( \{ x_{1} , x_{2} \}) = 0$.
Let $W = (x_{1}, v_{1}, x_{1}, x_{2}, v_{2})$ be a walk in $G_{\theta}$ and
observe $\mathsf{occ}_W( \{ x_{1}, v_{1} \}) = 2$ and $\mathsf{occ}_W ( \{
x_{1}, x_{2} \}) = 1$.
Hence, $\Psi_W ( x_{1} , f_{x_{1}} ) = (x_{1}, f_{x_{1}}^W)$ where $f_{x_{1}}^W
( \{ x_{1}, v_{1} \}) = 0$ and $f_{x_{1}}^W( \{ x_{1} , x_{2} \}) = 1$.
\end{example}

\begin{lemma}
\label{lemma:partial-iso}
Let $0 \leq i \leq \rho$ and let $\{ a_1, \dots, a_i \} \subseteq A$.
Let $W$ be a walk in $G_{\theta}$ from $s$ to a vertex in
  $\mathsf{avoid}_{\mathcal{M}}(\{ \pi_1(a_1), \dots, \pi_1(a_i) \} )$.
Then, the mapping $\Psi_W$ restricted to $\{ a_1, \dots, a_i \}$ is a partial
isomorphism between $\mathcal{A}$ and $\mathcal{B}$.
\end{lemma}

Using~\cref{lemma:partial-iso} as an invariant condition, we get the following
result.

\begin{lemma}
\label{lemma:winning-strat}
Duplicator has a winning strategy for the $\rho$-pebble game over $\mathcal{A}$
and $\mathcal{B}$.
\end{lemma}

Notice that $\rho$-many pebbles can hit at most $\rho$ vertices, and since the
bramble has an order greater than $\rho$, some bramble set will be unhit.
Thus, Duplicator can play in a way that maintains this invariant, by defining a
walk from $s$ to some unhit bramble set.
The proofs of~\cref{lemma:partial-iso} and~\cref{lemma:winning-strat} are
generalisations of the proofs of Lemma 14 and Lemma 16 in~\cite{BC19},
respectively.

To sum up, this section constructed structures $\mathcal{A}$ and $\mathcal{B}$
such that $\mathcal{A} \models_\mathsf{var} \varphi$, $\mathcal{B}
\not\models_\mathsf{var} \varphi$, and $\mathcal{A} \equiv^\rho_\mathsf{var}
\mathcal{B}$.
Thus, by~\cref{lemma:var-sorted} and the pebble-game characterisation recalled
in Section~2, we have that $\logwidth(\varphi) \geq \tw(H_{\theta}) + 1$, where
$\theta$ is the inner region of $\varphi$.

\section{Proof Sketch of~\cref{thm:main-exists-and}}
\label{sec:reductions}

This section illustrates why the width lower bound for simple QCQs is sufficient
to prove~\cref{thm:main-exists-and}.
The key idea is to define a sequence $\varphi_{1}, \varphi_{2}, \dots,
\varphi_{5}$ of formulae, where $\varphi_1$ is any self-join-free QCQ that
is a normal form of $[\varphi_{1}]_\mathbb{T}$ in the system $\mathscr{R}$ and $\varphi_{5}$ is a simple QCQ.
Moreover, we define this sequence such that for each $i \in \{ 2, \dots, 5 \}$
if there are structures $\mathcal{A}^{i}$ and $\mathcal{B}^{i}$ that
disagree on $\varphi_{i}$ but $\mathcal{A}^i \equiv^k_\mathsf{var} \mathcal{B}^i$,
then there are structures $\mathcal{A}^{i - 1}$ and
$\mathcal{B}^{i - 1}$ that disagree on $\varphi_{i - 1}$ but $\mathcal{A}^{i-1} \equiv^k_\mathsf{var} \mathcal{B}^{i-1}$.

For the following series of reductions, we require the following definition: For
a $\{ \forall \}$ region $\psi$ of some QCQ $\varphi$ with $|\free(\psi)| \geq
1$, the \emph{contraction of $\varphi$ at $\psi$}, denoted by
$\varphi^{\mathsf{ctr}}_\psi$ is defined as follows.
Let $F_\psi \df \{ z_{1}, \dots, z_{m} \}$ be the free variables of $\psi$,
given an arbitrary ordering denoted by the subscripts.
Introduce a new and unique binary relation $R_{z_{i}, z_{j}}^\psi$, for each
pair $z_{i}, z_{j} \in F_\psi$ with $i \leq j$.
Then, $\varphi^{\mathsf{ctr}}_\psi$ is obtained by replacing the occurrence of
the subformula rooted at
$\psi$ in $\varphi$ with $\bigland_{1 \leq i \leq j \leq m} R_{z_{i},
z_{j}}^\psi (z_{i}, z_{j})$.
Let $\varphi_{1}$ be a standardised self-join-free QCQ such that
$[\varphi_{1}]_\mathbb{T}$ is a normal form in the system $\mathscr{R}$, and let
$\theta$ be a $\{ \exists, \land \}$-region of $\varphi_1$ with maximum treewidth.
Our aim is to prove $\logwidth(\varphi_1) > \tw(H_\theta)$.
The reduction is structured as follows.
\begin{enumerate}[(1)]
\item
  Contract all $\{ \forall\}$ regions that occur below $\theta$ and that have
  an atomic child region.
  Set $\varphi_{2}$ to be this new self-join-free QCQ.
\item
  To obtain $\varphi_{3}$, replace all atoms of $\varphi_{2}$ of the form
  $R(z_{1}, \dots, z_{m})$ with $m > 2$ with the conjunction $\bigwedge_{1 \leq
  i < j \leq m} R_{i, j} (z_{i}, z_{j})$, where $R_{i, j}$ is a new relation
  symbol that is unique to $R$, $i$, and $j$.
  Thus, $\varphi_{3}$ is a self-join-free QCQ over a signature that only
  contains atoms with at most two variables.
\item \label{item:reduct-4}
  The next step in the reduction is to obtain $\varphi_{4}$ by contracting each
  non-atomic child region of~$\theta$.
  This is perhaps the most technical step in the reduction, since if
  $\psi(z_{1}, \dots, z_{r})$ is a child region of~$\theta$, then we wish to
  define the interpretation of relation symbols that occur in $\psi$ in such a
  way as to simulate the conjunction $\bigwedge_{1 \leq i \leq j\leq r}
  R_{z_{i}, z_{j}}^\psi (z_{i}, z_{j})$.
\item
  Finally, $\varphi_{5}$ is obtained from $\varphi_{4}$ by restricting
  $\varphi_{4}$ to a sub-signature such that every atom of $\varphi_{5}$
  contributes to the maximum treewidth region, and adheres to the definition of
  a simple QCQ.
  This final QCQ~$\varphi_{5}$ can then be shown to be a simple QCQ with an
  inner region that has the same treewidth as the maximum treewidth $\{ \exists, \land \}$-region of
  our original QCQ $\varphi_{1}$.
\end{enumerate}

By~\cref{lemma:disagree} and~\cref{lemma:winning-strat}, there are
variable-sorted structures $\mathcal{A}$ and $\mathcal{B}$ that disagree on
$\varphi_5$, but for which $\mathcal{A} \equiv^{\tw(H_\theta)}_\mathsf{var}
\mathcal{B}$.
Applying the reductions in reverse order essentially ``expands'' these two
structures to structures $\mathcal{A}^+$ and $\mathcal{B}^+$ that disagree on
$\varphi_1$ but for which $\mathcal{A}^+ \equiv^{\tw(H_\theta)}_\mathsf{var}
\mathcal{B}^+$.
Consequently, we achieve the desired lower bound.

We conclude this section by noting that the reductions performed in this section
require that $\varphi_1$ is a self-join-free QCQ.
To see why, consider~\cref{item:reduct-4} above.
That is, $\varphi_4$ is obtained by replacing each non-atomic child region $\psi(z_{1}, \dots, z_{r})$ of $\theta$
 with $\bigwedge_{1 \leq i \leq j\leq r}
  R_{z_{i}, z_{j}}^\psi (z_{i}, z_{j})$ where each $R_{z_i, z_j}^\psi$ is a fresh relation symbol.
We define the relation symbols that appear in $\psi(z_{1}, \dots, z_{r})$ to simulate
$\bigwedge_{1 \leq i \leq j\leq r}
  R_{z_{i}, z_{j}}^\psi (z_{i}, z_{j})$
  in order to transfer width lower bounds from $\varphi_4$ back to $\varphi_3$.
  Since we have a self-join-free QCQ, we may interpret the relation symbols in $\psi(z_1, \dots, z_r)$
  without affecting any other regions as each relation symbol
may only appear in at most one region.

\section{Conclusions}
\label{sec:conclusions}
\newcommand{\relb}{\mathbf{B}}

The main result of this article analysed self-join-free QCQs, and showed that there is an algorithm that, given a self-join-free QCQ $\phi$, outputs a second QCQ whose width is minimal over all first-order sentences that are logically equivalent to $\phi$.  Our analysis revealed that this second QCQ is obtainable from the original QCQ $\phi$ via the application of syntactic rewriting rules.

Beyond self-join-freeness being a natural syntactic restriction, there is a technical reason why we view the understanding of self-join-free QCQs as crucial.  In the case of \emph{conjunctive queries} (for short, \emph{CQs}), which we here define as QCQs where universal quantification is not used, the logical width has been analysed~\cite{BC19}.
For a given CQ $\phi$, the logical width can be computed by passing to the \emph{core} of $\phi$, and then taking the treewidth (plus one) of the core.
Indeed, one can apply the rewriting methods described in this article to minimise the core's width.
To prove that the treewidth (plus one) of a CQ's core is equal to the CQ's logical width, the mentioned analysis~\cite{BC19} can be viewed as an analysis of a self-join-free CQ obtained from the core by renaming relation symbols so that no relation symbol is repeated.\footnote{The construction of the two structures in Section 3.1 of \cite{BC19} can be seen to specify, for each atom $R(a_1, \ldots, a_r)$ of a core, relations $R^{\relb}$ and $R^{\relb'}$ whose elements have the form $((a_1, f_1), ..., (a_r, f_r))$.}
That is, for a given CQ $\phi$, the logical width can be characterised by first performing the semantic step of passing to the core of $\phi$, and then syntactically measuring the resulting core, where repeated relation symbols are not given any special treatment.

We regard obtaining an understanding of and an algorithm for logical width minimisation of general QCQs as the most immediate and notable issue left open by this article.
The situation for CQs suggests a parallel approach to understanding the logical width of general QCQs: we conjecture that the logical width of a given QCQ $\phi$ can be obtained by passing to a core $\psi$ of $\phi$, and then using this article's rewriting algorithm to find a minimum width QCQ among those in
$[\psi]_{\mathbb{A}}$.
This approach would involve introducing the notion of a core of a QCQ, but
the work of Bova and Mogavero~\cite{BM17} gives hope that such a core notion can
be defined and analysed, as it gives homomorphism-based characterisations of
QCQ entailment.

Another open issue is to understand the relaxation of QCQs where free variables are permitted.  Here, we can
mention that, in the case of CQs, the corresponding relaxation was analysed; the main argumentation went into understanding
sentences, and then the more general case of formulas could be readily derived
from the case of sentences~\cite{BC19}.

Finally, we wish to mention one more open issue.  The main construction of this article gives, for each self-join-free QCQ, a pair of structures that disagree on the QCQ but are locally indistinguishable.  This construction %
generalises a previous one for CQs~\cite{BC19}, and that previous construction can be viewed as a relational analogue of the standard Tseitin parity system~\cite{ItsyksonRiazanovSmirnov2022}, which is used widely in proof complexity.  As this article's construction generalises that previous construction to a setting where both quantifiers are present, we end by posing a question: can one perform a transfer back in the other direction, and present applications of our construction to proof complexity?

\begin{acks}
The authors would like to thank the anonymous reviewers for their numerous
helpful comments and suggestions, and
Stan P. Hauke for thorough proofreading
of this work.
H. Chen and S. M. Thompson acknowledge the support of the EPSRC grant
EP/V039318/1; the bulk of the research and article preparation was performed while S. M. Thompson was employed at King's College London.
H. Chen also acknowledges the support of the EPSRC grant
 EP/X03190X/1.
F. Mogavero acknowledges the support of the University of Naples grant FRA
FORMALG (CUP:E65F22000060001)
\end{acks}

\bibliographystyle{ACM-Reference-Format}
\bibliography{References}

\appendix

\section{Appendix}\label{sec:main-appendix}

This appendix contains all omitted details from~\cref{sec:qsg},
except for the proof of~\cref{lemma:winning-strat}, which is omitted
due to space constraints.

Let us start with the following result that states any satisfying assignment, in either $\mathcal{A}$ or $\mathcal{B}$, that maps some $x_i \in \mathcal{X}$ to some $4$-tuple $(v_r, f_{v_r}, e, x_i)$, must map all variables in $\mathcal{X}$ to a $4$-tuple with the same first three components.

\begin{lemma}\label{lemma:X-values}
Suppose $\subs \colon \mathcal{Z} \rightarrow A$ has $(\mathcal{A}, \subs) \models_\mathsf{var} \phi$.
If $\subs(x_i) = (v_r, f_{v_r}, e, x_i) \in A_{x_i}'$
for some $x_i \in \mathcal{X}$,
then $\subs(x_j) = (v_r, f_{v_r}, e, x_j) \in A_{x_j}'$ for all $x_j \in \mathcal{X}$.
The analogous result holds for~$\mathcal{B}$.
\end{lemma}
\begin{proof}
We prove this result for $\mathcal{A}$; the proof for $\mathcal{B}$ is totally analogous.
Fix $\subs \colon \mathcal{Z} \rightarrow A$ where $\subs(x_i) = (v_r, f_{v_r}, e, x_i) \in A_{x_i}'$ for some $x_i \in \mathcal{X}$, and suppose that $(\mathcal{A}, \subs) \models_\mathsf{var} \phi$.
For any $R(x_p, x_q) \in \phi$ with $x_p, x_q \in \mathcal{X}$, we have that $(\mathcal{A}, \subs) \models_\mathsf{var} R(x_p, x_q)$ and $\subs(x_p) = (v_r, f_{v_r}, e, x_p)$ if and only if $\subs(x_q) = (v_r, f_{v_r}, e, x_q)$; see~\cref{defn:R-inter}.
Now by \cref{defn:rb-pointed}(\ref{rbp3}), the primal graph
$G_\theta|_{\mathcal X}$ is connected.
Thus, for any pair $x_r,x_s\in\mathcal X$, there is a walk
$(y_0,y_1,\ldots,y_m)$ from $y_0=x_r$ to $y_m=x_s$ consisting entirely of
elements from $\mathcal X$.
For every $\ell\in[m]$, since $\{y_{\ell-1},y_\ell\}$ is an edge of $G_\theta$,
there is a hyperedge of $H_\theta$ containing both $y_{\ell-1}$ and $y_\ell$.
Since both variables belong to $\mathcal X$, this hyperedge cannot be
$\free(\theta)$; hence, by \cref{defn:rb-pointed}(1), there is an atom of $\phi$
containing both $y_{\ell-1}$ and $y_\ell$.
\end{proof}

\subsection{Proof of~\cref{lem:first-inconsistent-loses}}
\begin{proof}
We prove the statement for $\mathcal{A}$; the argument for $\mathcal{B}$ is identical.
Recall that
\[
\varphi = Q_1 v_1\; Q_2 v_2 \cdots Q_{m-1} v_{m-1}\; \forall v_m\; \exists x_1\cdots \exists x_k(\phi).
\]

Let $(r+1, \subs)$ be a position of the evaluation game with $r \leq m$ such that one player has played consistently while the other player has played inconsistently.
In particular, let $i$ and $j$ be values with $j < i \leq r$ where
$\subs(v_i) = (v_i, f_{v_i})$, where $\subs(v_j) = (v_j, f_{v_j})$, and where 
$f_{v_i} (\{ v_i, v_j \}) \neq f_{v_j} (\{ v_i, v_j\})$.
Since exactly one player played inconsistently, we split this proof into
the following two cases.

\proofsubparagraph{Case 1. Universal played inconsistently:} 
Thus, $Q_i = \forall$ and Universal chose $(v_i, f_{v_i})$ such that  
$f_{v_i}(\{ v_i, v_j \}) \neq f_{v_j}(\{ v_i, v_j \})$.
Existential's strategy is to first play consistently up to the position $(m+1, \subs)$.
From this position, we consider two subcases; 
one where Universal has picked $\subs(v_m) \in A_{v_m}'$, 
and one where Universal has picked $\subs(v_m) \in \mathcal{F}_{v_m}$.

First, consider the case where $\subs(v_m) \in A_{v_m}'$.
That is, $\subs(v_m) = (v_\ell, \{ v_\ell, v_\jmath \})$ where $v_\ell$ is existentially quantified, and $\jmath < \ell$.
For this case, Existential's winning strategy from position $(m+1, \subs)$ is to pick
$(v_\ell, f_{v_\ell}, \{ v_\ell, v_\jmath \}, x)$ for each $x \in \mathcal{X}$, 
where $\subs(v_\ell) = (v_\ell, f_{v_\ell})$.
We shall verify that all atoms of $\phi$ are satisfied when $\subs$ is the 
assignment resulting from the final position under this strategy.

If $R(x_p,x_q)\in\phi$ with $x_p,x_q\in\mathcal{X}$,
then $\bigl((v_\ell, f_{v_\ell}, \{ v_\ell, v_\jmath \}, x_p),(v_\ell, f_{v_\ell}, \{ v_\ell, v_\jmath \}, x_q)\bigr)\in R^{\mathcal{A}}$
by~\cref{defn:R-inter}(\ref{R-inter:two}).
If $P(v, x) \in \phi$ with $v \in \mathcal{V}$ and $x \in \mathcal{X}$, there are several subcases:
\begin{enumerate}
\item If $v=v_m$, then $\bigl( (v_\ell, \{ v_\ell, v_\jmath \}),  (v_\ell, f_{v_\ell}, \{ v_\ell, v_\jmath \}, x) \bigr) \in P^\mathcal{A}$ by~\cref{defn:P-inter}(\ref{item:2.5}).
\item If $v=v_\ell$, then $\bigl( (v_\ell, f_{v_\ell}), (v_\ell, f_{v_\ell}, \{ v_\ell, v_\jmath \}, x) \bigr) \in P^\mathcal{A}$ by~\cref{defn:P-inter}(\ref{item:2.3}).
\item If $v=v_\jmath$, then $\bigl( (v_\jmath, f_{v_\jmath}), (v_\ell, f_{v_\ell}, \{ v_\ell, v_\jmath \}, x) \bigr) \in P^\mathcal{A}$
by~\cref{defn:P-inter}(\ref{item:2.3})
along with the fact that $f_{v_\jmath}( \{ v_\ell, v_\jmath \}) = f_{v_\ell} ( \{ v_\ell, v_\jmath \})$, because $v_\ell$ is existentially quantified and Existential plays consistently.
\item If $v \notin \{ v_m, v_\ell, v_\jmath \}$, then $\bigl( (v, f_v), (v_\ell, f_{v_\ell}, \{ v_\ell, v_\jmath \}, x) \bigr) \in P^\mathcal{A}$, where $\subs(v) = (v, f_{v}) \in \mathcal{F}_v$,  by~\cref{defn:P-inter}(\ref{item:2.3}) along with the fact that $v \neq v_\ell$ and $\{ v_\ell, v_\jmath \} \neq \{ v, v_\ell \}$.
\end{enumerate}

Next, consider the case where Universal has picked $\subs(v_m) \in \mathcal{F}_{v_m}$.
For each $\ell \in [k]$, define Existential's choice as 
$\subs(x_\ell) \df (v_i, f_{v_i}, \{ v_i, v_j \}, x_\ell) \in A_{x_\ell}'$. 
We shall show that this is indeed a winning strategy from the position $(m+1, \subs)$.
Recall that $v_i$ is universally quantified, and Universal picked 
$\subs(v_i) = (v_i, f_{v_i})$ such that 
$f_{v_i}( \{ v_i, v_j \} ) \neq f_{v_j}(\{v_i, v_j \})$.
Again, we verify that all atoms in $\phi$ are satisfied under the assignment 
$\subs$ that results from the final position under this strategy.

If $R(x_p,x_q)\in\phi$ with $x_p,x_q\in\mathcal{X}$, then 
$\bigl( (v_i, f_{v_i}, \{ v_i, v_j \}, x_p) , (v_i, f_{v_i}, \{ v_i, v_j \}, x_q) \bigr)\in R^{\mathcal{A}}$ by~\cref{defn:R-inter}(\ref{R-inter:two}).
If $P(v, x) \in \phi$ with $v \in \mathcal{V}$ and $x \in \mathcal{X}$, there are several subcases:
\begin{enumerate}
\item If $v=v_i$, then $\bigl( (v_i , f_{v_i}) , (v_i, f_{v_i}, \{ v_i, v_j \}, x) \bigr)\in P^{\mathcal{A}}$ by Definition~\ref{defn:P-inter}(\ref{item:2.2}).
\item If $v=v_j$, then $( (v_j,f_{v_j}),  (v_i, f_{v_i}, \{ v_i, v_j \}, x_i) ) \in P^\mathcal{A}$
because $f_{v_i}( \{ v_i, v_j \} ) \neq f_{v_j}( \{ v_i, v_j \} )$; 
which is the condition required in the definition of $P^\mathcal{A}$ for this case; 
see~\cref{defn:P-inter}(\ref{item:2.2}).
\item If $v \notin \{ v_i, v_j \} $, then 
  $((v,f_{v}), (v_i, f_{v_i}, \{ v_i, v_j \}, x) ) \in P^\mathcal{A}$, 
  where $\subs(v) = (v, f_v)$, because $v \neq v_i$ and $\{v, v_i \}\neq\{ v_i , v_j \}$; 
  see~\cref{defn:P-inter}(\ref{item:2.2}).
\end{enumerate}

Thus, for both cases ($\subs(v_m) \in A_{v_m}'$ or $\subs(v_m) \in
\mathcal{F}_{v_m}$), Existential can play in such a way that the resulting
assignment is $\subs$ where $(\mathcal{A},\subs)\models_\mathsf{var}\phi$.

\proofsubparagraph{Case 2. Existential played inconsistently:} 
Thus, $Q_i = \exists$ and Existential chose $(v_i , f_{v_i})$ such that 
$f_{v_i}( \{ v_i, v_j \} ) \neq f_{v_j}( \{ v_i, v_j \} )$.
Universal's strategy is first to play consistently up to position $(m, \subs)$.
Then, for $\forall v_m$, let Universal pick 
$\subs(v_m)\df (v_i, \{ v_i, v_j \} )\in A_{v_m}'$.

Towards a contradiction, suppose that 
$(\mathcal{A}, \subs) \models_\mathsf{var} \exists x_1 \dots \exists x_k (\phi) $.
Since $\varphi$ is simple, there is an atom of the form $P(v_m, x) \in \phi$ 
for some $x \in \mathcal{X}$ and $P \in \tau$, 
see~\cref{defn:rb-pointed}(\ref{rbp5}) and~\cref{defn:rb-pointed}(\ref{rbp4}).
By the definition of $P^\mathcal{A}$, we have that 
$( (v_i, \{ v_i, v_j \} ),(v_i, f_{v_i}', \{ v_i, v_j \}, x) )$, 
for some $f_{v_i}' \colon E_{v_r} \rightarrow \{ 0, 1\}$ having even parity, 
are the only tuples in $P^\mathcal{A}$ with $(v_r, \{ v_j, v_r \} )$ 
being the first component,
see~\cref{defn:P-inter}(\ref{item:2.5}).
Furthermore, by~\cref{lemma:X-values}, it follows that each $x \in \mathcal{X}$ 
must then be witnessed by~$( v_i, f_{v_i}', \{ v_i, v_j \}, x)$.

Again, since $\varphi$ is simple, there exist some $T, Q \in \tau$ 
such that $T(v_i, x_\ell), Q(v_j, x_\jmath) \in \phi$ with 
$x_\ell, x_\jmath \in \mathcal{X}$.
We know that $x_\ell$ and $x_\jmath$ are witnessed by 
$( v_i, f_{v_i}', \{ v_i, v_j \}, x_\ell)$ and $( v_i, f_{v_i}', \{ v_i, v_j \}, x_\jmath)$ respectively.
Let us now look at $T^\mathcal{A}$ and $Q^\mathcal{A}$:
\begin{itemize}
\item $( (v_i, f_{v_i}), ( v_i, f_{v_i}', \{ v_i, v_j \}, x_\ell) ) \in T^\mathcal{A}$
implies
$f_{v_i}( \{ v_i, v_j \}) = f_{v_i}' ( \{ v_i, v_j \})$.
\item $( (v_j, f_{v_j}), ( v_i, f_{v_i}', \{ v_i, v_j \}, x_\jmath) ) \in Q^\mathcal{A}$
implies
$f_{v_j}( \{ v_i, v_j \}) = f_{v_i}' ( \{ v_i, v_j \})$.
\end{itemize}
Both of the above items hold due to~\cref{defn:P-inter}(\ref{item:2.3}).
We have thus reached a contradiction since 
$f_{v_i}( \{ v_i, v_j \}) \neq f_{v_j}( \{ v_i, v_j \})$.
Thus, if $(r+1, \subs)$ is a position where Existential played inconsistently and Universal has played consistently, then Universal has a winning strategy.

Combining the two cases ($Q_i= \forall$ and $Q_i = \exists$) concludes the proof.
\end{proof}

\subsection{Proof of~\cref{lem:U-dishonest}}
\begin{proof}
We prove this result by establishing~\cref{item-one-dishonest} and~\cref{item-two-dishonest} separately over $\mathcal{A}$, since the proof for $\mathcal{B}$ is analogous.

\proofsubparagraph{Regarding~\cref{item-one-dishonest}:}
Let $(m+1, \subs)$ be a position of the evaluation game where both players have played consistently,
but Universal has played dishonestly by choosing $\subs(v_m)=(v_r,e)\in A_{v_m}'$ with $e=\{v_j,v_r\}$ and $j<r$, where $v_r$ is existentially quantified.
Since play was consistent, we have that $f_{v_r}(e)=f_{v_j}(e)$.
Existential now chooses $\subs(x_\ell)=(v_r,f_{v_r},e,x_\ell)\in A_{x_\ell}'$ for every $x_\ell \in\mathcal{X}$.
By~\cref{defn:R-inter}, all $R(x_p,x_q)$-atoms with $x_p, x_q \in \mathcal{X}$ are satisfied, and all $P(v,x)$-atoms with $v \in \mathcal{V}$ and $x \in \mathcal{X}$ are satisfied due to the following reasoning:
\begin{itemize}
\item If $v = v_m$, then $( (v_r, e), (v_r,f_{v_r},e,x) ) \in P^\mathcal{A}$ from~\cref{defn:P-inter}(\ref{item:2.5}).
\item If $v = v_r$, then $( (v_r, f_{v_r}), (v_r,f_{v_r},e,x) ) \in P^\mathcal{A}$ from~\cref{defn:P-inter}(\ref{item:2.3}).
\item If $v = v_j$, then $( (v_j, f_{v_j}), (v_r,f_{v_r},e,x) ) \in P^\mathcal{A}$ from~\cref{defn:P-inter}(\ref{item:2.3}) along with the fact that the functions for $\subs(x_\ell)$ and $\subs(v)$ agree on $e$; that is, $f_{v_j}(e) = f_{v_r}(e)$.
\item Otherwise, $((v, f_v), (v_r,f_{v_r},e,x) ) \in P^\mathcal{A}$, with $\subs(v) = (v, f_v)$, follows from~\cref{defn:P-inter}(\ref{item:2.3}) along with the fact that $e \neq \{ v, v_r \}$ and $v \neq v_r$.
\end{itemize}
Hence, every atom of $\phi$ is satisfied by the resulting assignment $\subs$.
Thus, $(\mathcal{A},\subs)\models_\mathsf{var}\phi$, and Existential has a winning strategy from the position $(m+1,\subs)$.

\proofsubparagraph{Regarding~\cref{item-two-dishonest}:}
Let $(m+\ell+1,\subs)$ be a position of the game, where $\ell\in[k]$, such that
both players have played consistently in the first $m$ rounds, Universal has
played honestly, but Existential has played dishonestly, that is, $\subs(x_\ell)
=(v_r,f'_{v_r},\{v_j,v_r\},x_\ell) \in A'_{x_\ell}$, for some $j<r$.
Notice by~\cref{lemma:X-values}, it holds that Existential must pick $\subs(x) =
(v_r, f_{v_r}', \{ v_j, v_r \} , x)$ for each $x \in \mathcal{X}$ to have any
chance of winning.

Let $\subs \colon \mathcal{Z} \rightarrow A$ be the resulting assignment here.
We now show that assuming $(\mathcal{A}, \subs) \models_\mathsf{var} \phi$ results in a contradiction:
Consider the atom $P(v_m, x) \in \phi$ for some $x \in \mathcal{X}$ which must exist due to the fact that $\varphi$ is simple; see~\cref{defn:rb-pointed}(\ref{rbp5}) and~\cref{defn:rb-pointed}(\ref{rbp4}).
By our assumption that $(\mathcal{A}, \subs) \models_\mathsf{var} \phi$, it must hold that $( (v_m, f_{v_m}), (v_r, f_{v_r}',  \{ v_j, v_r \}, x) ) \in P^\mathcal{A}$.
By~\cref{defn:P-inter}, this only holds when $v_r$ is universally quantified since there are no tuples with $( a, (v_r, f_{v_r}',  \{ v_j, v_r \}, x) ) \in P^\mathcal{A}$ with $a \in \mathcal{F}_{v_m}$ and $v_r$ being existentially quantified.

Now consider the atoms $T(v_r, x)$ and $Q(v_j, x')$, for some $x, x' \in \mathcal{X}$ and $T, Q \in \tau$, which again must appear in $\phi$ since $\varphi$ is simple.
We have that $\subs(v_r) = (v_r, f_{v_r})$, $\subs(v_j) = (v_j, f_{v_j})$, and $\subs(x) = (v_r, f_{v_r}', \{ v_r, v_j \} , x)$ for each $x \in \mathcal{X}$.
Since $(\mathcal{A}, \subs) \models_\mathsf{var} \phi$ it follows
\begin{itemize}
\item $( (v_r, f_{v_r}), (v_r, f_{v_r}',  \{ v_j, v_r \}, x) ) \in T^\mathcal{A}$ which implies $f_{v_r} = f_{v_r}'$; see~\cref{defn:P-inter}(\ref{item:2.2}).
\item $( (v_j, f_{v_j}), (v_r, f_{v_r}',  \{ v_j, v_r \}, x) ) \in Q^\mathcal{A}$ which implies $f_{v_j}(\{ v_j, v_r\}) \neq f_{v_r}'(\{ v_j, v_r\})$; see~\cref{defn:P-inter}(\ref{item:2.2}).
\end{itemize}
However, we know that both players played consistently which means $f_{v_r}( \{ v_j, v_r \})= f_{v_j}( \{ v_j, v_r \})$.
Thus, we have reached a contradiction.
\end{proof}

\subsection{Proof of~\cref{lemma:disagree}}
Let us state the following well-known graph theory result, see Section 16.4 of~\cite{BM08b} for example.

\begin{lemma}\label{lemma:t-join}
Let $G \df (V, E)$ be a connected graph and let $K \subseteq V$ such that $|K|$ is even.
Then, there exists a subgraph of $G$ where every vertex in $K$ has odd degree, and every other vertex has even degree.
\end{lemma}

Let us now give the actual proof of~\cref{lemma:disagree}.

\begin{proof}
Recall that \[ \varphi \df Q_1 v_1 \, Q_2 v_2 \dots Q_{m-1} v_{m-1} \, \forall v_m \exists x_1 \dots \exists x_k ( \phi ). \]
Fix $(m+1,\subs)$ to be a position of the evaluation game where both players play consistently and Universal plays honestly.
That is, $\subs(v_i) = (v_i, f_{v_i}) \in \mathcal{F}_{v_i}$ for $1 \leq i \leq m$, and $f_{v_i} (\{ v_i, v_j \}) = f_{v_j}(\{ v_i, v_j \})$ for each $1 \leq i < j \leq m$.
We show that Existential has a winning strategy from this position for $\mathcal{A}$, and Universal has a winning strategy from this position for $\mathcal{B}$.

\proofsubparagraph{Regarding $\mathcal{A} \models_\mathsf{var} \varphi$:}
Recall that $\subs(v_m) = (v_m, f_{v_m}) \in \mathcal{F}_{v_m}$ and
\[ f_{v_i}(\{v_i,v_j\}) = f_{v_j}(\{v_i,v_j\}) \qquad \text{for all}  \qquad 1 \leq j < i \leq m, \]
since both players have played consistently and Universal has played honestly.

Let
\[
S \df \left\{ x \in \mathcal{X} \;\middle|\; \sum_{v_i \in \mathcal{V}\,\cap\, N_{G_\theta}(x)} f_{v_i}(\{v_i,x\}) \equiv 1 \pmod 2 \right\},
\]
where $N_{G_\theta}(x)$ denotes the neighborhood of $x$ in the primal graph $G_\theta$.

Let us establish that $|S|$ is even:
Since $f_{v_i}$ has even parity for every $v_i \in \mathcal{V}$, we have
\[
\sum_{v_i \in \mathcal{V}} \sum_{e \in E_{v_i}} f_{v_i}(e) \equiv 0 \pmod 2.
\]
For each edge $\{v_i,v_j\} \in \binom{\mathcal{V}}{2}$, set
$b_{v_i,v_j} \df f_{v_i}(\{v_i,v_j\})$.
Notice that $b_{v_i,v_j}=f_{v_j}(\{v_i,v_j\})$ since both players have played consistently.
It follows that
\[
\sum_{v_i \in \mathcal{V}} \sum_{e \in E_{v_i}} f_{v_i}(e)
\equiv
 \sum_{\{v_i,v_j\} \in \binom{\mathcal{V}}{2}} 2  b_{v_i,v_j} \;+\;
\sum_{x \in \mathcal{X}} \ \sum_{v_i \in \mathcal{V}\cap N_{G_\theta}(x)} f_{v_i}(\{v_i,x\}) \pmod 2.
\]
By the definition of $S$, for any $x \in \mathcal{X}$ we have that
\[\sum_{v_i \in \mathcal{V}\cap N_{G_\theta}(x)} f_{v_i}(\{v_i,x\}) \equiv 1 \pmod{2} \quad \iff \quad x \in S\]
Furthermore, it is immediate that
$\sum_{\{v_i,v_j\} \in \binom{\mathcal{V}}{2}} 2  b_{v_i,v_j}  \equiv 0 \pmod{2}$.
Therefore,
\[
0 \equiv \sum_{v_i \in \mathcal{V}} \sum_{e \in E_{v_i}} f_{v_i}(e)
\equiv |S| \pmod 2.
\]

Since $|S|$ is even and $G_\theta|_{\mathcal{X}}$ is connected (by~\cref{defn:rb-pointed}(\ref{rbp3})),
we may apply Lemma~\ref{lemma:t-join} to obtain a subgraph $H \df (V(H), E(H))$ of $G_\theta|_{\mathcal{X}}$ whose set of odd-degree vertices is exactly $S$.
For each $x_i \in \mathcal{X}$, define a function $f_{x_i}\colon E_{x_i}\to\{0,1\}$ by:
\begin{itemize}
\item for every $\{x_i,x_j\}\in E_\theta$ with $x_j \in \mathcal{X}$, set
$ f_{x_i}(\{x_i,x_j\}) = 1 \quad \text{iff} \quad \{x_i,x_j\}\in E(H)$.
\item for every $\{v_\ell,x_i\}\in E_\theta$ with $v_\ell\in \mathcal{V}$, set
$
f_{x_i}(\{v_\ell,x_i\}) = f_{v_\ell}(\{v_\ell,x_i\}).$
\end{itemize}

We claim that each $f_{x_i}$ has even parity.
The degree of $x_i$ in $H$, which we will denote by $\mathsf{deg}_H(x_i)$, is odd if and only if $x_i \in S$.
Moreover,
\[ \sum_{v_\ell\in \mathcal{V}\cap N_{G_\theta}(x_i)} f_{x_i}(\{v_\ell,x_i\}) = \sum_{v_\ell\in \mathcal{V}\cap N_{G_\theta}(x_i)} f_{v_\ell}(\{v_\ell,x_i\}) ,  \]
by the definition of $f_{x_i}$, which is also odd iff $x_i\in S$ by definition of $S$.
Therefore,
\[ \sum_{e \in E_{x_i}} f_{x_i} (e) \equiv \mathsf{deg}_H(x_i) + \sum_{v_\ell\in \mathcal{V}\cap N_{G_\theta}(x_i)} f_{v_\ell}(\{v_\ell,x_i\}) \equiv 0 \pmod{2} \]
and so $(x_i,f_{x_i})\in \mathcal{F}_{x_i}$ for all $i \in [k]$.

Now define an assignment $\subs' \colon \mathcal{Z}\to A$ by letting $\subs'(v_i)=\subs(v_i)$ for all $v_i\in\mathcal{V}$ and $\subs'(x_i)=(x_i,f_{x_i})$ for all $x_i\in\mathcal{X}$.
We show that $(\mathcal{A},\subs')\models_\mathsf{var} \phi$.
Let $Q(z,z')$ be any atom of $\phi$.
Since the query is simple, $Q(z,z')$ is either of the form $R(x_i,x_j)$ for some $x_i, x_j \in \mathcal{X}$ and $R \in \tau$, or it is of the form $P(v_r,x_i)$ for some $v_r \in \mathcal{V}$, $x_i \in \mathcal{X}$ and $P \in \tau$.
In the first case, by the construction of $f_{x_i}$, we have
$f_{x_i}(\{x_i,x_j\})=f_{x_j}(\{x_i,x_j\})$, hence $((x_i,f_{x_i}),(x_j,f_{x_j}))\in R^\mathcal{A}$.
In the second case, by construction
\[
f_{v_r}(\{v_r,x_i\}) = f_{x_i}(\{v_r,x_i\}),
\]
and so $((v_r,f_{v_r}),(x_i,f_{x_i}))\in P^\mathcal{A}$.
Thus every atom of $\phi$ is satisfied under $\subs'$, i.e. $(\mathcal{A},\subs')\models_\mathsf{var} \phi$,
and hence $(\mathcal{A},\subs)\models_\mathsf{var} \exists x_1\cdots \exists x_k(\phi)$.

\proofsubparagraph{Regarding $\mathcal{B} \not\models_\mathsf{var} \varphi$:}
Towards a contradiction, suppose that Existential has a winning strategy from $(m+1, \subs)$.
Let $\subs'$ be the final position such that $(\mathcal{B}, \subs') \models_\mathsf{var} \phi$ where both players have played consistently and honestly.
Thus $\subs'(z) = (z, f_{z}) \in \mathcal{F}_z'$ for all $z \in \mathcal{Z}$.
Furthermore, for any $z, z' \in \mathcal{Z}$ where $\{ z, z' \}$ is an edge of $G_\theta$, we have that $f_z ( \{ z, z' \}) = f_{z'}( \{ z, z' \})$; otherwise $( \subs'(z), \subs'(z') ) \notin T^\mathcal{B}$ where $T \in \tau$ with $\mathsf{ar}(T) = z {z'}$; recall~\cref{defn:R-inter} and~\cref{defn:P-inter}.
By definition, we have that
\[ \sum_{e \in E_z} f_z(e) \equiv
\begin{cases}
1 \pmod{2}, & \text{ if $z = s$} \\
0 \pmod{2}, & \text{ otherwise}
\end{cases}
 \]
Letting $b_{z, z' } \df f_z(\{z, z' \}) = f_{z'}(\{z, z' \})$ for any $\{z, z' \} \in E_\theta$, we get
\[
1  \equiv \sum_{e \in E_s} f_s(e) +  \sum_{z \in \mathcal{Z} \setminus \{ s\}} \sum_{e \in E_z} f_z(e) = \sum_{\{ z, z' \} \in E_\theta} 2 \cdot b_{z, z'}  \equiv 0 \pmod{2}
\]
and thus gives us the desired contradiction.
\end{proof}

\subsection{Proof of~\cref{lemma:partial-iso}}
\begin{proof}
Let $h$ be the mapping $\Psi_W$ restricted to $\{ a_1, \dots, a_i \}$ and fix $b_\ell \df h(a_\ell)$ for each $\ell \in [i]$.
We first establish the following claim:

\begin{claim}
$h$ is an injective partial mapping from $A$ to $B$.
\end{claim}
\begin{claimproof}
First, we show that $b_\ell \in B$ for those $\ell \in [i]$ where $a_\ell \notin A_{v_m}'$; the case where $a_\ell \in A_{v_m}'$ is obvious since $B_{v_m}' = A_{v_m}'$.

We distinguish between two cases:
\begin{description}
\item[Case 1, $\pi_1(a_\ell) = s$:] For this case, we have that
$\sum_{e \in E_s} \mathsf{occ}_W(e) \equiv 1 \pmod{2}$
since $W$ starts at $s$, ends at $t \neq s$, and $G_\theta$ does not contain self-loops.
Notice that $a_\ell$ is of the form $(s, f_s)$ where $f_s \colon E_s \rightarrow \{ 0, 1\}$ such that
$ \sum_{e \in E_s} f_s(e) \equiv 0 \pmod{2} $
because these tuples $(s, f_s)$ are the only elements of $A$ where the first component is $s$.
Consequently, we have that
$\sum_{e \in E_s} f_s^W(e) \equiv 1 \pmod{2}$
 and thus $b_\ell = (s, f_s^W) \in B$.
\item[Case 2, $\pi_1(a_\ell) \neq s$:] Let $\pi_1(a_\ell) = z$ for some $z \in \mathcal{Z}$.
We have that
$ \sum_{e \in E_z} \mathsf{occ}_W(e) \equiv 0 \pmod{2}$
since the walk $W$ does not start or end at $z$, and $G_\theta$ does not contain self-loops.
Thus,
$ \sum_{e \in E_z} \pi_2(a_\ell)^W(e) \equiv 0 \pmod{2} $
 which means that $b_\ell \in B$ by definition.
\end{description}
We have now established that $h$ is indeed a partial mapping from $A$ to $B$ and it is clear by definition that $h$ is injective.
Furthermore, $h$ clearly respects the sorts.
\end{claimproof}

All that remains is to establish the following claim:
\begin{claim}
For any $j, \ell \in [i]$ and any relation symbol $T \in \tau$, we have that
\[ (a_j, a_\ell) \in T^\mathcal{A} \iff (h(a_j), h(a_\ell)) \in T^{\mathcal{B}}. \]
\end{claim}
Let $j, \ell \in [i]$ be arbitrary but fixed values, and let $T \in \tau$ be an arbitrary but fixed relation symbol.
We prove the above claim for two cases:

\proofsubparagraph{Case 1, $a_j, a_\ell \notin A_{v_m}'$:}
It immediately follows that $b_j, b_\ell \notin B_{v_m}'$.
To ease some notation, let
\begin{itemize}
\item $z_j = \pi_1(a_j)$ and $f_j = \pi_2(a_j)$,
\item $z_\ell = \pi_1(a_\ell)$ and $f_\ell = \pi_2(a_\ell)$,
\item $z_j' = \pi_1(b_j)$ and $f_j' = \pi_2(b_j)$, and
\item $z_\ell' = \pi_1(b_\ell)$ and $f_\ell' = \pi_2(b_\ell)$.
\end{itemize}

Note that by~\cref{defn:R-inter} and~\cref{defn:P-inter}, if
\[ f_j( e) = f_\ell( e ) \iff f_j'( e ) = f_\ell'( e ). \]
for any $e \in \domain(f_j) \intersect \domain(f_\ell)$, then
\[ (a_j, a_\ell) \in T^\mathcal{A} \iff (b_j, b_\ell) \in T^{\mathcal{B}}  \]
for any $T \in \tau$.
For some intuition as to why this holds, the only difference between $a_j$ and $b_j$ (as well as $a_\ell$ and $b_\ell$) is that we have swapped out $f_j$ with $f_j'$ (respectively, $f_\ell$ with $f_\ell'$).
Therefore, the truth values for the antecedents of the conditional definitions in~\cref{defn:P-inter}, such as~\cref{defn:P-inter}(\ref{item:2.2}), remain unchanged.

($\Rightarrow$)
Assume $f_j( e) = f_\ell( e )$.
By definition, there is a walk $W$ from $s$ to $\mathsf{avoid}_{\mathcal{M}}( \{ \pi_1(a_1), \dots, \pi_1(a_i) \} )$ such that
\begin{align*}
f_j'(e) & \equiv f_j(e) + \mathsf{occ}_W(e) \pmod{2}, \\
f_\ell'(e) & \equiv f_\ell(e) + \mathsf{occ}_W(e) \pmod{2}
\end{align*}
for any edge $e \in E_\theta$.
Thus, it follows that $ f_j'( e ) = f_\ell'( e )$.

($\Leftarrow$)
Assume $f_j'( e) = f_\ell'( e )$.
Similarly, we have a walk $W$ from $s$ to $\mathsf{avoid}_{\mathcal{M}}(\pi_1(a_1), \dots, \pi_1(a_i))$ such that
\begin{align*}
f_j(e) & \equiv f_j'(e) - \mathsf{occ}_W(e) \pmod{2}, \\
f_\ell(e) & \equiv f_\ell'(e) - \mathsf{occ}_W(e) \pmod{2} 
\end{align*}
for any edge $e \in E_\theta$.
Thus, it follows that $ f_j( e ) = f_\ell( e )$.

Due to the fact that $v_j = v_j'$ and $v_\ell = v_\ell'$ and moreover, if $\pi_3(a_j)$ or $\pi_4(a_j)$ is defined, then $\pi_3(a_j) = \pi_3(b_j)$ and $\pi_4(a_j) = \pi_4(b_j)$ it follows (from the definition of the interpretation of relations in $\mathcal{A}$ and $\mathcal{B}$)
\[ (a_j, a_\ell) \in T^\mathcal{A} \iff (b_j, b_\ell) \in T^{\mathcal{B}}. \]
Thus, we are done for this case.

\proofsubparagraph{Case 2, $a_j \in A_{v_m}'$ or $a_\ell \in A_{v_m}'$:}
Suppose first that $a_j \in A_{v_m}'$.
By definition of the mapping $h$, we have $b_j = a_j$, since elements of $A_{v_m}'$ are left unchanged.
Thus, it suffices to show that for every $a_\ell$ and every relation symbol $T\in\tau$,
\[
(a_j,a_\ell)\in T^{\mathcal A} \iff (a_j,b_\ell)\in T^{\mathcal B}.
\]

By construction of the structures, elements of $A_{v_m}'$ can only appear as the
first component of some relation $P^\mathcal{A}$ where $P(v_m,x_i)$ occurs in $\phi$.
More precisely, if $a_j=(v_r,e)\in A_{v_m}'$, then $\bigl((v_r,e),(v_r,f_{v_r},e,x_i)\bigr)\in P^{\mathcal A}$ by~\cref{defn:P-inter}(\ref{item:2.5}),
and no other tuples in $P^{\mathcal A}$ have first component $(v_r,e)$.
The same holds in $\mathcal B$, since the interpretation of $P$ is defined analogously.

Hence, $(a_j,a_\ell)\in P^{\mathcal A}$ if and only if $a_\ell$ is some $(v_r,f_{v_r},e,x_i) \in A_{x_i}'$ for some $x_i \in \mathcal{X}$.
By the definition of $h$, we have that $h( (v_r, f_{v_r}, e, x_i) ) = (v_r, f_{v_r}^W, e, x_i)$ and thus we immediately have that
$(a_j, (v_r, f_{v_r}^W, e, x_i)) \in P^\mathcal{B}$ by~\cref{defn:P-inter}(\ref{item:2.5}).
Likewise, if $( h(a_j), h(a_\ell)) \in P^\mathcal{B}$ with $a_j \in A_{v_m}'$, then $a_j = h(a_j)$ and $h(a_\ell) = (v_r, f_{v_r}^W, e, x_i)$ and thus $(a_j, a_\ell) \in P^\mathcal{A}$.

The case where $a_\ell \in A_{v_m}'$ is obvious:
elements of $A_{v_m}'$ never occur as the second component of any relation.

Consequently, for the case where at least one of $a_j,a_\ell$ lies in $A_{v_m}'$, we have
\[
(a_j,a_\ell)\in T^{\mathcal A} \iff (b_j,b_\ell)\in T^{\mathcal B}.
\]
This completes the proof that $h$ is a partial isomorphism.
\end{proof}

\section{Omitted Details From~\cref{sec:prelims}}
\subsection{Proof of~\cref{lemma:var-sorted}}
\begin{proof}
We shall define the one-sorted structure $\bar{\mathcal{A}}$ given the variable-sorted structure $\mathcal{A}$, the definition for $\bar{\mathcal{B}}$ follows analogously.
Let $x_1, \dots, x_m$ be the variables that appear in $\varphi$, and let $A_{x_i}$ denote the domain for each $x_i$ in the structure $\mathcal{A}$.
Recall that the domains for variables are pairwise disjoint in a variable-sorted structure.

We may assume $\varphi$ is in prenex normal form for the following reason: since $\varphi$ is standardised, we have that $(Qx \alpha) \land \beta$ is logically equivalent to $Qx (\alpha \land \beta)$ as $x \notin \mathsf{free}(\beta)$; likewise, $\alpha \land (Qx \beta)$ is logically equivalent to $Qx (\alpha \land \beta)$.
Thus, let $\varphi = Q_1 x_1 \cdots Q_m x_m \phi$ where $Q_i \in \{ \forall, \exists \}$ for each $i \in [m]$ and $\phi$ is a self-join-free conjunction of atoms.
Further assume, without loss of generality, that $x_i \in \mathsf{free}(\phi)$ for each $i \in [m]$.

Let $\bar{A} \df \bigsqcup_{i \in [m]} A_{x_i}$ be the domain for $\bar{\mathcal{A}}$.
For each atom $R(x_{i_1}, x_{i_2}, \dots,  x_{i_r})$ that appears in $\varphi$, we shall define $R^{\bar{\mathcal{A}}}$ based on $R^{{\mathcal{A}}}$ along with how $x_{i_1}, \dots, x_{i_r}$ are quantified.
Let 
\begin{align*}
E_R & \df \{ p \in [r] \mid x_{i_p} \text{ is existentially quantified} \},\\
U_R & \df \{ p \in [r] \mid x_{i_p} \text{ is universally quantified} \} .
\end{align*}
Since $\varphi$ is standardised and self-join-free, $E_R$ and $U_R$ are well-defined.
For $(a_1, \dots, a_r) \in \bar{A}^r$, let $(a_1, \dots, a_r) \in R^{\bar{\mathcal{A}}}$ if and only if:
\begin{itemize}
\item For every $p \in E_R$, we have that $a_{p} \in A_{x_{i_p}}$, and either
\begin{enumerate}
\item  $a_{q} \in A_{x_{i_q}}$ for every $q \in U_R$, and $(a_1, \dots, a_r) \in R^\mathcal{A}$; or

\item $a_{q} \notin A_{x_{i_q}}$ for some $q \in U_R$.
\end{enumerate}
\end{itemize}

We now show that $\mathcal{A} \models_{\mathsf{var}} \varphi$ if and only if $\bar{\mathcal{A}} \models \varphi$.
The argument for $\mathcal{B}$ and $\bar{\mathcal{B}}$ is identical.

Let us proceed via an evaluation game argument (refer  back to~\cref{defn:eval-game} for the formal definition of the evaluation game).
When restricting the game to variable-sorted logic, we simply add the additional constraint that both Universal and Existential must pick an element from the correct domain (i.e., if they are picking an element to assign to $x_i$, then that element belongs to $A_{x_i}$).
Since $\varphi$ is fixed, we simply say ``the game over $\mathcal{A}$'' to denote the evaluation game over $\mathcal{A}$ and $\varphi$, and likewise with $\bar{\mathcal{A}}$.

($\mathcal{A} \models_\mathsf{var} \varphi \implies \bar{\mathcal{A}} \models \varphi$)
Suppose Existential has a winning strategy over the variable-sorted structure $\mathcal{A}$.
Let us define a winning strategy for Existential over $\bar{\mathcal{A}}$ using their winning strategy over $\mathcal{A}$ as a guide.
First, for each $\ell \in [m]$ with $Q_\ell = \forall$, fix some arbitrary element $d_\ell \in A_{x_\ell}$.

In the game over $\bar{\mathcal{A}}$, for any $i \in [m]$ with $Q_i = \forall$, suppose Universal picks $a$ for $x_i$.
We simulate this move in the game over $\mathcal{A}$ by assigning $x_i$ to $a$ if $a \in A_{x_i}$, and to $d_i$ otherwise.
Existential then plays the game over $\bar{\mathcal{A}}$ according to their strategy in $\mathcal{A}$.
Thus, for $j \in [m]$ with $Q_j = \exists$, we have that the value chosen for $x_j$ belongs to $A_{x_j}$.

Let $\bar\subs$ be the assignment resulting from the final position in the game over $\bar{\mathcal{A}}$ where Existential plays the above-defined strategy, and let $\subs$ be the assignment resulting from the final position in the game over $\mathcal{A}$ where Existential plays with their winning strategy.
Since Existential has a winning strategy over $\mathcal{A}$, we have that $(\mathcal{A}, \subs) \models_\mathsf{var} \phi$.
Let $R(x_{i_1}, \dots, x_{i_r})$ be an arbitrary atom in $\phi$.
If every universally quantified variable among $x_{i_1}, \dots, x_{i_r}$ is assigned by $\bar\subs$ to its intended domain, then $\bar\subs$ and $\subs$ agree on $x_{i_1}, \dots, x_{i_r}$.
Since $(\mathcal A,\subs)\models_{\mathsf{var}} R(x_{i_1}, \dots, x_{i_r})$, it follows from the definition of $R^{\bar{\mathcal A}}$ that $(\bar{\mathcal A},\bar\subs)\models R(x_{i_1}, \dots ,x_{i_r})$.
Otherwise, at least one universally quantified variable $x_{i_j}$ with $1 \leq j \leq r$ is assigned outside its intended domain by $\bar\subs$. 
Since Existential's choices all lie in their intended domains, the definition of $R^{\bar{\mathcal A}}$ again implies that $(\bar{\mathcal A},\bar\subs)\models R(x_{i_1}, \dots, x_{i_r})$.

Therefore, the defined strategy for Existential in $\bar{\mathcal{A}}$ is indeed a winning strategy, since every atomic formula in $\phi$ is satisfied by the resulting variable assignment.

($ \bar{\mathcal{A}} \models \varphi \implies \mathcal{A} \models_\mathsf{var} \varphi$)
Assume that Existential has a winning strategy in the game over the one-sorted structure $\bar{\mathcal{A}}$.
Let us define a winning strategy for Existential in $\mathcal{A}$ using the game over $\bar{\mathcal{A}}$ as a guide.
For each choice Universal makes in the game over $\mathcal{A}$, mirror that choice in the game over $\bar{\mathcal{A}}$, and let Existential respond to both games using their winning strategy over $\bar{\mathcal{A}}$.
Notice that in the game on $\bar{\mathcal{A}}$, if $Q_i = \exists$, then Existential must pick some $a \in A_{x_i}$ for $x_i$.
This is because no variable assignment $\subs$ with $\subs(x_i) \notin A_{x_i}$ can satisfy any atom $R(x_{i_1}, \dots,  x_{i_r}) \in \phi$ with $x_i = x_{i_j}$ for some $j \in [r]$.
Therefore, if Existential picks some $a \notin A_{x_i}$, then they will lose. 
Thus, any pick adhering to Existential's winning strategy in the game over $\bar{\mathcal{A}}$ is also a valid choice for Existential in $\mathcal{A}$.
Consequently, Existential's winning strategy in the game over $\bar{\mathcal{A}}$ immediately transfers to a winning strategy in the game over $\mathcal{A}$.

We have established that $\mathcal{A} \models_\mathsf{var} \varphi \iff \bar{\mathcal{A}} \models \varphi$.
The symmetric reasoning means $\mathcal{B} \models_\mathsf{var} \varphi \iff \bar{\mathcal{B}} \models \varphi$.
Indeed, if there exist two variable-sorted structures $\mathcal{A}$ and $\mathcal{B}$ such that $\mathcal{A} \models_{\mathsf{var}} \varphi$ and $\mathcal{B} \not\models_{\mathsf{var}} \varphi$, then there exist one-sorted structures $\bar{\mathcal{A}}$ and $\bar{\mathcal{B}}$ such that $\bar{\mathcal{A}} \models \varphi$ and $\bar{\mathcal{B}} \not\models \varphi$.

($\mathcal{A} \equiv^k_\mathsf{var} \mathcal{B} \implies \bar{\mathcal{A}} \equiv^k \bar{\mathcal{B}}$)
Assume $\mathcal{A} \equiv^k_\mathsf{var} \mathcal{B}$.
We show that Duplicator has a winning strategy for the $k$-pebble game over the one-sorted structures $\bar{\mathcal{A}}$ and $\bar{\mathcal{B}}$.

Duplicator uses their winning strategy from the variable-sorted game for the one-sorted game.
Since the domains of the variables are disjoint, every element $a \in \bar A$ belongs to exactly one set $A_{x_i}$, and likewise every $b \in \bar B$ belongs to exactly one $B_{x_i}$.
Thus, whenever Spoiler pebbles an element $a \in \bar A$, Duplicator treats this like Spoiler pebbling $a \in A_{x_i}$.
Duplicator then responds in the game over $\mathcal{A}$ and $\mathcal{B}$ by pebbling $b \in B_{x_i}$, and for the game over $\bar{\mathcal{A}}$ and $\bar{\mathcal{B}}$, Duplicator mirrors this choice by pebbling the same $b \in \bar B$.
The case where Spoiler pebbles an element of $\bar B$ is symmetric.

We now show that this is a winning strategy for Duplicator.
Let $(\alpha, \beta)$ be an arbitrary position of the game over $\mathcal{A}$ and $\mathcal{B}$ where Duplicator plays with their winning strategy; thus implying $(\alpha,\beta)$ forms a partial isomorphism between $\mathcal{A}$ and $\mathcal{B}$.
By Duplicator's strategy for the game over $\bar{\mathcal{A}}$ and $\bar{\mathcal{B}}$, this is also a position for the one-sorted game.
We shall show that $(\alpha, \beta)$ forms a partial isomorphism between $\bar{\mathcal{A}}$ and $\bar{\mathcal{B}}$.
It is immediate that $\{ (\alpha(p), \beta(p) ) \mid p \in \domain(\alpha) \}$ defines an injective partial mapping.

Notice that for every pebble $p \in \domain(\alpha)$, we have that $\alpha(p) \in A_{x_i}$ if and only if $\beta(p) \in B_{x_i}$; we call this \emph{domain preservation} for this proof.
Recall the sets $E_R$ and $U_R$ denote which components of $(x_{i_1}, \dots, x_{i_r})$ are existentially quantified and universally quantified.

Let $R( x_{i_1}, \dots, x_{i_r} )$ be an arbitrary atom of $\varphi$.
Let $(a_1, \dots, a_r)$ be a tuple of pebbled elements of $\bar A$, and let $(b_1, \dots, b_r)$ be the corresponding tuple of pebbled elements of $\bar B$.
By domain preservation, we have that $a_p \in A_{x_{i_p}}$ if and only if $b_p \in B_{x_{i_p}}$.
We now establish that:
\[ (a_1, \dots, a_r) \in R^{\bar{\mathcal{A}}} \implies (b_1, \dots, b_r) \in R^{\bar{\mathcal{B}}}. \]
The converse direction is symmetric.

Suppose $(a_1, \dots, a_r) \in R^{\bar{\mathcal{A}}}$.
By the definition of $R^{\bar{\mathcal{A}}}$, we have that $a_p \in A_{x_{i_p}}$ for every $p \in E_R$.
Therefore, by domain preservation, $b_p \in B_{x_{i_p}}$ for every $p \in E_R$.
We now consider two cases:
\begin{enumerate}
\item There is some $p \in U_R$ such that $a_p \notin A_{x_{i_p}}$. By domain preservation, we have that $b_p \notin B_{x_{i_p}}$ and therefore $(b_1, \dots, b_r) \in R^{\bar{\mathcal{B}}}$ by definition.
\item For all $p \in U_R$ we have $a_p \in A_{x_{i_p}}$. Thus, each component of $(a_1, \dots, a_r)$ and $(b_1, \dots, b_r)$ lies in its intended domain.
This implies $(a_1, \dots, a_r) \in R^\mathcal{A}$ and thus $(b_1, \dots, b_r) \in R^\mathcal{B}$.
Therefore, by definition, $(b_1, \dots, b_r) \in R^{\bar{\mathcal{B}}}$.
\end{enumerate}

Consequently, $(\alpha,\beta)$ forms a partial isomorphism between $\bar{\mathcal{A}}$ and $\bar{\mathcal{B}}$ and thus Duplicator has a winning strategy for the $k$-pebble game between $\bar{\mathcal{A}}$ and $\bar{\mathcal{B}}$ as required.
\end{proof}

\section{Omitted Details From~\cref{sec:main}}

\subsection{Proof of~\cref{main:thm}}
\begin{proof}
Given a self-join-free QCQ $\varphi$, apply the algorithm of~\cref{lemma:algo} to obtain $\varphi^+ \in [\varphi]_{\mathbb{A}}$ where $\varphi^+$ has the minimum width among all formulae in $[\varphi]_\mathbb{A}$.
Thus, by~\cref{thm:width-treewidth}, we have that $\mathsf{width}(\varphi^+) = \mathsf{logwidth}(\varphi)$.
\end{proof}

\subsection{Proof of~\cref{cor:complexity}}
\begin{proof}
  For membership in NP, one first computes the normal form used by the algorithm
  of~\cref{lemma:algo}, together with its region hypergraphs.
  This can be done in polynomial time~\cite[Theorem 4.2]{CM26} before the calls
  for minimum tree decompositions.
  As described after~\cref{lemma:algo}, the regions are then minimised
  independently, and their minimum widths are given by their respective
  treewidths plus one.
  Hence, $\logwidth(\phi) \leq k$ precisely when every such region hypergraph
  has treewidth at most $k - 1$.
  Therefore, it suffices to guess and verify a tree decomposition of width at
  most $k - 1$ for each region hypergraph.
  Since there are polynomially many such hypergraphs, each of polynomial size,
  these decompositions form a polynomial-size certificate.

  For NP-hardness, let $G = (V, E)$ be a graph on $n$ vertices, with $V = \{
  v_1, \ldots, v_n \}$.
  We may assume that $G$ has at least one edge, since the edgeless case is
  trivial.
  Introduce a fresh binary relation symbol $R_{i, j}$ for every edge $\{ v_i,
  v_j \} \in E$ with $i < j$.
  Now, consider the self-join-free CQ
  \[
    \phi_G \df
    \exists v_1 \cdots \exists v_n
    \bigwedge_{i < j, \{ v_i, v_j \} \in E} R_{i, j}(v_i, v_j).
  \]
  The primal graph of its unique $\{ \exists, \land \}$-region is obtained from
  $G$ by removing its isolated vertices, and hence has the same treewidth.
  Thus, by~\cref{thm:width-treewidth,lemma:algo}, $\logwidth(\phi_G) = \tw(G) +
  1$.
  In particular, $\tw(G) \leq t$ iff $\logwidth(\phi_G) \leq t + 1$.
  NP-hardness thus immediately follows from the NP-completeness of the treewidth
  decision problem~\cite{ACP87}.

  The same arguments yield the polynomial-time Turing equivalence.
  After the polynomial-time normalisation discussed above, the logical width is
  obtained by computing the treewidth of polynomially many region hypergraphs
  and taking their maximum plus one.
  Conversely, the construction above gives $\tw(G) = \logwidth(\phi_G) - 1$.
\end{proof}

\subsection{Proof of~\cref{thm:forall-width-treewidth}}
Let us proceed with the following lemma.
\begin{lemma}\label{lemma:forall}
Let $\phi$ be a self-join-free QCQ, let $\phi^+$ be a self-join-free QCQ with minimum width among all formulae in $[\phi]_{\mathbb{A}}$, and
let $\psi \llbracket \theta \rrbracket$ be an $\{\forall\}$-organised formula
that is a subformula of $\phi^+$.
Then, the following conditions hold:
\begin{enumerate}
\item\label{forall-thm:one}
If $\theta$ is not an atom, then $\mathsf{tw}(H_\psi) \leq \mathsf{tw}(H_\theta)$. 

\item\label{forall-thm:two} 
If $\theta$ is an atom, then $\mathsf{logwidth}(\phi) \geq \mathsf{tw}(H_\psi)+1$.

\end{enumerate}
\end{lemma}
\begin{proof}
First, consider the case given in~\cref{forall-thm:one}.
Interpreting $\psi$ as a holey $\{ \forall\}$-formula, we have that $\psi = \forall y_1 \dots \forall y_k 1$, with an association $a$ such that $a(1) = \{ y_1, \dots, y_k, z_1, \dots, z_n\}$, where $\mathsf{free}(\psi) = \{ z_1, \dots, z_n \}$.
Thus, the hypergraph of $\psi$ contains the hyperedges $\{ y_1, \dots, y_k, z_1, \dots, z_n \}$ and $\{ z_1, \dots, z_n \}$.
Now consider the hypergraph of $\theta$, where $\theta$ is interpreted as a holey $\{ \exists, \land \}$-formula that is the child region of $\psi$.
Clearly, the free variables of $\theta$ are $\{ y_1, \dots, y_k, z_1, \dots, z_n \}$, and by definition $H_\theta$ has a hyperedge that contains all the free variables of $\theta$.
Thus, $\mathsf{tw}(H_\psi) \leq \mathsf{tw}(H_\theta)$.

The rest of this proof is focused on~\cref{forall-thm:two}.
Let $\psi\llbracket \theta \rrbracket = \forall y_1 \cdots \forall y_k  R(x_1, \dots, x_r)$.
To establish this case, we define two variable-sorted structures $\mathcal{A}$ and $\mathcal{B}$ such that $\mathcal{A} \models_\mathsf{var} \phi^+$, $\mathcal{B} \not\models_\mathsf{var} \phi^+$, and $\mathcal{A} \equiv_\mathsf{var}^t \mathcal{B}$, where $t = \mathsf{tw}(H_\psi)$.
Notice that $t  = | \{ x_1, \dots, x_r \}| - 1 $ since $\{ x_1, \dots, x_r\}$ is a hyperedge of $H_\psi$ that contains all the vertices in $H_\psi$.

For every variable $u$ of $\phi^+$, let $A_u = B_u = \{ \ast_u \}$.
For every atom $T(w_1, \dots, w_p)$ that occurs in $\phi^+$ with $T \neq R$, let $T^\mathcal{A} \df \{ ( \ast_{w_1}, \dots , \ast_{w_p} ) \}$ and let $T^\mathcal{B} \df \{ ( \ast_{w_1}, \dots , \ast_{w_p} ) \}$.
For the relation symbol $R$, such that $R(x_1, \dots, x_r)$ appears in $\psi \llbracket \theta \rrbracket$, let $R^\mathcal{A} \df \{ ( \ast_{x_1} , \dots, \ast_{x_r} ) \}$ and let $R^\mathcal{B} = \emptyset$.
Thus, for both $\mathcal{A}$ and $\mathcal{B}$, there is a single variable assignment $\subs$ for the variables of $\phi^+$ that respects domains: That is, $\subs(u) = \ast_u$ for every variable $u$ that appears in $\phi^+$.

Clearly, $\mathcal{A} \models_\mathsf{var} \phi^+$ since every variable is assigned to the unique element of its domain, and with this assignment every atomic formula in $\phi^+$ is satisfied. 
To show $\mathcal{B} \not\models_\mathsf{var} \phi^+$, notice that for the unique assignment $\subs$ resulting from the evaluation game, we get $(\mathcal{B}, \subs) \not\models R(x_1, \dots, x_r)$ since $R^\mathcal{B} = \emptyset$.
Thus, Universal has a winning strategy over $\mathcal{B}$ and $\phi^+$.

All that remains is to show $\mathcal{A} \equiv_\mathsf{var}^t \mathcal{B}$.
That is, Duplicator wins the $t$-pebble game over $\mathcal{A}$ and $\mathcal{B}$.
If Spoiler pebbles $\ast_u$ for some variable $u$, then Duplicator's only choice is to pick $\ast_u$ in the other structure.
Let $(\alpha, \beta)$ be an arbitrary position of the game where Duplicator plays the above-defined strategy.
Immediately from Duplicator's strategy, $\{ (\alpha(p), \beta(p)) \mid p \in \domain(\alpha) \}$ defines an injective partial mapping.
Moreover, since all relation symbols $T \neq R$ are interpreted the same in both $\mathcal{A}$ and $\mathcal{B}$, we only need to consider $R$.

Recall that $R^\mathcal{A}  = \{ (\ast_{x_1} , \ast_{x_2}, \dots, \ast_{x_r}) \}$.
The number of distinct elements occurring in the tuple $(\ast_{x_1}, \dots, \ast_{x_r})$ is $| \{ x_1, \dots, x_r \}| = t+1$.
Since we are considering the $t$-pebble game, no reachable position can contain all these elements simultaneously.
Hence, for any tuple of pebbles $\bar p \in \domain(\alpha)^r$, we have that $\alpha(\bar p)  \notin R^\mathcal{A}$ and $\beta(\bar p ) \notin R^\mathcal{B}$ simply because $\alpha(\bar p)$ and $\beta(\bar p)$ do not have enough distinct components.

Thus $\mathcal{A} \models_\mathsf{var} \phi^+$, $\mathcal{B} \not\models_\mathsf{var} \phi^+$ and $\mathcal{A} \equiv_\mathsf{var}^{\mathsf{tw}(H_\psi)} \mathcal{B}$.
By invoking~\cref{lemma:var-sorted} and~\cref{thm:pebble-game-char}, we have $\mathsf{logwidth}(\phi) \geq \mathsf{tw}(H_\psi)+1$.
\end{proof}
Then, the actual proof of~\cref{thm:forall-width-treewidth} is a straightforward application of~\cref{lemma:forall}:
\begin{proof}
We derive this result by establishing two inequalities.

($\geq$):
By~\cref{lemma:forall}(\ref{forall-thm:one}), it must hold that any such region with maximum treewidth must be of the form $\forall \bar y R(\bar z)$ where $R \in \tau$.
Let $\theta$ be a $\{\forall\}$-region with maximum treewidth.
Invoking~\cref{lemma:forall}(\ref{forall-thm:two}), any first-order formula equivalent to $\phi^+$ necessarily uses at least $\mathsf{tw}(H_\theta)+1$ variables.

($\leq$): This direction follows from~\cref{lemma:algo}.
\end{proof}

\subsection{Proof of~\cref{lemma:connected-and-tight}}
\begin{proof}
We prove this result in two parts.

\proofsubparagraph{Connectivity Property.}
Towards a contradiction, assume that $\phi$ is a QCQ where $[\phi]_{\mathbb{T}}$ is a normal form of the system $\mathscr{R}$ and there exists a $\{ \exists, \land \}$-region $\psi$ of $\phi$ which does not have the connectivity property.
Using the pushup rule, we can assume that $\psi$ is in prenex normal-form. 
Set $\psi = \exists x_1 \cdots \exists x_k \varphi$ where $\varphi$ is a quantifier-free holey formula, and let $a$ be an association for each hole in $\varphi$.
We shall use $X$ for $\{ x_1, \dots, x_k\}$ for the remainder of this proof.
Partition $X$ into $X_1 \sqcup X_2$ such that the hypergraph $H_\psi |_X$ is the disjoint union of $H_\psi |_{X_1}$ and $H_\psi |_{X_2}$. 
That is, $H_\psi |_{X_1}$ and $H_\psi |_{X_2}$ have no vertices or edges in common, but the union of the vertices and edges of these two hypergraphs accounts for all the vertices and edges in $H_\psi |_X$.

Without loss of generality, we may assume $X_1 = \{ x_1, \dots, x_r \}$ and $X_2 = \{ x_{r+1}, \dots, x_k \}$ where $1 \leq r < k$.
Then, using the associativity and commutativity rules, let $\psi_2 = \exists x_1 \cdots \exists x_k ( \varphi_1 \land \varphi_2)$ such that $\bigcup_{i \text{ is a hole of } \varphi_1} a(i)= X_1$, and $\bigcup_{j \text{ is a hole of } \varphi_2} a(j) = X_2$; note that this can be done since there is a one-to-one correspondence between hyperedges in  $H_\psi |_X$ and $a(i)$ for some hole $i$ of $\varphi$.
Observe that $\psi_2 $ belongs to the equivalence class $[\psi]_{\mathbb{T}}$ since we have only used the pushup rule, the commutativity rule, and the associativity rule.
We now have that $\psi_2 \overset{*}{\rightarrow}_{P \downarrow} \psi_3$ where  $\psi_3 = \exists x_1 \cdots \exists x_r (\varphi_1) \land \exists x_{r+1} \cdots \exists x_k (\varphi_2)$.
Again, observe that $\psi_3$ belongs to $[\varphi]_{\mathbb{T}}$ since we have used the pushdown rule, which belongs to $\mathbb{T}$.

Notice that we can take $\phi$ and replace the subformula $\psi$ with $\psi_3$ to obtain a new QCQ $\phi'$ which belongs to $[\phi]_{\mathbb{T}}$.
Since $\psi$ is a non-top region of $\phi$, there exists some subformula $\theta$ of $\phi$ such that
\[ \theta = \forall y_1 \cdots \forall y_p ( \exists x_1 \cdots \exists x_r (\varphi_1) \land \exists x_{r+1} \cdots \exists x_k (\varphi_2) ) . \]
We are now able to perform the splitdown rule on $\theta$ which contradicts our assumption that $[\phi]_{\mathbb{T}}$ is a normal form of the system $\mathscr{R}$.

\proofsubparagraph{Tightness Property.}
Towards a contradiction, assume that $\phi$ is a QCQ where $[\phi]_{\mathbb{T}}$ is a normal form of the system $\mathscr{R}$ and there exists a non-top $\{ \exists, \land \}$-region $\psi$ of $\phi$ which does not have the tightness property.
Using the pushup rule, we can assume that $\psi$ is in prenex normal-form. 
Set $\psi = \exists x_1 \cdots \exists x_k \bigwedge_{1 \leq i \leq n} i$ to be a holey formula and let $a$ be the corresponding association.
Since $\psi$ does not have the tightness property, there exists a hyperedge $e$ of $H_\psi$, that is not the single hyperedge containing all the free variables, but where $e \subseteq \mathsf{free}(\psi)$.
Set $r \in [n]$ such that $a(r) = e$.

Using the commutativity rule and the associativity rule, we have that $\psi \overset{*}{\rightarrow}_{AC} \psi_2$ where 
\[\psi_2 = \exists x_1 \cdots \exists x_k \left( \bigwedge_{i \in [n] \setminus \{ r \}} i \land r \right). \]
We now have that $\psi_2 \overset{*}{\rightarrow}_{P \downarrow} \psi_3$ where 
\[ \psi_3 =  \left( \exists x_1 \cdots \exists x_k \left( \bigwedge_{i \in [n] \setminus \{ r \}} i \right) \right) \land r. \]
However, since $\psi$ is a non-top region of $\phi$, there is a $\{\forall\}$-region with $\psi$ as its child region.
We can therefore perform the splitdown rule on $\phi'$ which contradicts our assumption that $[\phi]_{\mathbb{T}}$ is a normal form of the system $\mathscr{R}$.
\end{proof}

\section{Omitted Details From~\cref{sec:qsg}}
Recall that all omitted details from~\cref{sec:qsg} apart from the proof of~\cref{lemma:winning-strat} can be found in~\cref{sec:main-appendix}.

\subsection{Proof of~\cref{lemma:winning-strat}}
\begin{proof}
For any $a \in A$, any walk $W$ in $G_\theta$, and any function $f \colon E_a \rightarrow \{ 0, 1 \}$, recall that $f^W \colon E_a \rightarrow \{0,1\}$ is the function defined as
\[ f^{W}(e) \df f(e) + \mathsf{occ}_W(e) \pmod{2} \]
for each $e \in E_a$.
Duplicator maintains the following invariant after each round:

\proofsubparagraph{Invariant:}
\emph{Suppose $0 \leq i \leq \rho$ pebbles are placed on $a_1, \dots, a_i \in A $ and $b_1, \dots, b_i \in B$.
Then, there exists a walk $W$ in $G_\theta$ from $s$ to $t \in \mathsf{avoid}_{\mathcal{M}}\bigl(\{ \pi_1(a_1),\dots,\pi_1(a_i) \} \bigr)$ such that $b_\ell = \Psi_W(a_\ell)$ for every $\ell \in [i]$.}

By Lemma~\ref{lemma:partial-iso}, any position satisfying this invariant
is a partial isomorphism between $\mathcal{A}$ and $\mathcal{B}$.
Hence, it suffices to show that Duplicator can always respond so as to preserve the invariant.

For the initial position, where no pebbles are placed, choose any bramble set $M \in \mathcal{M}$ with $s \notin M$.
Since $G_\theta$ is connected, there exists a walk $W$ from $s$ to some $t \in M$ and thus the invariant holds.

Now assume the invariant holds for $i < \rho$ pebbles with respect to some walk $W$ ending in some $t \in \mathsf{avoid}_{\mathcal{M}}\bigl( \{ \pi_1(a_1),\dots,\pi_1(a_i) \} \bigr)$.
Suppose Spoiler places a new pebble on some element $a_{i+1} \in A$ (the case where Spoiler plays in $\mathcal{B}$ is symmetric).

Since $\mathcal{M}$ has order greater than $\rho$, the set $\{\pi_1(a_1),\dots,\pi_1(a_{i+1}) \} $ cannot hit $\mathcal{M}$.
Hence, there exists $M' \in \mathcal{M}$ such that $M' \subseteq \mathsf{avoid}_{\mathcal{M}}\bigl( \{ \pi_1(a_1),\dots,\pi_1(a_{i+1}) \} \bigr)$.
Choose some $t' \in M'$.

Since bramble sets pairwise touch, there exists a walk
$W'$ from $t$ to $t'$ whose vertices lie entirely in $M \cup M'$.
Let $W''$ be the concatenation of the walks $W$ and $W'$.
Then, $W''$ is a walk from $s$ to $t'$ with
$t' \in \mathsf{avoid}_{\mathcal{M}}\bigl(\{ \pi_1(a_1),\dots,\pi_1(a_{i+1}) \} \bigr)$.

Duplicator now responds with $b_{i+1}$ such that if $a_{i+1} \in A_{v_m}'$, then $b_{i+1} = a_{i+1}$; otherwise, $b_{i+1}$ is the tuple $\Psi_{W''}(a_{i+1})$.

For $\ell \leq i$, note that $W'$ avoids $\pi_1(a_1),\dots,\pi_1(a_i)$.
Hence, for every edge $e \in E_{\pi_1(a_\ell)}$, we have that
$\mathsf{occ}_{W''}(e) = \mathsf{occ}_W(e)$.
Thus the invariant holds with the new walk $W''$.

Consequently, Duplicator has a winning strategy since the invariant can be maintained for at most $\rho$ pebbles, and every position satisfying the invariant is a partial isomorphism (\cref{lemma:partial-iso}).
\end{proof}

\section{Omitted Details From~\cref{sec:reductions}}
This section establishes~\cref{thm:main-exists-and}, by reducing the problem to simple QCQs.
That is, for any self-join-free QCQ $\varphi$ and a non-top $\{ \exists, \land \}$-region $\theta$ of maximum treewidth, to establish that $\mathsf{logwidth}(\varphi) > \mathsf{tw}(H_\theta)$, it suffices to show that $\mathsf{logwidth}(\phi) > \mathsf{tw}(H_\theta)$ for a certain simple QCQ $\phi$ obtained from $\varphi$.
The case where our $\{ \exists, \land \}$-region with maximum treewidth is a top region  shall be dealt with separately in~\cref{section:edge-case} (although most of the machinery defined in this section will also be used there).

\begin{notation}
Throughout this section, we fix $\varphi$ to be an arbitrary standardised self-join-free QCQ such that $[\varphi]_{\mathbb{T}}$ is a normal form in the  system $\mathscr{R}$.
We assume there is a non-top $\{ \exists, \land \}$-region of $\varphi$ with maximum treewidth, and we fix $\theta$ to denote such a region.
\end{notation}

For this section, we require some definitions:

\begin{definition}
An $\{ \exists, \land \}$-region $\theta'$ of $\varphi$ has the \emph{clasped condition} if for every $u \in \mathsf{free}(\theta')$, there exists a variable $x$ that is existentially quantified in $\theta'$ such that $u$ and $x$ appear in some hyperedge of $H_{\theta'}$.
\end{definition}

Recall that since $[\varphi]_{\mathbb{T}}$ is a normal form in the system $\mathscr{R} \df (\mathsf{QCQ} , \rightarrow_{\{ S \downarrow, M \}} ) / \overset{*}{\leftrightarrow}_{\mathbb{T}}$, every non-top $\{ \exists, \land \}$-region of $\varphi$ has the tightness property (see~\cref{lemma:connected-and-tight}), and therefore immediately has the clasped condition.

Notice that since $\varphi$ is a standardised and self-join-free QCQ, every region occurring in $\varphi$ is unique.
Indeed, every atomic region of $\varphi$ must occur exactly once, since $\varphi$ is self-join-free.
Moreover, every non-top and non-atomic region of $\varphi$ contains a quantifier; otherwise, it would be a $\{ \exists, \land\}$-region only containing conjunction and occurring below a $\{ \forall \}$-region, and thus would permit a splitdown, contradicting $[\varphi]_{\mathbb{T}}$ being a normal form of $\mathscr{R}$.
Therefore, since $\varphi$ is standardised, every region containing a quantification is unique.
The top region either does not contain a quantifier (and thus is unique) or contains the quantification of a variable unique to that region.

Since each region of $\varphi$ has a unique occurrence, we may say that a region occurs \emph{below} another region without ambiguity.
Moreover, we may define the subformula rooted at a region:

\begin{definition}\label{defn:sub-root}
Let $\rho$ be a region of $\varphi$.
We define the \emph{subformula of $\varphi$ rooted at $\rho$}, denoted $\mathsf{sub}_\varphi(\rho)$,
to be the unique subformula of $\varphi$
whose top region is $\rho$.
\end{definition}

Since every region has a unique occurrence, this is well-defined.
Next, we define the contraction of a region:

\begin{definition}\label{defn:contraction}
For a $\{ \forall \}$ region $\rho$ of $\varphi$ where $|\mathsf{free}(\rho)| \geq 1$,
the \emph{contraction of $\varphi$ at $\rho$}, denoted by $\varphi^{\mathsf{ctr}}_\rho$ is defined as follows:
Let $F_\rho \df \{ z_1, \dots, z_m \}$ be the free variables of $\rho$, given in an arbitrary order.
Introduce a new and unique binary relation $R_{z_i, z_j}^\rho$ for each pair $z_i, z_j \in F_\rho$ where $i \leq j$.
Then, $\varphi^{\mathsf{ctr}}_\rho$ is obtained by replacing $\mathsf{sub}_\varphi(\rho)$ with $\bigland_{1 \leq i \leq j \leq m} R_{z_i, z_j}^\rho (z_i, z_j)$.
\end{definition}

Intuitively speaking,
we shall use~\cref{defn:contraction} to contract all regions that occur below our distinguished $\{ \exists, \land \}$-region with maximum treewidth.
Once this has been done, we are able to assume our distinguished region only has atomic child regions.

\subsection{Contracting Basic Regions}\label{subsec:red-1}
We say that a $\{ \forall\}$-region $\rho$ of $\varphi$ is a \emph{$\theta$-basic region} if
$\rho$ that occurs below $\theta$, and $\rho$ has a single child region that is an atom $\alpha$.
The first technical step is to contract all $\theta$-basic regions.

\begin{lemma}\label{lemma:contract-easy}
Let $\rho$ be a $\theta$-basic region of $\varphi$ where $|\mathsf{free}(\rho)| \geq 1$.
Let $\varphi' \df \varphi^\mathsf{ctr}_\rho$.
If there exist variable-sorted structures $\mathcal{A}$ and $\mathcal{B}$ over the signature of $\varphi'$ such that
\[ \mathcal{A} \models_\mathsf{var} \varphi' , \; \mathcal{B} \not\models_\mathsf{var} \varphi' , \; \mathcal{A} \equiv^k_\mathsf{var} \mathcal{B} ,\]
then, there exist variable-sorted structures $\mathcal{A}^+$ and $\mathcal{B}^+$ over the signature of $\varphi$ such that
\[ \mathcal{A}^+ \models_\mathsf{var} \varphi , \; \mathcal{B}^+ \not\models_\mathsf{var} \varphi , \; \mathcal{A}^+ \equiv^k_\mathsf{var} \mathcal{B}^+  . \]
\end{lemma}
\begin{proof}
Let $F_\rho=\mathsf{free}(\rho)=\{ z_1,\dots,z_m \}$, where $m \geq 1$, and let $\psi \df \mathsf{sub}_{\varphi}(\rho)$, that is:
\[ \psi(z_1,\dots,z_m) = \forall y_1\cdots \forall y_\ell\, R(t_1,\dots,t_r) \]
for some atom $R(t_1, \dots, t_r)$ and some variables $y_1, \dots, y_\ell$ with $\ell \geq 1$.
For convenience, let $R^{\mathcal C}_{i,j} \df \bigl(R^\rho_{z_i,z_j}\bigr)^{\mathcal C}$ for each $\mathcal C\in\{\mathcal A,\mathcal B\}$ and each $1\leq i \leq j\leq m$.

We now define, for each $\mathcal C\in\{\mathcal A,\mathcal B\}$, a variable-sorted structure $\mathcal C^+$ over the signature of $\varphi$.
Since $\varphi$ is standardised, the variables $y_1,\dots,y_\ell$ occur only inside $\psi$.
For each $j\in[\ell]$, set $C^+_{y_j}\df \{\ast_{y_j}\}$.
For every other variable $u$ of $\varphi$, set $C^+_u\df C_u$, where $C_u$ is the domain for $u$ in $\mathcal{C}$.

For every relation symbol $Q$ occurring in $\varphi$ with $Q\neq R$, set $Q^{\mathcal C^+}\df Q^{\mathcal C}$.
This is well-defined since $R$ occurs in $\psi$ and $\varphi$ is self-join-free.
Moreover, every relation symbol $Q\neq R$ occurring in $\varphi$ also occurs in $\varphi'$, and, since $\varphi$ is standardised, its domains are inherited from $\mathcal C$.

For each $i\in[m]$, fix one position $p_i\in[r]$ such that $t_{p_i}=z_i$, which must exist since $z_1, \dots , z_m$ are the free variables of $\psi$.
It remains to define the interpretation of $R$ in $\mathcal{C}^+$.
For every tuple $(c_1,\dots,c_r)\in C^+_{t_1}\times\cdots\times C^+_{t_r}$ where $C^+_{t_q}$ denotes the domain of the variable $t_q$, let
\[ (c_1,\dots,c_r)\in R^{\mathcal C^+} \iff (c_{p_i},c_{p_j})\in R^{\mathcal C}_{i,j} \text{ for every }1\leq i \leq j\leq m \]
The components occupied by the variables $y_1,\dots,y_\ell$ simply take the value from their singleton domain.

We claim that for every assignment $\subs$ defined on $F_\rho$, we have
\begin{equation}\label{equation:basic-region}
 (\mathcal C^+,\subs)\models_{\mathsf{var}} \psi \iff (\mathcal C,\subs)\models_{\mathsf{var}} \bigwedge_{1\leq i \leq j\leq m} R^\rho_{z_i,z_j}(z_i,z_j).
 \end{equation}
Since each $C^+_{y_j}$ is a singleton, there is exactly one extension of $\subs$ that includes the variables $y_1,\dots,y_\ell$.
Let $\subs'$ be such an extension, then $(\mathcal{C}^+, \subs') \models_\mathsf{var} R(t_1,\dots,t_r)$ if and only if $(\subs(z_i),\subs(z_j))\in R^{\mathcal C}_{i,j}$ for every $1\leq i \leq j\leq m$, by the definition of $R^{\mathcal C^+}$.
Thus,~\cref{equation:basic-region} holds.

Since $\varphi'$ is obtained from $\varphi$ by replacing $\psi$ with
$\bigwedge_{1\leq i \leq j\leq m} R^\rho_{z_i,z_j}(z_i,z_j)$, along with~\cref{equation:basic-region}, a straightforward induction on the syntax tree of $\varphi$, gives us
$\mathcal C^+\models_{\mathsf{var}} \varphi $ if and only if $\mathcal C \models_{\mathsf{var}}\varphi' $ for each $\mathcal C\in\{\mathcal A,\mathcal B\}$.
Hence, $\mathcal{A}^+ \models_\mathsf{var} \varphi$ and $\mathcal{B}^+ \not\models_\mathsf{var} \varphi$.

It remains to be shown that $\mathcal A^+\equiv^k_{\mathsf{var}}\mathcal B^+$.
Duplicator transfers their winning strategy from the $k$-pebble game over $\mathcal A$ and $\mathcal B$ to the $k$-pebble game over $\mathcal A^+$ and $\mathcal B^+$ as follows:
If Spoiler plays an element that inherited from $\mathcal{A}$ or $\mathcal{B}$, Duplicator responds according to the original winning strategy over $\mathcal A$ and $\mathcal B$.
If Spoiler plays the unique element $\ast_{y_j}$ in one of the new singleton domains, Duplicator plays $\ast_{y_j}$ on the other side.
The strategy is symmetric for moves made in $\mathcal B^+$ by Spoiler.

We verify that every position reached by this strategy is a partial isomorphism.
First, immediately from Duplicator's strategy, every position reached results in an injective partial mapping.
Every relation symbol $Q\neq R$ is dealt with by the original strategy, since $Q^{\mathcal A^+}=Q^{\mathcal A}$ and $Q^{\mathcal B^+}=Q^{\mathcal B}$.

It remains to consider $R$.
Suppose that $a_1, \dots, a_r$ are pebbled elements from $\mathcal{A}^+$ and $b_1, \dots, b_r$ are the corresponding pebbled elements from $\mathcal{B}^+$ such that  $(a_1,\dots,a_r)\in R^{\mathcal A^+}$.
The components corresponding to variables among $y_1,\dots,y_\ell$ are elements from their singleton domain.
For the chosen coordinates $p_1, \dots, p_m$ corresponding to $z_1,\dots,z_m$, membership in $R^{\mathcal A^+}$ gives $(a_{p_i},a_{p_j})\in R^{\mathcal A}_{i,j}$ for every $1\leq i \leq j\leq m$.
Since $R^\rho_{z_i,z_j}$ is a relation symbol of $\varphi'$, by Duplicator's strategy we have that $(b_{p_i},b_{p_j})\in R^{\mathcal B}_{i,j}$ for every $1\leq i \leq j\leq m$.
Hence, by the definition of $R^{\mathcal B^+}$, we have that $(b_1,\dots,b_r)\in R^{\mathcal B^+}$.
The converse implication, from $\mathcal B^+$ to $\mathcal A^+$, is identical.

Thus, every reachable position under Duplicator's strategy is a partial isomorphism. Therefore $\mathcal A^+\equiv^k_{\mathsf{var}}\mathcal B^+$.
\end{proof}

\begin{notation}\label{notation:formula-2}
For the remainder of this section,
fix $\varphi_2$ to denote the self-join-free QCQ obtained from $\varphi$ by iteratively and exhaustively contracting all $\theta$-basic regions.
\end{notation}

\begin{lemma}\label{lemma:reduction-one}
If there exist two variable-sorted structures $\mathcal{A}$ and $\mathcal{B}$ such that
\[ \mathcal{A} \models_\mathsf{var} \varphi_2 , \quad \mathcal{B} \not\models_\mathsf{var} \varphi_2 , \quad \mathcal{A} \equiv^k_\mathsf{var}  \mathcal{B}, \]
then $\mathsf{logwidth}(\varphi) > k$.
\end{lemma}
\begin{proof}
Suppose that there exist two variable-sorted structures $\mathcal{A}$ and $\mathcal{B}$ such that
\[ \mathcal{A} \models_\mathsf{var} \varphi_2 , \quad \mathcal{B} \not\models_\mathsf{var} \varphi_2 , \quad \mathcal{A} \equiv^k_\mathsf{var}  \mathcal{B}. \]
Let $\Psi_1, \Psi_2, \dots, \Psi_r$ be the intermediate formulae that are obtained on the way to $\varphi_2$.
That is, let $\Psi_1 = \varphi$, $\Psi_r = \varphi_2$, and where, for $i \in [r-1]$, the formula $\Psi_{i+1}$ is obtained from $\Psi_i$ by contracting a $\theta$-basic region.
By~\cref{lemma:contract-easy} (which applies verbatim, even though it is stated in terms of $\varphi$) for $i \in [r-1]$, if there exist variable-sorted structures $\mathcal{A}_{i+1}$ and $\mathcal{B}_{i+1}$ such that
\[ \mathcal{A}_{i+1} \models_\mathsf{var} \Psi_{i+1}, \quad \mathcal{B}_{i+1} \not\models_\mathsf{var} \Psi_{i+1}, \quad \mathcal{A}_{i+1} \equiv^k_\mathsf{var}  \mathcal{B}_{i+1}, \]
then there exist variable-sorted structures $\mathcal{A}_i$ and $\mathcal{B}_i$ such that
\[ \mathcal{A}_{i} \models_\mathsf{var} \Psi_{i}, \quad \mathcal{B}_{i} \not\models_\mathsf{var} \Psi_{i}, \quad \mathcal{A}_{i} \equiv^k_\mathsf{var}  \mathcal{B}_{i}.  \]
Thus, with $\mathcal{A} = \mathcal{A}_r$ and $\mathcal{B} = \mathcal{B}_r$, there exist structures $\mathcal{A}_1$ and $\mathcal{B}_1$ with
\[ \mathcal{A}_1 \models_\mathsf{var} \varphi , \quad \mathcal{B}_1 \not\models_\mathsf{var} \varphi , \quad \mathcal{A}_1 \equiv^k_\mathsf{var} \mathcal{B}_1 . \]
Finally, invoking~\cref{lemma:var-sorted}, we have that $\mathsf{logwidth}(\varphi) > k$.
\end{proof}

\subsection{Binary Relation Encoding}\label{subsec:red-2}
In this section, we take $\varphi_2$ and construct $\varphi_3$ which is a self-join-free QCQ over a binary signature (that is, all atomic formulae that appear in $\varphi_3$ are applications of binary, unary, or nullary relation symbols).
We call $\varphi_3$ the \emph{binary relation encoding of $\varphi_2$}, and is defined as follows:

\begin{notation}
Fix $\varphi_3$ to be the formula obtained from $\varphi_2$ by replacing each atomic formula of the form $R(x_1, \dots, x_r)$, where $r >2$, with $\bigwedge_{1 \leq i < j \leq r} R_{i,j} (x_i, x_j)$ where $R_{i,j}$ is a new relation symbol for each $i,j$ and each such relation symbol $R$.
\end{notation}

Notice that in the binary relation encoding, we simply replace atomic formulae with a conjunction of atomic formulae.
Therefore, there is a one-to-one mapping between a $\{ \exists, \land \}$-region in $\varphi_2$ and the corresponding $\{ \exists, \land \}$-region (with which the original region is replaced) in $\varphi_3$.

\begin{observation}\label{obs:iso-primal-graphs}
Let $\chi$ be an arbitrary $\{ \exists, \land \}$-region of $\varphi_2$, and let $\chi'$ be the corresponding $\{ \exists, \land \}$-region in $\varphi_3$.
Then, the primal graphs of $H_\chi$ and $H_{\chi'}$ are isomorphic.
\end{observation}
\begin{proof}
Take any two variables $x$ and $y$ that appear in $\varphi_2$.
Note that the construction of $\varphi_3$ does not add or remove any variables, and therefore $x$ and $y$ appear in $\varphi_3$.
Let $G_{\chi}$ and $G_{\chi'}$ be the primal graphs of $H_{\chi}$ and $H_{\chi'}$ respectively.
If $\{ x,y \}$ appears as an edge of $G_{\chi}$, then $x$ and $y$ appear in some atomic formula of $\varphi_2$ (or are both free variables of $\chi$) and by the construction of $\varphi_3$ must also appear in some atomic formula of $\varphi_3$ (or are also free variables of $\chi'$), implying $\{x,y\}$ appears as an edge of $G_{\chi'}$.
The other direction is symmetric.
\end{proof}

\begin{notation}
Fix $\bar\theta$ to be the $\{ \exists, \land \}$-region of $\varphi_3$ that corresponds to the region $\theta$ of $\varphi_2$.
Recall that $\theta$ is the region of $\varphi_2$ with maximum treewidth.
\end{notation}

\begin{lemma}\label{lemma:bin-rel-enc}
If there exist variable-sorted structures $\mathcal{A}^+$ and $\mathcal{B}^+$ such that
\[ \mathcal{A}^+ \models_\mathsf{var} \varphi_3 , \quad  \mathcal{B}^+ \not\models_\mathsf{var} \varphi_3, \quad \mathcal{A}^+ \equiv^k_\mathsf{var} \mathcal{B}^+ , \]
then there exist variable-sorted structures $\mathcal{A}$ and $\mathcal{B}$ such that
\[ \mathcal{A} \models_\mathsf{var} \varphi_2 , \quad \mathcal{B} \not\models_\mathsf{var} \varphi_2 , \quad \mathcal{A} \equiv^k_\mathsf{var} \mathcal{B} . \]
\end{lemma}
\begin{proof}
Suppose the existence of variable-sorted structures $\mathcal{A}^+$ and $\mathcal{B}^+$ such that $\mathcal{A}^+ \models_\mathsf{var} \varphi_3$, $\mathcal{B}^+ \not\models_\mathsf{var} \varphi_3$ and $\mathcal{A}^+ \equiv^k_\mathsf{var} \mathcal{B}^+$.

We shall define $\mathcal{A}$ over the signature of $\varphi_2$.
Let $\mathcal{A}$ have the same universe as $\mathcal{A}^+$ for every variable.
For any nullary, unary, or binary relation  symbol $P$, let $P^\mathcal{A} \df P^{\mathcal{A}^+}$.

Now let $R(x_1, \dots, x_r)$ be an atomic formula that appears in $\varphi_2$ where $r > 2$.
Define $R^{\mathcal{A}}$ such that:
\[  (a_1, \dots, a_r) \in R^\mathcal{A} \iff (a_i, a_j) \in R_{i,j}^{\mathcal{A}^+} \text{ for all } 1 \leq i < j \leq r. \]

Next, we show that $\mathcal{A}^+ \models_\mathsf{var} \varphi_3 \iff \mathcal{A} \models_\mathsf{var} \varphi_2$.
Notice that for any variable assignment $f$ on $x_1, \dots, x_r$ we have that:
\[ (\mathcal{A}, f) \models_\mathsf{var} R(x_1, \dots, x_r) \iff (\mathcal{A}^+, f ) \models_\mathsf{var} \bigwedge_{1 \leq i < j \leq r} R_{i,j} (x_i, x_j).  \]
Moreover, $\varphi_3$ is obtained by replacing each atom $R(x_1, \dots, x_r)$ with $r > 2$ with $\bigwedge_{1 \leq i < j \leq r} R_{i,j} (x_i, x_j)$.
Thus, $\mathcal{A}^+ \models_\mathsf{var} \varphi_3 \iff \mathcal{A} \models_\mathsf{var} \varphi_2$ by a straightforward induction on the syntax tree.
The analogous statement holds for $\mathcal{B}^+$ and $\mathcal{B}$.
Assuming $\mathcal{A}^+ \models_\mathsf{var} \varphi_3$ and $\mathcal{B}^+ \not\models_\mathsf{var} \varphi_3$, we get $\mathcal{A} \models_\mathsf{var} \varphi_2$ and $\mathcal{B} \not\models_\mathsf{var} \varphi_2$.

All that remains is to prove $\mathcal{A}^+ \equiv^k_\mathsf{var} \mathcal{B}^+ \implies \mathcal{A} \equiv^k_\mathsf{var} \mathcal{B}$.
Consider any position $(\alpha, \beta)$ of the $k$-pebble game over $\mathcal{A}^+$ and $\mathcal{B}^+$ that results from Duplicator's winning strategy.
Let $a_1, \dots, a_m \in A$ and $b_1, \dots, b_m \in B$ for some $m \leq k$ such that $\alpha(p_i) = a_i$ and $\beta(p_i) = b_i$ for pebbles $p_1, \dots, p_m$.
Immediately, $(\alpha, \beta)$ forms a partial isomorphism between $\mathcal{A}^+$ and $\mathcal{B}^+$.
We claim that $(\alpha, \beta)$ also forms a partial isomorphism between $\mathcal{A}$ and $\mathcal{B}$.

Suppose $(a_{i_1}, \dots, a_{i_r} ) \in R^{\mathcal{A}}$ for any relation symbol $R$ where $r > 2$, and $i_1, \dots, i_r \in \{ 1, \dots, m \}$.
By the definition of $R^{\mathcal{A}}$ it holds that $\bigwedge_{1 \leq p < q \leq r} R^{\mathcal{A}^+}(a_{i_p}, a_{i_q})$ and since $(\alpha, \beta)$ forms a partial isomorphism, we have that $\bigwedge_{1 \leq p < q \leq r} R_{p,q}^{\mathcal{B}^+}(b_{i_p}, b_{i_q})$.
Immediately from the definition of $R_{p,q}^{\mathcal{B}^+}$ for $1 \leq p < q \leq r$, it follows that $(b_{i_1}, \dots, b_{i_r}) \in R^{\mathcal{B}}$.
The converse direction (from $R^{\mathcal{B}}$ to $R^{\mathcal{A}}$) is symmetric.
The proof concludes by noticing that unary and binary relations remain unchanged between $\mathcal{A}$ and $\mathcal{A}^+$, and $\mathcal{B}$ and $\mathcal{B}^+$.
\end{proof}

Recall  that $\varphi_3$ is a self-join-free QCQ obtained from $\varphi$ by contracting all the $\theta$-basic regions, and then performing the binary relation encoding.

\begin{lemma}\label{lemma:reduction-two}
If there exist two variable-sorted structures $\mathcal{A}$ and $\mathcal{B}$ such that
\[ \mathcal{A} \models_\mathsf{var} \varphi_3 , \quad \mathcal{B} \not\models_\mathsf{var} \varphi_3 , \quad \mathcal{A} \equiv^k_\mathsf{var}  \mathcal{B}, \]
then $\mathsf{logwidth}(\varphi) > k$.
\end{lemma}
\begin{proof}
This follows by combining~\cref{lemma:bin-rel-enc} and~\cref{lemma:contract-easy}.
\end{proof}

\subsection{Contracting More Regions}\label{subsec:red-3}

We say that a $\{\forall\}$-region $\rho$ of $\varphi_3$ is a \emph{lowest $\{ \forall \}$-region} of $\varphi_3$
if $\mathsf{sub}_{\varphi_3}(\rho)$ does not have a non-top $\{ \forall \}$-region.
In other words, $\rho$ is the unique $\{\forall\}$-region in $\mathsf{sub}_{\varphi_3}(\rho)$.

\begin{lemma}\label{lemma:lowest-forall}
For any lowest $\{ \forall \}$-region $\rho$ that occurs below $\theta$, we have that $\mathsf{sub}_{\varphi_3}(\rho) = \rho \llbracket \chi \rrbracket$ where $\chi$ is an $\{ \exists, \land \}$-region such that:
\begin{enumerate}
\item The top node of the parse tree of $\chi$ is an existential quantifier,
\item all child regions of $\chi$ are atomic,
\item $\chi$ has the connectivity property, and
\item $\chi$ has the clasped condition.
\end{enumerate}
\end{lemma}
\begin{proof}
Note that in our original QCQ $\varphi$, the subformulae rooted at any lowest $\{ \forall \}$-regions are of the form $(\forall y_1' \cdots \forall y_\jmath' 1) \llbracket \chi \rrbracket$, where either:
\begin{enumerate}
\item $\chi$ is an atomic formula, or
\item $\chi$ is an $\{ \exists, \land \}$-formula with an existential quantifier as the top node of the parse tree.
\end{enumerate}
This holds by the following reasoning:
The only alternative for $\chi$ is that it is an $\{ \exists, \land \}$-formula with conjunction as the top node of the parse tree.
However, we can then perform the splitdown rule on $\varphi$, contradicting the fact that $[\varphi]_\mathbb{T}$ is a normal form in the system $\mathscr{R}$.

We know that $\chi$ has the connectivity property and the tightness property if $\chi$ is a non-atomic $\{ \exists,\land\}$-region; recall~\cref{lemma:connected-and-tight}.
Further recall that to obtain $\varphi_3$, we first contracted all the $\theta$-basic regions (after which, all subformulae rooted at a lowest $\{ \forall\}$ region below $\theta$ have a non-atomic $\{ \exists, \land \}$ child region), then performed a binary relation encoding.
Therefore, there are no $\{ \forall\}$-regions occurring below $\bar\theta$ whose subformula has the form $\rho \llbracket \chi \rrbracket$ with $\chi$ being an atom.
Recall that $\bar\theta$ is the region of $\varphi_3$ that corresponds to the distinguished region $\theta$ of $\varphi$.

Thus, for any lowest $\{ \forall\}$-region $\rho$ occurring below $\bar\theta$, we have that $\mathsf{sub}_{\varphi_3}(\rho) $ is of the form
$ (\forall y_1' \cdots \forall y_\jmath' 1 ) \llbracket \chi' \rrbracket$ where $\chi'$ is a non-atomic $\{ \exists, \land \}$-region.
The binary relation encoding simply replaces atoms with a conjunction of binary atoms.
Therefore, the top node of the parse tree of $\chi'$ must be an existential quantifier.
Clearly all child regions of $\chi'$ must be atomic, since otherwise they are a $\{ \forall \}$-region, contradicting $\rho$ being a lowest $\{ \forall \}$-region.
The fact that $\chi'$ has the connectivity property and the clasped condition follows immediately from~\cref{obs:iso-primal-graphs}, and the fact that the hyperedge representing a $\theta$-basic region has been replaced with an equivalent clique when passing from $\varphi$ to~$\varphi_3$.
\end{proof}

Next, we work towards a result analogous to~\cref{lemma:contract-easy} for the more difficult case where we contract a subformula of the form $\forall \bar y \exists \bar x \chi$, where $\chi$ is a conjunction of atomic formulae.

Fix $\rho$ to be an arbitrary region below $\bar\theta$ such that $\mathsf{sub}_{\varphi_3}(\rho)$ is of the form
\[ \psi(z_1,\dots, z_m) = \forall y_1 \cdots \forall y_\ell \exists x_1 \cdots \exists x_n \chi\]
where $\chi$ is a conjunction of binary atomic formulae, and where $m, \ell, n \geq 1$.
Fix $\zeta$ to be the inner region $\exists x_1 \cdots \exists x_n \chi$ of $\psi$.

Recall that when contracting $\rho$, we replace $\psi$ with $\bigland_{1 \leq i \leq j \leq m} R^\rho_{z_i, z_j} (z_i, z_j)$.

Let $\mathcal{C}$ be an arbitrary variable-sorted structure over the signature of $(\varphi_3)_\rho^\mathsf{ctr}$.
We define the structure $\mathcal{C}^\rho$ over the signature of $\varphi_3$.
For convenience, let $R^{\mathcal C}_{i,j} \df \bigl(R^\rho_{z_i,z_j}\bigr)^{\mathcal C}$ for each $1\leq i \leq j\leq m$.

\subsection*{Defining the Structure $\mathcal{C}^\rho$}

\para{Domains}
For each variable $u$ of $\varphi_3$, let $C_u$ be the domain of $u$ in $\mathcal{C}$.
We define the domain $C_u^\rho$ of $u$ in $\mathcal{C}^\rho$ as follows:
\begin{itemize}
\item For every variable $u$ of $\varphi_3$ that is not quantified inside $\psi$, set $C_u^\rho \df C_u$.
\item For each $t < \ell$, set $C_{y_t}^\rho  \df \{ \ast_{y_t} \}$.
\item Set $C_{y_\ell}^\rho  \df \{ s_{z_i} \mid i \in [m] \}$.
\item Set $C_{x_p}^\rho \df \{ (z_i, a, x_p) \mid i \in [m], a \in C_{z_i} \}$ for every $p \in [n]$.
\end{itemize}

\para{Interpretation of Relation Symbols}
For every relation symbol $Q$ that appears in $\varphi_3$ but does not appear in $\psi$, let $Q^{\mathcal{C}^\rho} \df Q^\mathcal{C}$.
This is well-defined since $\varphi_3$ is standardised and self-join-free.
Let us define the interpretation of relation symbols according to the corresponding atom in $\chi$, up to reversing the order of the variables:

\begin{itemize}
\item For each relation symbol $S$ such that $S(z_i, x_p) \in \chi$ for some $i \in [m]$ and $p \in [n]$, let
\[
S^{\mathcal{C}^\rho}
\df
\Big\{ \big( a,(z_j,b,x_p) \big) \;\Big|\;  j \in [m] ,\; b \in C_{z_j} , \;
\begin{cases}
(a,b) \in R_{i,j}^\mathcal{C} & \text{if } i \leq j,\\
(b,a) \in R_{j,i}^\mathcal{C} & \text{if } i > j
\end{cases}
\Big\}.
\]

\item For each relation symbol $P$ such that $P(y_\ell, x_p) \in \chi$ for some $p \in [n]$, let
\[ P^{\mathcal{C}^\rho} \df  \{ (s_{z_i}, (z_i, a, x_p ) ) \mid i \in [m], a \in C_{z_i} \} . \]

\item For each relation symbol $E$ such that $E(x_p, x_q) \in \chi$ for some $p, q \in [n]$, let
\[ E^{\mathcal{C}^\rho} \df  \{  ( (z_i, a, x_p), (z_i, a, x_q) ) \mid i \in [m], a \in C_{z_i} \}. \]
\end{itemize}

For every other atom $F(u,v) \in \chi$, we let $F^{\mathcal{C}^\rho} \df C_u^\rho \times C_v^\rho$.

\begin{lemma}\label{lemma:contract-equiv}
Let $\varphi' \df (\varphi_3)_\rho^\mathsf{ctr}$.
For any variable-sorted structure $\mathcal{C}$ over the signature of $\varphi'$, we have that $\mathcal{C} \models_\mathsf{var} \varphi'$ if and only if $\mathcal{C}^\rho \models_\mathsf{var} \varphi_3$.
\end{lemma}
\begin{proof}
For convenience, let $R_{i,j}^\mathcal{C}$ be shorthand for $( R_{z_i, z_j}^\rho )^\mathcal{C}$ for each $1 \leq i \leq j \leq m$.
We start by establishing
\begin{equation}\label{eq:equiv}
(\mathcal{C}^\rho, \subs) \models_\mathsf{var} \psi \iff ( \subs(z_i), \subs(z_j) ) \in R_{i,j}^\mathcal{C} \text{ for every $1 \leq i \leq j \leq m$}
\end{equation}
for any variable assignment $\subs$ on $\{ z_1, \dots, z_m \}$.
We proceed by proving both directions independently.

($\Rightarrow$:)
Fix an arbitrary variable assignment $\subs$ on $z_1, \dots, z_m$ such that $(\mathcal{C}^\rho, \subs) \models_\mathsf{var} \psi$.
Towards a contradiction, assume that there is some $i, j \in [m]$ with $i \leq j$ such that $( \subs(z_i), \subs(z_j) ) \notin R_{i,j}^\mathcal{C}$.
Let $\subs'$ extend $\subs$ by $\subs'(y_t) = \ast_{y_t}$ for each $t \in [\ell -1]$ and $\subs'(y_\ell) = s_{z_j}$.
Since $(\mathcal{C}^\rho, \subs) \models_\mathsf{var} \psi$, there exists an extension $\subs''$ of $\subs'$ to the variables $x_1,\dots,x_n$ such that $(\mathcal{C}^\rho, \subs'') \models_\mathsf{var} \chi$.
Fix such an assignment $\subs''$.

By~\cref{lemma:lowest-forall}, we know that the inner region $\zeta$ of $\psi$ has the clasped condition and thus contains the following atomic formulae:
\begin{itemize}
\item $P(y_\ell, x_p)$ for some $p \in [n]$ and some relation symbol $P$;
\item $S(z_i, x_{p'})$ for some $p' \in [n]$ and some relation symbol $S$; and
\item $S_2 (z_j, x_{p''})$ for some $p'' \in [n]$ and some relation symbol $S_2$.
\end{itemize}

Since $(\mathcal{C}^\rho, \subs'') \models_\mathsf{var} P(y_\ell, x_p)$ where $\subs''(y_\ell) = s_{z_j}$, it follows from the definition of $P^{\mathcal{C}^\rho}$ that $\subs''(x_p)$ must be $(z_j, b', x_p)$ for some $b' \in C_{z_j}$.
Moreover, since the hypergraph of $\psi$ induced on $\{ x_1, \dots, x_n\}$ is connected, along with the definition of the relations $E^{\mathcal{C}^\rho}$ where $E(x_q,x_{q'}) \in \chi$ for some $q,q' \in [n]$, it follows that $\subs''(x_q) = (z_j, b', x_q)$ for each $q \in [n]$
(the proof of~\cref{lemma:X-values} gives a proof of an analogous result, the reasoning of which can be applied here).

Now consider the fact that $(\mathcal{C}^\rho, \subs'') \models_\mathsf{var} S_2(z_j, x_{p''})$ where $\subs''(z_j) = b$ for some $b \in C_{z_j}$ and that $\subs''(x_{p''}) = (z_j, b', x_{p''})$.
However, $(b, (z_j, b', x_{p''} )) \in S_2^{\mathcal{C}^\rho}$ if and only if $b = b'$.

Next, consider the fact that $(\mathcal{C}^\rho, \subs'') \models_\mathsf{var} S(z_i, x_{p'})$ where $\subs''(z_i) = a$ and $\subs''(x_{p'}) = (z_j, b, x_{p'})$ such that $(a,b) \notin (R_{z_i, z_j}^\rho)^\mathcal{C}$.
This gives a contradiction, as $(a , (z_j, b, x_{p'} )) \in S^{\mathcal{C}^\rho}$ if and only if $(a,b) \in R_{i,j}^\mathcal{C}$.

($\Leftarrow$:)
Fix an arbitrary assignment $\subs$ for $z_1, \dots, z_m$ such that $( \subs(z_i), \subs(z_j) ) \in R_{i,j}^\mathcal{C}$ for each $1 \leq i \leq j \leq m$.
Our aim here is to establish that $( \mathcal{C}^\rho, \subs ) \models_\mathsf{var} \psi$.

Let $\subs'$ be an extension of $\subs$ for the variables $y_1, \dots, y_\ell$, and note that since $C_{y_i}^\rho = \{ \ast_{y_i} \}$ for each $i < \ell$, we have that $\subs'(y_i) = \ast_{y_i}$ for each $i < \ell$.
Fix an arbitrary $k \in [m]$ such that $\subs'(y_\ell) = s_{z_k}$.
We define the witness for each $x_p$ with $p \in [n]$ as $(z_k, \subs(z_k), x_p)$.
Let $\subs''$ be the extension of $\subs'$ such that $\subs''(x_p) = (z_k, \subs(z_k), x_p)$ for each $p \in [n]$ and notice that it suffices to establish that $(\mathcal{C}^\rho, \subs'') \models_\mathsf{var} \chi$.
Furthermore, as $\chi$ is a conjunction of atomic formulae, it suffices to check each such atomic formula.

\begin{itemize}

\item For $E(x_p, x_q) \in \chi$ with $p,q \in [n]$, we have that
\[ (\subs''(x_p), \subs''(x_q) ) = ( (z_k, \subs(z_k), x_p ), (z_k, \subs(z_k), x_q) ) \in E^{\mathcal{C}^\rho} \]
 by the definition of $E^{\mathcal{C}^\rho}$.

\item For $P(y_\ell, x_p) \in \chi$ with $p \in [n]$, we have that
\[ (\subs''(y_\ell), \subs''(x_p) ) = ( s_{z_k}, (z_k,  \subs(z_k), x_p) ) \in P^{\mathcal{C}^\rho} \]
 by the definition of $P^{\mathcal{C}^\rho}$.

\item For $S(z_i, x_p) \in \chi$ with $i \in [m]$ and $p \in [n]$, we have that
\[ ( \subs''(z_i), \subs''(x_p) ) = ( \subs(z_i), (z_k, \subs(z_k), x_p ) ) \in S^{\mathcal{C}^\rho} \]
 due to the following reasoning:
\begin{itemize}
\item If $i \leq k$, then  $(\subs(z_i), (z_k, \subs(z_k), x_p) ) \in S^{\mathcal{C}^\rho}$ by the definition of $S^{\mathcal{C}^\rho}$ along with the fact that $(\subs(z_i), \subs(z_k)) \in R_{i,k}^\mathcal{C}$.
\item If $i>k$, then $(\subs(z_k), \subs(z_i) ) \in R_{k,i}^\mathcal{C}$ and therefore $(\subs(z_i), (z_k, \subs(z_k), x_p) ) \in S^{\mathcal{C}^\rho}$ by definition.
\end{itemize}

\item Finally, for any other $F(u,v) \in \chi$, we have that $( \subs''(u), \subs''(v)) \in F^{\mathcal{C}^\rho}$ since $F^{\mathcal{C}^\rho} = C_u^\rho \times C_v^\rho$.
\end{itemize}
Since $\subs'$ was chosen as an arbitrary extension of $\subs$, it follows that $({\mathcal{C}^\rho}, \subs) \models_\mathsf{var} \psi$.
\end{proof}

\begin{lemma}\label{lemma:contract-game-conversion}
$\mathcal{A} \equiv^k_\mathsf{var} \mathcal{B} \implies \mathcal{A}^\rho \equiv^k_\mathsf{var} \mathcal{B}^\rho$ for any two variable-sorted structures $\mathcal{A}, \mathcal{B}$ over the signature of $(\varphi_3)_\rho^\mathsf{ctr}$.
\end{lemma}
\begin{proof}
Let Duplicator use their winning strategy for the $k$-pebble game on $\mathcal A$ and $\mathcal B$ as a guide.
\begin{itemize}
\item If Spoiler plays an element $a\in A_u=A^\rho_u$ for some variable $u$ not quantified inside $\psi$, then Duplicator responds with the element
$b\in B_u=B^\rho_u$ using their winning strategy on $\mathcal A,\mathcal B$.

\item If Spoiler plays $(z_i,a,x_p)\in A^\rho_{x_p}$, Duplicator first treats the parameter $a\in A_{z_i}$ as a move in the game on $\mathcal A,\mathcal B$.
Let $b\in B_{z_i}$ be the response in the game over $\mathcal{A}$ and $\mathcal{B}$.
Duplicator then plays $(z_i,b,x_p)\in B^\rho_{x_p}$.

\item If Spoiler plays an element of the form $\ast_{y_t}$ or $s_{z_i}$, then Duplicator plays the identically named element on the other side.
\end{itemize}

The strategy for moves from $\mathcal B^\rho$ to $\mathcal A^\rho$ is symmetric.

We now show that every position reached by this strategy is a partial isomorphism.
Let $(\alpha^\rho,\beta^\rho)$ be a position reached in the $k$-pebble game on
$\mathcal A^\rho$ and $\mathcal B^\rho$ where Duplicator plays their above-defined strategy.

We define the \emph{projected position} $(\alpha,\beta)$ in the game over $\mathcal A$ and $\mathcal B$ as follows.
For every pebble $p$ in the domain of $\alpha^\rho$ and $\beta^\rho$ let:
\[
\alpha(p) \df
\begin{cases}
\alpha^\rho(p) & \text{if } \alpha^\rho(p)\in A_u \text{ for some variable } u
              \text{ not quantified inside } \psi,\\
a & \text{if } \alpha^\rho(p)=(z_i,a,x_q),\\
\text{undefined} & \text{if } \alpha^\rho(p)=\ast_{y_t} \text{ or } s_{z_i},
\end{cases}
\]
and analogously, let
\[
\beta(p) \df
\begin{cases}
\beta^\rho(p) & \text{if } \beta^\rho(p)\in B_u \text{ for some variable } u
              \text{ not quantified inside } \psi,\\
b & \text{if } \beta^\rho(p)=(z_i,b,x_q),\\
\text{undefined} & \text{if } \beta^\rho(p)=\ast_{y_t} \text{ or } s_{z_i}.
\end{cases}
\]

Ignoring pebbles whose projections are undefined, the projected pair $(\alpha,\beta)$ is a reachable position in the original $k$-pebble game when Duplicator follows their winning strategy.
Thus, $(\alpha,\beta)$ is a partial isomorphism.

We now prove that $(\alpha^\rho,\beta^\rho)$ is a partial isomorphism.
First, $\{ ( \alpha^\rho(p), \beta^\rho(p) ) \mid p \in \domain(\alpha^\rho) \}$ defines an injective partial mapping, since $(\alpha, \beta)$ is a partial isomorphism.
In particular, if two pebbled elements in $\mathcal A^\rho$ have the same projection in $\mathcal A$, then their responses have the same projection in $\mathcal B$.

Next, consider a relation symbol $Q$ that does not occur inside $\psi$.
By definition, $Q^{\mathcal A^\rho}=Q^{\mathcal A}$ and $Q^{\mathcal B^\rho}=Q^{\mathcal B}$.
Thus, $Q$ is dealt with by the fact that $(\alpha,\beta)$
is a partial isomorphism between $\mathcal A$ and $\mathcal B$.

It remains to consider relation symbols occurring inside $\psi$.
To that end, let $p$ and $p'$ be two arbitrary pebbles in $\domain(\alpha^\rho)$.

\begin{itemize}
\item Consider $P(y_\ell,x_q)\in \chi$. If $(\alpha^\rho(p),\alpha^\rho(p'))\in P^{\mathcal A^\rho}$,
then, by definition of $P^{\mathcal A^\rho}$, there are $i \in [m]$ and
$a\in A_{z_i}$ such that
\[ \alpha^\rho(p)=s_{z_i} \qquad\text{and}\qquad \alpha^\rho(p')=(z_i,a,x_q). \]
Immediately from Duplicator's strategy, we have that
\[ \beta^\rho(p)=s_{z_i} \qquad\text{and}\qquad \beta^\rho(p')=(z_i,b,x_q) \]
for some $b\in B_{z_i}$.
Hence, $(\beta^\rho(p),\beta^\rho(p'))\in P^{\mathcal B^\rho}$.
The converse direction is symmetric.

\item Next, consider $E(x_p,x_q)\in\chi$.
If $(\alpha^\rho(p),\alpha^\rho(p'))\in E^{\mathcal A^\rho}$, then there are $i \in [m]$ and $a \in A_{z_i}$ such that
\[ \alpha^\rho(p)=(z_i,a,x_p) \qquad\text{and}\qquad \alpha^\rho(p')=(z_i,a,x_q). \]
Immediately from Duplicator's strategy, we have that
\[ \beta^\rho(p)=(z_i,b,x_p) \qquad\text{and}\qquad \beta^\rho(p')=(z_i,b,x_q), \]
and hence $ (\beta^\rho(p),\beta^\rho(p'))\in E^{\mathcal B^\rho}$.
Again, the converse direction is symmetric.

\item Now, consider $S(z_i,x_q)\in\chi$.
Assume $(\alpha^\rho(p),\alpha^\rho(p'))\in S^{\mathcal A^\rho}$.
Then $\alpha^\rho(p)=a\in A_{z_i}$ and $\alpha^\rho(p')=(z_j,a',x_q)$ for some $j \in [m]$ and $a'\in A_{z_j}$.
Further suppose that $i \leq j$, the case where $i > j$ is immediate.
By definition of $S^{\mathcal{A}^\rho}$, we have $(a, a') \in (R^\rho_{z_i, z_j})^{\mathcal A}$.
Since $(\alpha,\beta)$ is a partial isomorphism between $\mathcal A$ and
$\mathcal B$, and since $R^\rho_{z_i,z_j}$ is a relation symbol of
$\varphi'$, we get
$(b,b')\in (R^\rho_{z_i,z_j})^{\mathcal B}$.
Hence
$(b,(z_j,b',x_q))\in S^{\mathcal B^\rho}$.
The converse implication, from $\mathcal B^\rho$ to $\mathcal A^\rho$, is symmetric.
Therefore, $S$ is dealt with.

\item Every remaining relation symbol $F$, with $F(u,v)\in\chi$, is interpreted as $F^{\mathcal{A}^\rho} \df A_{u}^\rho \times A_v^\rho$, and likewise, $F^{\mathcal{B}^\rho} \df B_u^\rho \times B_v^\rho$.
Thus, it is immediate that such a relation symbol $F$ is dealt with.
\end{itemize}

Hence, every position reachable by Duplicator's strategy is a partial isomorphism between $\mathcal A^\rho$ and $\mathcal B^\rho$.
Therefore, Duplicator has a winning strategy in the $k$-pebble game over $\mathcal A^\rho$ and $\mathcal B^\rho$, and so $\mathcal A^\rho \equiv^k_\mathsf{var} \mathcal B^\rho $.
\end{proof}

\begin{notation}
Let $\varphi_4$ be obtained by iteratively and exhaustively contracting all lowest $\{ \forall \}$-regions occurring below $\bar{\theta}$.
\end{notation}

Thus, in $\varphi_4$, our distinguished region with maximum treewidth $\bar\theta$ only has atomic child regions.

\begin{lemma}\label{lemma:induct-lowest}
Let $\Phi$ be an intermediate formula obtained from $\varphi_3$ by contracting some collection of lowest $\{\forall\}$-regions that occur below $\bar{\theta}$.
Then, for every lowest $\{\forall\}$-region $\rho$ of $\Phi$ occurring below $\bar\theta$, the formula $\mathsf{sub}_{\Phi}(\rho)$ has the form $\rho \llbracket \chi \rrbracket$ where $\chi$ is an $\{\exists,\land\}$-region such that:
\begin{enumerate}
\item the top node of the parse tree of $\chi$ is an existential quantifier;
\item all child regions of $\chi$ are atomic;
\item $\chi$ has the connectivity property; and
\item $\chi$ has the clasped condition.
\end{enumerate}
\end{lemma}
\begin{proof}
We proceed by induction on the number of contractions used to obtain $\Phi$ from $\varphi_3$.

The base case is \cref{lemma:lowest-forall}.
Now suppose the claim holds for an intermediate formula $\Phi$, and let $\rho$
be a lowest $\{\forall\}$-region of $\Phi$ that occurs below $\bar{\theta}$.
Let $\Phi' \df \Phi^{\mathrm{ctr}}_{\rho}$.
We show that every subformula of $\mathsf{sub}_{\Phi'}(\bar\theta)$ rooted at a lowest $\{\forall\}$-region has the required form.

The contraction of $\rho$ replaces the corresponding subformula by $\bigwedge_{1 \leq i \leq j \leq m} R^{\rho}_{z_i,z_j}(z_i,z_j)$, where $z_1,\ldots,z_m$ are the free variables of $\rho$.
In particular, the contraction neither introduces nor removes any quantifiers.

Let $\eta$ be a lowest $\{\forall\}$-region of $\Phi'$ that occurs in $\bar\theta$.
There are two cases:
\begin{enumerate}
\item Suppose that $\eta$ already occurred as a lowest $\{\forall\}$-region of $\Phi$.
Then, $\eta$ is disjoint from the contracted occurrence of $\rho$.
Therefore, the contraction does not change $\eta$, and the desired properties follow from the induction hypothesis.

\item Suppose that $\eta$ was not lowest in $\Phi$.
Since the contraction creates no new $\{\forall\}$-regions, the region $\eta$ must already have occurred in $\Phi$, and it becomes lowest in $\Phi'$ because $\rho$ occurs below $\eta$ and was contracted.
The subformula rooted at $\eta$ is $\eta \llbracket \chi \rrbracket$, where $\chi$ is the unique child $\{\exists,\land\}$-region of $\eta$.
The top node of $\chi$ is still an existential quantifier, since contraction only replaces proper $\{\forall\}$-organised subformulae of $\chi$ by a conjunction of atomic formulae.
Moreover, since $\eta$ is lowest in $\Phi'$, every child region of $\chi$ in $\Phi'$ is atomic.

Consider the effect of contracting a proper lowest $\{\forall\}$-subregion $\omega$ below $\chi$ with $\mathsf{free}(\omega) = \{ z_1, \dots, z_m \}$.
Before the contraction, this child region contributes the hyperedge $\{z_1,\ldots,z_m\}$ to the hypergraph of $\chi$.
After the contraction, it is replaced by $\bigwedge_{1 \leq i \leq j \leq m} R^{\omega}_{z_i,z_j}(z_i,z_j)$.
Thus, in the primal graph of $H_\chi$, the contraction replaces the clique on $\{z_1,\ldots,z_m\}$ by the same clique.
Consequently, the primal graph of $H_\chi$ is unchanged by the contraction and thus, the connectivity property and the clasped condition are preserved.
\end{enumerate}

Therefore, every lowest $\{ \forall \}$-region occurring below $\bar\theta$ has the required form, completing the induction.
\end{proof}

By the above lemma, at every intermediate step of constructing $\varphi_4$ from $\varphi_3$, every subformula of $\bar\theta$ rooted at a lowest $\{ \forall \}$-region has the correct form for invoking~\cref{lemma:contract-equiv} and~\cref{lemma:contract-game-conversion}.

\begin{lemma}\label{lemma:complex-contractions}
If there exist variable-sorted structures $\mathcal{A}$ and $\mathcal{B}$ such that:
\[ \mathcal{A} \models_\mathsf{var} \varphi_4, \quad \mathcal{B} \not\models_\mathsf{var} \varphi_4, \quad \mathcal{A} \equiv^k_\mathsf{var} \mathcal{B}, \]
then, there exist variable-sorted structures $\mathcal{A}'$ and $\mathcal{B}'$ such that
\[ \mathcal{A}' \models_\mathsf{var} \varphi_3, \quad \mathcal{B}' \not\models_\mathsf{var} \varphi_3, \quad \mathcal{A}' \equiv^k_\mathsf{var} \mathcal{B}' \]
\end{lemma}
\begin{proof}
Let $\varphi_3 = \Phi_0 , \Phi_1, \dots, \Phi_r = \varphi_4$ be the sequence of QCQs obtained during the construction of $\varphi_4$.
That is, for each $i \in [r]$, there is some lowest $\{\forall\}$-region $\rho_{i-1}$ of $\Phi_{i-1}$ that occurs below $\bar\theta$
and where $\Phi_i = (\Phi_{i-1})^\mathsf{ctr}_{\rho_{i-1}}$.

By~\cref{lemma:induct-lowest}, each $\mathsf{sub}_{\Phi_{i-1}} (\rho_{i-1})$ has the correct form to invoke~\cref{lemma:contract-equiv} and~\cref{lemma:contract-game-conversion}.
Assume there exists $\mathcal{A}^r$ and $\mathcal{B}^r$ such that
\[  \mathcal{A}^r \models_\mathsf{var} \Phi_r , \quad \mathcal{B}^r \not\models_\mathsf{var} \Phi_r, \quad \mathcal{A}^r \equiv^k_\mathsf{var} \mathcal{B}^r . \]
For each $0 \leq i < r$, let $\mathcal{A}^i \df (\mathcal{A}^{i+1})^{\rho_{i}}$ be obtained using the expansion construction associated with the contraction of $\rho_{i}$.
Let $\mathcal{B}^i \df (\mathcal{B}^{i+1})^{\rho_{i}}$ be defined analogously.
By~\cref{lemma:contract-equiv} and~\cref{lemma:contract-game-conversion}, for each $i$ from $r-1$ to $0$, we have that:
\[ \mathcal{A}^{i+1} \models_\mathsf{var} \Phi_{i+1} , \quad \mathcal{B}^{i+1} \not\models_\mathsf{var} \Phi_{i+1} , \quad \mathcal{A}^{i+1} \equiv^k_\mathsf{var} \mathcal{B}^{i+1} \]
implies
\[ \mathcal{A}^{i} \models_\mathsf{var} \Phi_{i} , \quad \mathcal{B}^{i} \not\models_\mathsf{var} \Phi_{i} , \quad \mathcal{A}^{i} \equiv^k_\mathsf{var} \mathcal{B}^{i} . \]
Thus, by induction, we have established our desired claim.
\end{proof}

\begin{lemma}\label{lemma:reduction-three}
If there exist two variable-sorted structures $\mathcal{A}$ and $\mathcal{B}$ such that
\[ \mathcal{A} \models_\mathsf{var} \varphi_4 , \quad \mathcal{B} \not\models_\mathsf{var} \varphi_4 , \quad \mathcal{A} \equiv^k_\mathsf{var}  \mathcal{B}, \]
then $\mathsf{logwidth}(\varphi) > k$.
\end{lemma}
\begin{proof}
This follows by combining~\cref{lemma:complex-contractions} and~\cref{lemma:reduction-two}.
\end{proof}

So far, we have reduced our QCQ $\varphi$ to $\varphi_4$ where every atom in $\varphi_4$ has an arity of at most size two, and where we have a $\{ \exists, \land \}$-region $\bar\theta$ of $\varphi_4$ such that:
\begin{itemize}
\item Every child region of $\bar\theta$ is an atomic formula;
\item $\mathsf{tw}(H_{\bar\theta}) = \mathsf{tw}(H_\theta)$;
\item $\bar\theta$ has the connectivity property; and
\item $\bar\theta$ has the clasped condition.
\end{itemize}

\subsection{Ignoring Some Atomic Formulae}\label{subsec:red-4}
\begin{definition}
Let $(\tau, \mathcal{V})$ be a variable-sorted signature, and let $\tau = X \sqcup Y$.

For a variable-sorted structure $\mathcal{A}$ over $(\tau, \mathcal{V})$, let $\mathcal{A} |_Y$ be the $(Y, \mathcal{V})$-structure with the same domains for each variable as $\mathcal{A}$, and consists of a relation $R^{\mathcal{A} |_Y} \df R^\mathcal{A}$ for every $R \in Y$.

For a QCQ $\varphi$, we define $\varphi |_Y$ recursively as follows:
\begin{itemize}
\item If $R \in Y$, then $R(\bar z) |_Y \df R(\bar z)$, and if $R \in X$, then $R(\bar z) |_Y \df \top$.

\item $(\alpha \land \beta) |_Y \df ( \alpha |_Y \land \beta |_Y )$ for QCQs $\alpha, \beta$

\item $( Qx \alpha) |_Y \df Qx ( \alpha |_Y)$ for a QCQ $\alpha$ and $Q \in \{ \forall, \exists \}$.
\end{itemize}
Then, we remove $\top$ by first exhaustively applying $Qx \top = \top$ and then exhaustively applying $\alpha\land \top = \top \land \alpha = \alpha$.
\end{definition}

Next, we show that we can simply ``ignore'' certain relation symbols in the following way:
\begin{lemma}\label{lemma:forget-relations}
Let $(\tau, \mathcal{V})$ be a variable-sorted signature, and let $\tau = X \sqcup Y$.
Let $\mathcal{A}$ and $\mathcal{B}$ be variable-sorted $(\tau, \mathcal{V})$-structures where for every $R \in X$ with $\mathsf{ar}(R) = s_1 \cdots s_m$, we have that $R^\mathcal{A} = A_{s_1} \times \cdots \times A_{s_m}$ and $R^\mathcal{B} = B_{s_1} \times \cdots \times B_{s_m}$.
Let $\phi$ be a QCQ over $(\tau, \mathcal{S})$.
Then $\mathcal{C} \models_\mathsf{var} \phi$ if and only if $\mathcal{C} |_Y \models_\mathsf{var} \phi |_Y$ for $\mathcal{C} \in \{ \mathcal{A}, \mathcal{B} \}$.
Moreover,  $\mathcal{A} |_Y \equiv^k_\mathsf{var} \mathcal{B} |_Y$ implies $\mathcal{A} \equiv^k_\mathsf{var} \mathcal{B}$.
\end{lemma}
\begin{proof}
We prove a slightly stronger claim by induction on the structure of $\phi$:
for every $\mathcal C \in \{\mathcal A,\mathcal B\}$, every subformula $\phi'$ of $\phi$, and every assignment
$\sigma$ of the free variables of $\phi'$ to elements of their respective domains, we have
\[ (\mathcal C,\sigma)\models_{\mathsf{var}}\phi' \iff (\mathcal C|_Y,\sigma)\models_{\mathsf{var}}\phi' |_Y . \]

The atomic case is immediate.
If the atom is $R(z_1, \dots, z_m)$ with $R\in X$, then it is true under every assignment (with the correct sorts) because $R^\mathcal C \df C_{z_1} \times \cdots \times C_{z_m}$; hence:
\[ (\mathcal{C}, \subs) \models_\mathsf{var} R(z_1, \dots, z_m ) \iff (\mathcal{C} |_Y, \subs) \models_\mathsf{var} \top. \]
If $R\in Y$, then the interpretation of $R$ is unchanged, and therefore
\[  (\mathcal{C}, \subs) \models_\mathsf{var} R(z_1, \dots, z_m ) \iff (\mathcal{C} |_Y, \subs) \models_\mathsf{var} R(z_1, \dots, z_m) \]
is immediate.
The conjunction and quantifier cases follow directly from the induction hypothesis, using the fact that $\mathcal C$ and $\mathcal C|_Y$ have the same domains for each variable.

The rest of the proof is dedicated to showing $\mathcal{A} |_Y \equiv^k_\mathsf{var} \mathcal{B} |_Y$ implies $\mathcal{A} \equiv^k_\mathsf{var} \mathcal{B}$.
Let $P$ be the set of pebbles and let $\alpha \colon P \rightarrow A$ and $\beta \colon P \rightarrow B$ be partial mappings.
We show that $(\alpha, \beta)$ is a partial isomorphism between $\mathcal{A} |_Y$ and $\mathcal{B} |_Y$ implies $(\alpha, \beta)$ is a partial isomorphism between $\mathcal{A}$ and $\mathcal{B}$.
Thus, Duplicator can simply copy their winning strategy over $\mathcal{A} |_Y$ and $\mathcal{B} |_Y$ to get a winning strategy between $\mathcal{A}$ and $\mathcal{B}$.

Suppose $(\alpha, \beta)$ is a partial isomorphism between $\mathcal{A} |_Y$ and $\mathcal{B} |_Y$.
Pick some arbitrary $F \in Y$, and since $F^\mathcal{A} = F^{\mathcal{A} |_Y}$, along with $F^\mathcal{B} = F^{\mathcal{B} |_Y}$, clearly
\[ \alpha( \bar p)  \in F^\mathcal{A} \iff  \beta( \bar p)  \in F^\mathcal{B} . \]
for any $\bar p \in \domain(\alpha)^{|\mathsf{ar}(F)|}$.

It remains to check relation symbols in $X$.
Pick some arbitrary $R \in X$.
Suppose $p_{i_1}, \dots, p_{i_\ell} \in \domain(\alpha)$ such that $\alpha(p_{i_j}) \in A_{s_j}$, and $\beta(p_{i_j}) \in B_{s_j}$ and $\mathsf{ar}(R) = s_1 \cdots s_\ell$, and notice that any set of pebbles from $\domain(\alpha)$ where the previously mentioned condition does not hold cannot describe a tuple from $R^\mathcal{A}$ or $R^\mathcal{B}$.
Clearly, $(\alpha(p_{i_1}), \dots, \alpha(p_{i_\ell}) ) \in R^\mathcal{A}$ and $( \beta(p_{i_1}), \dots, \beta(p_{i_\ell}) ) \in R^\mathcal{B}$ for any such $p_{i_1}, \dots, p_{i_\ell} \in P$.
This is because $R^\mathcal{A} = A_{s_1} \times \cdots \times A_{s_\ell}$ and $R^\mathcal{B} = B_{s_1} \times \cdots \times B_{s_\ell}$.
Thus, $(\alpha, \beta)$ is a partial isomorphism between $\mathcal{A}$ and $\mathcal{B}$.
\end{proof}

\begin{notation}
Let $\tau \df X_{\bar\theta} \sqcup X_\mathsf{else}$ be a partition, where $X_{\bar\theta}$ contains all relation symbols $R$, such that $R(u,v)$ occurs in $\bar\theta$ (up to reordering of the variables) where $u \neq v$ and:
\begin{enumerate}
\item $u \in \mathsf{free}(\bar\theta)$ and $v \notin \mathsf{free}(\bar\theta)$, or
\item $u, v \notin \mathsf{free}(\bar\theta)$.
\end{enumerate}

Then, set $\varphi_4' \df \varphi_4 |_{X_{\bar\theta}}$, and then $\varphi_5$ is obtained from $\varphi_4'$ by exhaustive applications of $Qx \alpha \rightarrow_M \alpha$ where $x \notin \mathsf{free}(\alpha)$.
\end{notation}

We say $Q x \alpha$ where $x \notin \mathsf{free}(\alpha)$ and $Q \in \{ \exists, \forall \}$ is a \emph{vacuous quantifier}.

\begin{lemma}\label{lemma:ignore-relations-translation}
If there exist variable-sorted structures $\mathcal{A}$ and $\mathcal{B}$ such that
\[ \mathcal{A} \models_\mathsf{var} \varphi_5 , \quad \mathcal{B} \not\models_\mathsf{var} \varphi_5, \quad \mathcal{A} \equiv^k_\mathsf{var} \mathcal{B} ,\]
then, there exist variable-sorted structures $\mathcal{C}$ and $\mathcal{D}$ such that
\[ \mathcal{C} \models_\mathsf{var} \varphi_4 , \quad \mathcal{D} \not\models_\mathsf{var} \varphi_4, \quad \mathcal{C} \equiv^k_\mathsf{var} \mathcal{D} . \]
\end{lemma}
\begin{proof}
Suppose that there exist variable-sorted structures $\mathcal{A}$ and $\mathcal{B}$ such that $\mathcal{A} \models_\mathsf{var} \varphi_5$, $\mathcal{B} \not\models_\mathsf{var} \varphi_5$, and $\mathcal{A} \equiv^k_\mathsf{var} \mathcal{B} $.

Let $\mathsf{Rem}$  be the set of variables whose quantification is removed in the reduction from $\varphi_4'$ to $\varphi_5$.
Let $\mathcal A^+$ and $\mathcal B^+$ be variable-sorted structures defined from $\mathcal{A}$ and $\mathcal{B}$ by keeping the same domains and relation interpretations on the variables and relation symbols occurring in $\varphi_5$, and by setting
$A_u^+ \df \{a_u^\ast\}$ and $B_u^+ \df \{b_u^\ast\}$ for each $u \in \mathsf{Rem}$.
For every $u \in \mathsf{Rem}$, we have that $A_u^+ \neq \emptyset \neq B_u^+$ and therefore adding vacuous quantifiers does not affect truth value, implying:
\[ \mathcal{A}^+ \models_\mathsf{var} \varphi_4' , \quad \mathcal{B}^+ \not\models_\mathsf{var} \varphi_4' , \quad \mathcal{A}^+ \equiv^k_\mathsf{var} \mathcal{B}^+ . \]
Here, Duplicator's strategy for the $k$-pebble game over $\mathcal{A}^+$ and $\mathcal{B}^+$ is identical to their $k$-pebble strategy over $\mathcal{A}$ and $\mathcal{B}$, unless Spoiler picks an element from $A_u^+$ (or $B_u^+$) for some $u \in \mathsf{Rem}$, in which case Duplicator responds with the unique element in the corresponding domain.

Now define $\mathcal C$ and $\mathcal D$ as extensions of $\mathcal A^+$ and $\mathcal B^+$ to the signature $\tau$. For every relation symbol
$R\in X_\mathsf{else}$ with $\mathsf{ar}(R)=s_1\cdots s_m$, set
\[
R^\mathcal C \df C_{s_1}\times\cdots\times C_{s_m},
\qquad
R^\mathcal D \df D_{s_1}\times\cdots\times D_{s_m}.
\]
For every $R\in X_{\bar\theta}$, let $R^\mathcal C=R^{\mathcal A^+}$ and
$R^\mathcal D=R^{\mathcal B^+}$.

Then $\mathcal C|_{X_{\bar\theta}}=\mathcal A^+$ and $\mathcal D|_{X_{\bar\theta}}=\mathcal B^+$. Hence, by \cref{lemma:forget-relations}, we get
\[ \mathcal C \models_\mathsf{var} \varphi_4 \iff \mathcal A^+ \models_\mathsf{var} \varphi_4', \]
and
\[ \mathcal D \models_\mathsf{var} \varphi_4 \iff \mathcal B^+ \models_\mathsf{var} \varphi_4'. \]
Therefore
\[ \mathcal C \models_\mathsf{var} \varphi_4, \quad \mathcal D \not\models_\mathsf{var} \varphi_4.
\]
Since $\mathcal A^+ \equiv^k_\mathsf{var} \mathcal B^+$, the ``moreover'' part of \cref{lemma:forget-relations} gives us $\mathcal{C} \equiv^k_\mathsf{var} \mathcal{D}$.
\end{proof}

\begin{lemma}\label{lemma:reduction-four}
If there exist two variable-sorted structures $\mathcal{A}$ and $\mathcal{B}$ such that
\[ \mathcal{A} \models_\mathsf{var} \varphi_5 , \quad \mathcal{B} \not\models_\mathsf{var} \varphi_5 , \quad \mathcal{A} \equiv^k_\mathsf{var}  \mathcal{B}, \]
then $\mathsf{logwidth}(\varphi) > k$.
\end{lemma}
\begin{proof}
This follows by combining~\cref{lemma:ignore-relations-translation} and~\cref{lemma:reduction-three}.
\end{proof}

Recall that our original QCQ $\varphi$ is a normal form of $\mathscr{R} \df (\mathsf{QCQ} , \rightarrow_{\{ S \downarrow, M \}} ) / \overset{*}{\leftrightarrow}_{\mathbb{T}}$, and contains the distinguished non-top $\{ \exists, \land \}$-region $\theta$  of maximum treewidth.

\begin{lemma}
$\varphi_5$ is a simple QCQ, with an inner region $\bar\theta'$, where $\mathsf{tw}(H_{\bar\theta'}) = \mathsf{tw}(H_{\theta})$.
\end{lemma}
\begin{proof}
Let $\bar\theta$ denote the region of $\varphi_4$ corresponding to the distinguished region $\theta$ of $\varphi$.
Recall that $\varphi_5$ is obtained from $\varphi_4$ by deleting all atoms whose relation symbols lie outside of $X_{\bar\theta}$ and then removing vacuous quantifiers.
Let $\bar\theta'$ be the region induced by $\bar \theta$ after this deletion.

First, let us verify that $\varphi_5$ is a simple QCQ.
Since $\varphi$ is self-join-free and all reductions used to obtain $\varphi_5$ do not introduce repeating relation symbols, $\varphi_5$ is also self-join-free.
Moreover, by the definition of $X_{\bar\theta}$, every atom that remains in $\varphi_5$ is a binary atom applied to two distinct variables.
This is because all atoms that are nullary, unary, or of the form $R(u,u)$ are
deleted, and all atoms with an arity of more than two have been replaced by
their binary relation encoding.

All atoms outside of $\bar\theta$ are deleted when passing from $\varphi_4$ to $\varphi_5$.
Thus, after removing vacuous quantifiers, the remaining QCQ has the form
\[ Q_1 v_1 \cdots Q_{m-1} v_{m-1} \forall v_m \exists x_1 \cdots \exists x_k (\psi), \]
where $\exists x_1 \cdots \exists x_k(\psi)$ is the subformula rooted at $\bar\theta'$.
Since $\theta$ is a non-top region, $\bar\theta$ must have at least one free variable.
Moreover, $[\varphi]_{\mathbb{T}}$ is a normal form of $\mathscr{R}$, and thus no quantifier in $\varphi$ is vacuous.
Hence, when obtaining $\varphi_5$, at least one universal quantifier precedes $\bar\theta'$.
Thus, $m \geq 1$.

Let us now check the connectivity property and the tightness property.
The connectivity property holds because when passing from $\varphi_4$ to $\varphi_5$, we do not remove atoms whose two distinct variables are quantified in $\bar\theta$.
Therefore, the hypergraph induced by $H_{\bar\theta'}$ on the variables quantified in $\bar\theta'$ is isomorphic to the corresponding hypergraph induced by $H_{\bar\theta}$ on the variables quantified in $\bar\theta$ and this latter hypergraph is connected.
For the tightness property, let $e$ be an arbitrary hyperedge of $H_{\bar\theta'}$.
If this hyperedge corresponds to an atom, then this atom has not been deleted to obtain $\varphi_5$ and thus contains at least one variable that is existentially quantified in $\bar\theta'$.

Next, let us show that $\mathsf{tw}(H_{\bar\theta}) = \mathsf{tw}(H_\theta)$.
The transformations from $\theta$ to $\bar\theta$ preserve the primal graph of the region:
contracting subregions replaces a child with a conjunction of binary atoms, inducing the same clique on its associated variables, and the binary encoding of higher-arity atoms replace the clique induced by that atom with the same clique.
Therefore, $\mathsf{tw}(H_{\bar\theta}) = \mathsf{tw}(H_\theta)$.

Now consider passing from $\bar\theta$ to $\bar\theta'$.
This deletes atoms outside of $\bar\theta$, nullary and unary atoms, as well as atoms of the form $R(u,u)$, and atoms $R(u,v)$ with $u, v \in \mathsf{free}(\bar\theta)$.
Deleting the first three types of atoms clearly does not affect the primal graph, and the last kind does not affect it because $\mathsf{free}(\bar\theta)$ is already a hyperedge.
Therefore, the primal graph, and therefore the treewidth, is unchanged when passing from $\bar\theta$ to $\bar\theta'$.

Consequently, $\varphi_5$ is a simple QCQ with an inner region $\bar\theta'$
with $\mathsf{tw}(H_{\bar\theta'}) = \mathsf{tw}(H_\theta)$.
\end{proof}

\begin{lemma}\label{lemma:final-reduction}
Let $\bar\theta'$ be the inner region of $\varphi_5$.
There exist two variable-sorted structures $\mathcal{A}$ and $\mathcal{B}$ such that $\mathcal{A}\models_\mathsf{var} \varphi_5$, $\mathcal{B}\not\models_\mathsf{var} \varphi_5$, and $\mathcal{A} \equiv^{\rho}_\mathsf{var} \mathcal{B}$, where $\rho = \mathsf{tw}(H_{\bar\theta'})$.
\end{lemma}
This immediately follows from~\cref{lemma:disagree} and~\cref{lemma:winning-strat}.
Combining this with~\cref{lemma:reduction-four}:

\begin{theorem}
$\mathsf{logwidth}(\varphi) > \mathsf{tw}(H_\theta)$.
\end{theorem}

Recall that we fixed $\theta$ to be a non-top $\{ \exists, \land \}$-region of $\varphi$.
Thus, all that remains is to deal with the case where the maximum treewidth region is a top $\{ \exists , \land \}$-region.

\subsection{Edge Case: $\theta$ is a top region of $\varphi$}\label{section:edge-case}
Let $\varphi$ be an arbitrary standardised self-join-free QCQ such that $[\varphi]_{\mathbb{T}}$ is a normal form in the system $\mathscr{R}$.
Now consider the remaining case in which the maximum-treewidth $\{ \exists, \land\}$-region $\theta$ is a top region.
Since $\theta$ is top, $\mathsf{free}(\theta) = \emptyset$, thus the hyperedges of $H_\theta$ are determined precisely by the association with the holes that appear in $\theta$.
We apply to $\theta$ the same transformations as
in~\cref{subsec:red-1}-\ref{subsec:red-3}.
That is,
\begin{enumerate}
\item Contract $\{ \forall\}$-subregions of $\theta$ that have an atomic child region.
\item Replace higher-arity atoms by their binary encoding.
\item Contract the remaining $\{ \forall \}$-subregions of $\theta$.
\end{enumerate}

By the same primal graph-preservation argument as in~\cref{obs:iso-primal-graphs}, the resulting existential-positive sentence $\varphi'$ has an associated hypergraph $H_{\varphi'}$ with $\mathsf{tw}(H_\theta) = \mathsf{tw}(H_{\varphi'})$.
The formula $\varphi'$ is in the correct form to invoke Theorem 6 of~\cite{BC19}.
Hence, there exist finite structures $\mathcal{A}$ and $\mathcal{B}$ such that $\mathcal{A} \models \varphi'$, $\mathcal{B} \not\models \varphi'$ and $\mathcal{A} \equiv^k \mathcal{B}$ with $k = \mathsf{tw}(H_{\varphi'})$.
Equivalently, $\mathsf{logwidth}(\varphi') \geq \mathsf{tw}(H_{\varphi'})+1$.

It remains to lift this lower bound back to $\varphi$.
We start with the following:
\begin{lemma}
\label{lemma:var-sorted-from-one}
Let $\varphi$ be a standardised and self-join-free QCQ.
If there exist one-sorted structures $\mathcal{A}$ and $\mathcal{B}$ such that
\[ \mathcal{A} \models \varphi , \quad \mathcal{B} \not\models \varphi, \quad \mathcal{A} \equiv^k \mathcal{B} ,\]
then, there exist variable-sorted structures $\mathcal{A}'$ and $\mathcal{B}'$ such that
\[ \mathcal{A}' \models_\mathsf{var} \varphi, \quad \mathcal{B}' \not\models_\mathsf{var} \varphi , \quad \mathcal{A}' \equiv^k_\mathsf{var} \mathcal{B}' . \]
\end{lemma}
\begin{proof}
Let $A$ be the domain of $\mathcal{A}$ and let $B$ be the domain of $\mathcal{B}$.
For each $x$ that appears in $\varphi$, let $A_x' \df  A \times \{ x \}$ and let $B_x' \df B \times \{ x \}$ be the domains for $x$ in $\mathcal{A}'$ and $\mathcal{B}'$, respectively.
For each atom $R(x_1, \dots, x_r)$ that appears in $\varphi$, let
\[  ( (a_1, x_1), \dots, (a_r, x_r) ) \in R^{\mathcal{A}'} \iff (a_1, \dots, a_r) \in R^\mathcal{A}.  \]
The relation symbols in $\mathcal{B}'$ are interpreted analogously.
Relation symbols that do not appear in $\varphi$ are interpreted arbitrarily.

Immediately, we get $\mathcal{C} \models \varphi$ if and only if $\mathcal{C}' \models_\mathsf{var} \varphi$ for $\mathcal{C} \in \{ \mathcal{A}, \mathcal{B} \}$.
This is because a variable assignment $\subs$ is uniquely coupled with $\subs'$ such that $\subs(x) =a$ if and only if $\subs'(x) = (a,x)$.
Thus, any subformula of $\varphi$ is true under $\subs$ if and only if it is
true under $\subs'$, by the definition of relations in $\mathcal{C}'$.

Moreover, $\mathcal{A} \equiv^k \mathcal{B}$ implies $\mathcal{A}' \equiv^k \mathcal{B}'$ by augmenting Duplicator's strategy with responding in the corresponding domain.
That is, if Spoiler picks some $(a,x)$ in the game over $\mathcal{A}'$ and
$\mathcal{B}'$, Duplicator consults their winning strategy over $\mathcal{A}$
and $\mathcal{B}$ with Spoiler picking $a$.
If Duplicator's one-sorted winning strategy says to respond with $b$, then
Duplicator responds in the variable-sorted game with $(b,x)$.
\end{proof}

By~\cref{lemma:var-sorted-from-one}, we may assume Theorem 6 of~\cite{BC19} yields variable-sorted structures $\mathcal{A}$ and $\mathcal{B}$ with $\mathcal{A} \models_\mathsf{var} \varphi'$, $\mathcal{B} \not\models_\mathsf{var} \varphi'$ and $\mathcal{A} \equiv^k_\mathsf{var} \mathcal{B}$; where $k= \mathsf{tw}(H_{\varphi'})$.

For each transformation used to obtain $\varphi'$, apply the corresponding lemma
from~\cref{subsec:red-1}-\ref{subsec:red-3} to obtain structures $\mathcal{A}'$
and $\mathcal{B}'$ such that $\mathcal{A}' \models_\mathsf{var} \varphi$,
$\mathcal{B}' \not\models_\mathsf{var} \varphi$, and $\mathcal{A}'
\equiv^k_\mathsf{var} \mathcal{B}'$.
Thus, $\mathsf{logwidth}(\varphi) > k = \mathsf{tw}(H_\theta)$.

\end{document}